\documentclass[twocolumn,tighten]{aastex631}
\usepackage{color}

\begin{document}

\shortauthors{Luhman}

\shorttitle{Census of the $\beta$ Pic Moving Group and Other Associations}

\title{A Census of the $\beta$ Pic Moving Group and Other Nearby Associations 
with Gaia\footnote{Based on observations made with the Gaia 
mission, the Two Micron All Sky Survey, the Wide-field Infrared Survey Explorer,
the NASA Infrared Telescope Facility, Cerro Tololo Inter-American Observatory,
the Southern Astrophysical Research Telescope, Gemini Observatory, the 
and the eROSITA instrument on the Spectrum-Roentgen-Gamma mission.}}

\author{K. L. Luhman}
\affiliation{Department of Astronomy and Astrophysics,
The Pennsylvania State University, University Park, PA 16802, USA;
kll207@psu.edu}
\affiliation{Center for Exoplanets and Habitable Worlds, The
Pennsylvania State University, University Park, PA 16802, USA}

\begin{abstract}

I have used the third data release of the Gaia mission
to improve the reliability and completeness of membership samples in 
the $\beta$~Pic moving group (BPMG) and other nearby associations with ages 
of 20--50 Myr (Sco Body, Carina, Columba, $\chi^1$ For, Tuc-Hor, IC 2602,
IC 2391, NGC 2547). I find that Carina, Columba, and $\chi^1$ For
are physically related and coeval, and that Carina is the
closest fringe of a much larger association. Similarly, Tuc-Hor and IC 2602
form a coeval population that is spatially and kinematically continuous.
Both results agree with hypotheses from \citet{gag21}.
I have used the new catalogs to study the associations in terms
of their initial mass functions, X-ray emission, ages, and circumstellar disks.
For instance, using the model for Li depletion from \citet{jef23},
I have derived an age of 24.7$^{+0.9}_{-0.6}$ Myr for BPMG, which is
similar to estimates from previous studies.
In addition, I have used infrared photometry from the
Wide-field Infrared Survey Explorer to check for excess emission
from circumstellar disks among the members of the associations, which
has resulted in a dramatic increase in the number of known disks around M 
stars at ages of 30--50 Myr and a significant improvement in measurements of
excess fractions for those spectral types and ages.
Most notably, I find that the W3 excess fraction for M0--M6 initially 
declines with age to a minimum in BPMG ($<$0.015), increases 
to a maximum in Carina/Columba/$\chi^1$ For ($0.041^{+0.009}_{-0.007}$, 
34 Myr), and declines again in the oldest two associations (40--50 Myr). 
The origin of that peak and the nature of the M dwarf disks at $>$20 Myr 
are unclear.

\end{abstract}

\section{Introduction}
\label{sec:intro}

The discovery of a large debris disk around the nearby A star $\beta$~Pic
by the Infrared Astronomical Satellite \citep{smi84,aum85} prompted an
interest in measuring the age of the system. To do so, \citet{bar99}
sought to identify additional stars that were born in the same molecular
cloud as $\beta$~Pic by examining the kinematics of stars that appeared to
be comoving with it. Using data from the Hipparcos
mission and other sources, they concluded that the M stars GJ~799 and GJ~803 
were likely members of a moving group associated with $\beta$~Pic (BPMG).
Based on a comparison of the two stars to theoretical isochrones in a 
color-magnitude diagram (CMD), \citet{bar99} estimated an age of $20\pm10$~Myr, 
which was consistent with other age indicators for the stars.
Through similar analysis of kinematics and signatures of youth,
\citet{zuc01c} expanded the proposed membership of BPMG to 18 systems
and subsequent studies have reached $\sim$100 candidate members
\citep{zuc01c,son03,kai04,zuc04,tor06,lep09,tei09a,kis11,schl10,schl12a,schl12b,schl12c,liu13,mal13,moo13,bin14,bin16,gag14a,gag15c,gag18b,rie14,shk17,gag18d,lee19,lee22,sta22,sch23,lee24}.
The internal kinematics, lithium abundances, and CMDs of those samples have 
been used to estimate the age of the association, where most values have
ranged between 20--25~Myr
\citep{zuc01c,ort02,ort04,son02,men08,mac10,yee10,bin14,bin16,mal14b,mam14,bel15,her15,nie16,shk17,cru19,mir20,gal22,lee22,cou23,jef23,lee24}.
Given their youth and proximity, members of BPMG are amenable to
studies of primordial and debris disks 
\citep{spa01,kal04,liu04,che05,che08,sch06,reb08,pla09,sch09,riv14,moo16,moo20,bin17,sis18,tan20,ada21,paw21,cro22,cro23}
and surveys for brown dwarf and planetary companions 
\citep{low00,kas07,bil10,lag09,lag10,lag19,mug10,mac15,pla20,mar21,wit22,wit23,der23,fra23,mes23}.

The Gaia mission is providing high-precision photometry, proper motions, and
parallaxes for stars as faint as $G\sim20$ \citep{per01,deb12,gaia16b}, which 
corresponds to substellar masses in nearby young associations.
As a result, Gaia data have made it possible to significantly improve
the reliability and completeness of membership catalogs for those associations.
In the case of BPMG, \citet{gag18d} and \citet{lee22} have used Gaia to 
search for new members, each identifying more than 50 candidates.
\citet{lee24} further assessed the membership of candidates from \citet{gag18d} 
and other surveys using Gaia data and radial velocities.
Many of the candidates from Gaia lack measurements of age diagnostics
or velocities, which are needed for confirmation of membership.
In addition, although the available samples of proposed members
share a core group of a few dozen stars, the samples can otherwise differ
substantially. 

In this paper, I attempt to perform a thorough census of BPMG using 
the third data release of Gaia \citep[DR3,][]{bro21,val23} in conjunction with
photometry, spectroscopy, and radial velocities from other sources.
The resulting catalog of adopted members is used to characterize
the association's initial mass function (IMF), age, and disk population.
To facilitate the interpretation of the disk census in BPMG, 
I also present new membership catalogs for additional nearby associations with
ages of 20--50~Myr, consisting of Sco Body, Carina, Columba, $\chi^1$ For, 
Tuc-Hor, IC 2602, IC 2391, and NGC 2547.

\section{Search for New Members of BPMG}

\subsection{Identification of Candidate Members}
\label{sec:identbp}

The following data from Gaia DR3 have been employed in my survey of BPMG:
photometry in bands at 3300--10500~\AA\ ($G$), 3300--6800~\AA\ ($G_{\rm BP}$),
and 6300-10500~\AA\ ($G_{\rm RP}$); proper motions and parallaxes
($G\lesssim20$); and radial velocities ($G\lesssim15$).
I have adopted the geometric values of parallactic distances derived by
\citet{bai21} from the Gaia DR3 parallaxes. 
My survey also utilizes $JHK_s$ photometry from the Point Source Catalog 
of the Two Micron All Sky Survey \citep[2MASS,][]{skr03,skr06} and
photometry at 3.4, 4.6, 12, and 22~$\mu$m (W1--W4) 
from the WISE All-Sky Source Catalog \citep{cut12a},
the AllWISE Source Catalog, and the AllWISE Reject Catalog \citep{cut13a,wri13}
of the Wide-field Infrared Survey Explorer \citep[WISE,][]{wri10}.
I use the photometry from AllWISE for most Gaia sources, but a small 
fraction of the latter have close matches only in one of the other two WISE
catalogs \citep{luh22disks}. 
Because of potentially large errors in W1 and W2 for bright stars
\citep{cut12b,cut13b}, I have adopted data at W1$<$8 and W2$<$7 from the
All-Sky Catalog instead of AllWISE and I have ignored all W2 data at W2$<$6,
as done in \citet{esp18}.

In my previous surveys of young associations
\citep[e.g.,][]{esp17,luh20u,luh22sc}, I have analyzed Gaia astrometry 
in terms of a ``proper motion offset",
which is defined as the difference between the observed proper 
motion of a star and the motion expected at the celestial coordinates 
and parallactic distance of the star for a specified space velocity.
This metric serves to reduce projection effects, which is important for
an association like BPMG that is extended across a large area of sky.
In my survey of BPMG, the proper motion offsets are calculated relative to
the motions expected for a velocity of $U, V, W = -10, -16, -9$~km~s$^{-1}$,
which is similar to the median velocity for my final catalog of BPMG
members (Section~\ref{sec:uvw}).
If a star's velocity matches the median velocity and its proper motion and
parallax measurements are accurate, it will have a proper motion offset of 
zero. However, as in any association, the members of BPMG exhibit a
spread in velocities.
A given deviation from the median velocity will correspond to larger offsets
at smaller distances. As a result, since the members of BPMG span a large
range of distances (a factor of $\sim$10), the spread in their offsets can 
vary significantly with distance. To remove this effect, I have
included a factor of ``distance/50 pc" in the proper motion offsets.
Since my selection of BPMG candidates is based on clustering of the proper
motion offsets, it does not depend on the value of the constant in that factor.

To perform my census of BPMG, I began by compiling 
proposed members from previous studies \citep[e.g.,][]{bel15,shk17,gag18b}.
Within this sample, I identified likely members based on clustering in 
(1) proper motion offsets calculated from Gaia DR3 data, (2) ages inferred from
Gaia CMDs, and (3) $U$, $V$, and $W$ velocities as a function of $X$, $Y$, 
and $Z$ Galactic Cartesian positions, respectively.  The $UVW$ velocities 
were available only for stars that have measurements of radial velocities 
from Gaia DR3 or other sources (Section~\ref{sec:uvw}). 
I then searched Gaia DR3 for additional candidate members that 
resemble the likely members in terms of the proper motion offsets, CMD ages, 
and $UVW$ velocities (when available). Since stars with disks can appear
underluminous in CMDs, I allowed the selection of underluminous stars if
they exhibited infrared (IR) excess emission in WISE photometry.
I only considered stars with $\sigma_{\pi}/\pi<0.1$ and accurate photometry
($\sigma<0.1$ mag) in both $G_{\rm BP}$ and $G_{\rm RP}$ or in both $G$ 
and $G_{\rm RP}$. For fainter candidates that lacked photometry in $G_{\rm BP}$
and thus could be plotted in only a single Gaia CMD, I required that their
positions in IR CMDs from 2MASS and WISE were consistent with membership.

Initially, my search for BPMG candidates was performed for a spatial volume 
directly surrounding the likely members from previous studies. 
The search volume was then expanded in most directions until few 
additional candidates were identified. During this process, I found that 
the BPMG candidates extended into a volume inhabited by a neighboring 
group called Sco Body that is located in front of the Sco-Cen OB association 
\citep{rat23}\footnote{Portions of Sco Body have also been identified 
in other Gaia surveys \citep[e.g.,][]{ker21}, as discussed in \citet{rat23}.}.
The two associations have similar kinematics but different
ages (Sections~\ref{sec:uvw} and \ref{sec:ages}). 
Based on close inspection of the CMD ages and spatial distribution of
young low-mass stars near the interface of BPMG and Sco Body,
my BPMG candidates should have little contamination from the latter
at $X<100$~pc, so that was selected as a boundary for my survey.
Beyond that boundary, it becomes increasingly difficult to reliably
separate members of BPMG and Sco Body. The final volume for my survey was 
defined by $-$80~pc$<X<$100~pc, $-$50~pc$<Y<$50~pc, and $-$70~pc$<Z<$20~pc.

For the BPMG candidates, I searched for spectroscopic measurements of age 
diagnostics (e.g., Li) and radial velocities in data archives
and previous studies. I pursued new spectroscopy for some candidates 
to measure spectral types and age diagnostics (Section~\ref{sec:spec}). 
The $UVW$ velocities calculated from radial velocities and Gaia astrometry
and the spectroscopic constraints on age were used to further refine the
sample of candidates (Sections~\ref{sec:spec} and \ref{sec:uvw}).
This process produced 180 objects that I have adopted as members of BPMG. 
To that sample, I have added three previously proposed members that lack
parallax measurements from Gaia DR3 but that have parallax data from other
sources, consisting of HD 161460, GJ 3076, and PSO J318.5338$-$22.8603.
I searched for companions to the adopted members that are in Gaia DR3 
and that did not satisfy the selection criteria for BPMG membership by
retrieving DR3 sources that are within $5\arcsec$ of the members
and whose available kinematic and photometric data are consistent with 
companionship, which resulted in the addition of 10 likely companions.
These objects were not selected initially because they lacked either parallax
measurements or photometry in $G_{\rm BP}$ and $G_{\rm RP}$.
Thus, the final catalog of adopted members contains 193 objects, all of which
have entries in Gaia DR3 with the exception of the L dwarf
PSO J318.5338$-$22.8603 \citep{liu13}.
The catalog does not contain separate entries for the substellar
companions to the following stars since they lack Gaia
detections: $\beta$ Pic \citep{lag09,lag10,lag19}, PZ Tel \citep{bil10,mug10},
51 Eri \citep{mac15}, AU Mic \citep{pla20,mar21}, and AF Lep 
\citep{mes23,fra23,der23}.

To illustrate the clustering of the adopted BPMG members in proper motion
offsets, I have plotted the latter in Figure~\ref{fig:pp}.
The outliers are labeled with their names. They are retained as members
because their discrepant motions can be attributed to poor astrometric fits,
as indicated by the large values ($>$1.6) of their renormalized unit weight 
errors \citep[RUWEs,][]{lin18}. All of the outliers have close companions or
consist of a close pair that is unresolved by Gaia \citep[GJ 3305,][]{kas07}, 
which makes them susceptible to poor fits.  In addition to the adopted members, 
Figure~\ref{fig:pp} includes a sample of previous candidates that have been 
excluded from my catalog, which are discussed in Section~\ref{sec:prev}.

In Figure~\ref{fig:cmd1}, I show the adopted members of BPMG
in CMDs consisting of $M_{G_{\rm RP}}$ versus $G_{\rm BP}-G_{\rm RP}$ 
and $G-G_{\rm RP}$. I have omitted photometry with errors greater than 0.1 mag.
Each CMD includes a fit to the single-star sequence of the Pleiades cluster 
\citep[$\sim120$~Myr,][]{sta98,dah15}. The one star
that appears below the sequence in the $G_{\rm BP}-G_{\rm RP}$ CMD is
Gaia DR3 6806301370521783552. It is much closer to the lower envelope
of the sequence for BPMG (corresponding to single stars) when using photometry 
from the second data release of Gaia (DR2), so its DR3 photometry is probably 
erroneous, perhaps because of the presence of a companion at a separation of 
$1\farcs45$. As often observed in nearby associations \citep{luh23twa}, 
the CMD in $G-G_{\rm RP}$ for BPMG contains a group of red stars that 
are much too bright at a given color to be unresolved binaries. Those stars 
are likely to be marginally resolved binaries in which the $G$-band 
fluxes are underestimated \citep{rie21}.

\subsection{Spectroscopy of Candidates}
\label{sec:spec}

To measure and refine spectral classifications for candidate members
of BPMG, I have obtained spectra for 84 sources
and I have analyzed archival spectra for four additional stars.
For 58 of these sources, spectroscopy is reported for the first time 
in this study. The new observations were performed with SpeX \citep{ray03}
at the NASA Infrared Telescope Facility (IRTF), the Goodman spectrograph
\citep{cle04}
at the Southern Astrophysical Research Telescope (SOAR), FLAMINGOS-2 at the
Gemini South telescope \citep{eik04}, and the Cerro 
Tololo Ohio State Multi-Object Spectrograph (COSMOS)\footnote{COSMOS
is based on an instrument described by \citet{mar11}.} at the 4~m Blanco
telescope at the Cerro Tololo Inter-American Observatory (CTIO).
The archival data were collected with the RC spectrograph at the CTIO 1.5~m 
telescope through program UR10B-05 (E. Mamajek) and with 
the Gemini Multi-Object Spectrograph \citep[GMOS;][]{hoo04} at the 
Gemini South telescope through programs GS-2012B-Q-92 (D. Rodriguez) 
and GS-2012B-Q-70 (J. Gagn{\'e}). The instruments and 
observing modes are summarized in Table~\ref{tab:log}.
The instruments and observing dates for individual targets are
provided in Table~\ref{tab:spec}.

\begin{figure}
\epsscale{1.3}
\plotone{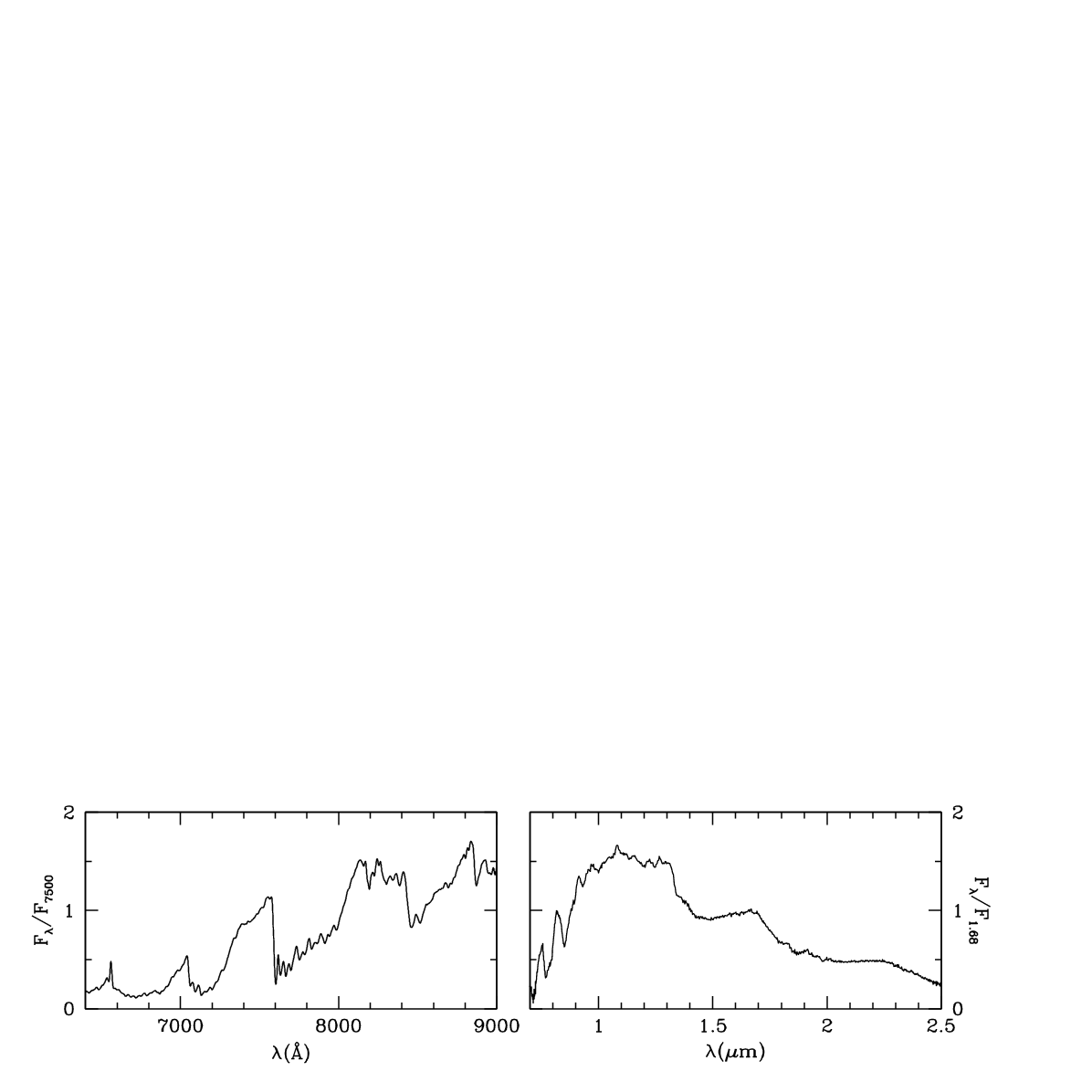}
\caption{ 
Examples of optical and IR spectra of adopted members of BMPG
(Gaia DR3 5809284739228623232 and Gaia DR3 6754492932379552896).
Each object has a spectral type of M6.
The optical spectrum is displayed at a resolution of 13~\AA.
}
\label{fig:spec}
\end{figure}

The spectra from SpeX were reduced with the Spextool package \citep{cus04}, 
which included a correction for telluric absorption \citep{vac03}.
The data from the remaining instruments were reduced with routines 
within IRAF. Examples of the reduced optical and IR spectra are 
presented in Figure~\ref{fig:spec}. The reduced spectra are available in an 
electronic file associated with Figure~\ref{fig:spec}.

For all of the spectroscopic targets, the age diagnostics (e.g., Li, Na, FeH, 
$H$-band continuum) are consistent with those of other BPMG members that 
I have adopted.  The same is true for the known companions to these sources. 
As a result, all of the spectroscopic targets are among the adopted members
of BPMG in Section~\ref{sec:uvw}.
I measured spectral types from the optical spectra through comparison
to field dwarf standards for $<$M5 \citep{hen94,kir91,kir97} and
averages of dwarf and giant standards for $\geq$M5 \citep{luh97,luh99}.
Classifications of the IR spectra are based on comparisons to
standard spectra for young stars and brown dwarfs \citep{luh17}.
The resulting spectral types are included in Table~\ref{tab:spec}.

\subsection{$UVW$ Velocities}
\label{sec:uvw}

For the BPMG candidates from Section~\ref{sec:identbp} that have radial 
velocity measurements, their membership has been further constrained using
$UVW$ velocities. I have compiled measurements of radial velocities 
from Gaia DR3 and all other available sources.
I have only considered velocities that have errors smaller than 5 km~s$^{-1}$.
If a star has multiple velocity measurements, I adopt the
velocity with the smallest error. For spectroscopic binaries, I use
the system velocities if they have been measured. I have not adopted
velocities for the following stars since each has a wide range
of values and lacks a measurement of the system velocity:
Gaia DR3 6747467431032539008, Gaia DR3 6833291426043854976,
RX J0050.2+0837, RX J2137.6+0137, HD 199143, and HD 172555B. 
I have used the adopted radial velocities in conjunction
with proper motions from Gaia DR3 and parallactic distances based on DR3
parallaxes \citep{bai21} to calculate $UVW$ velocities \citep{joh87}.
The velocity errors were derived in the manner described by \citet{luh20u}
using the python package {\tt pyia} \citep{pri21}.

To illustrate their spatial distribution, I have plotted the adopted
members of BPMG that have parallax measurements (i.e., excludes five
companions that lack parallaxes) in diagrams of $XYZ$ Galactic
Cartesian positions in the top row of Figure~\ref{fig:uvw}. The members are 
concentrated in two parallel filaments that are elongated in the $X$ direction. 
In the bottom row of Figure~\ref{fig:uvw}, the measurements of $U$, $V$, and 
$W$ are plotted versus $X$, $Y$, and $Z$, respectively, for the 134 stars 
that have parallax measurements and adopted radial velocities. 
(One companion lacks parallax data but has a radial velocity measurement).
As shown in Figure~\ref{fig:uvw}, the adopted members are well clustered in
$UVW$ versus $XYZ$. I have excluded candidates from Section~\ref{sec:identbp} 
that are discrepant relative to those distributions unless the discrepancies
can be explained by poor astrometric fits for companions.  The velocity 
components are correlated with their corresponding spatial dimensions,
which indicates the presence of expansion.
For the 134 adopted members that have measured $UVW$ velocities,
the median velocity is $U, V, W = -9.3, -15.6, -9.0$~km~s$^{-1}$.

\begin{figure}
\epsscale{1.2}
\plotone{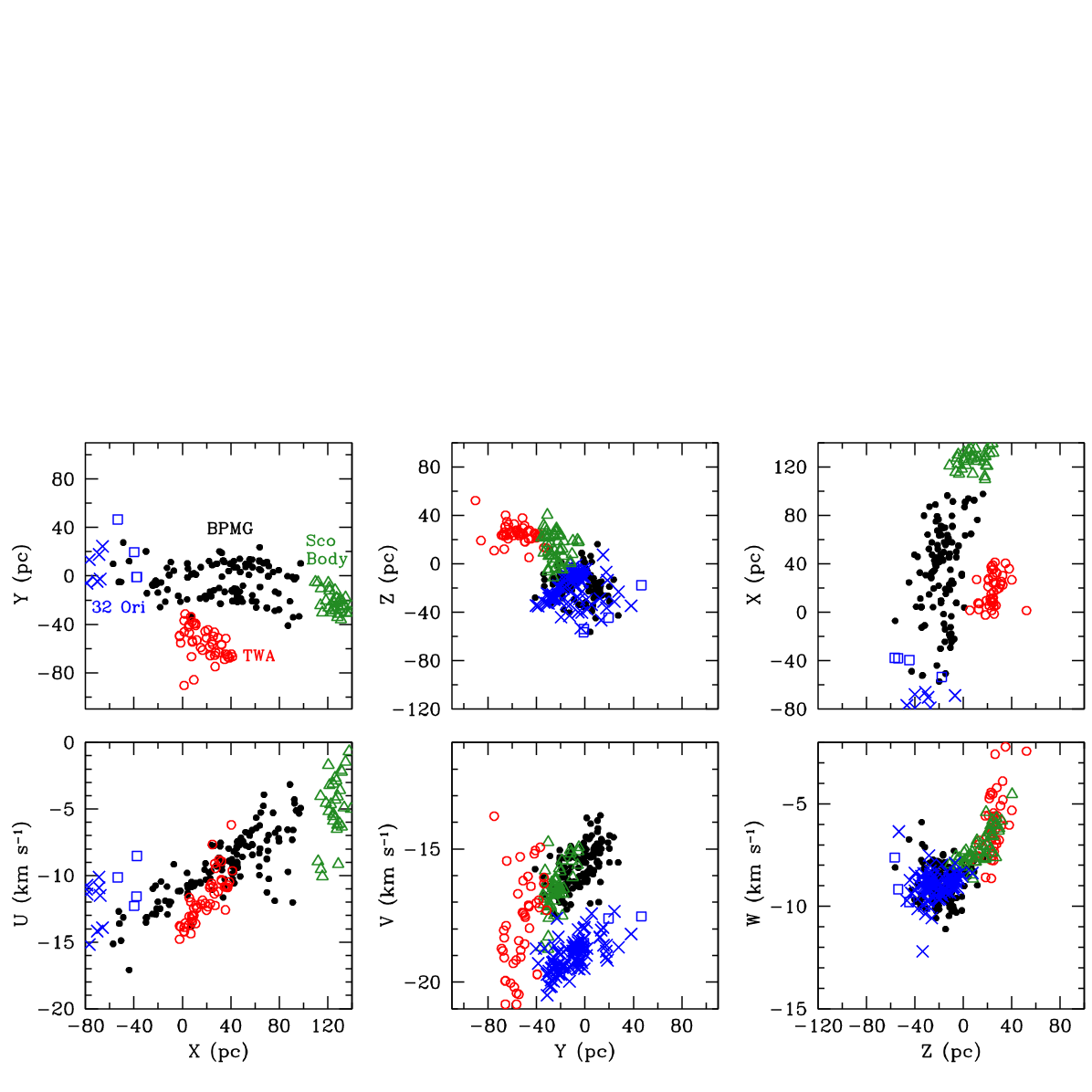}
\caption{
Galactic Cartesian coordinates and $UVW$ velocities for members 
of BPMG (black points), TWA (red circles), 32 Ori (blue crosses), 
and Sco Body (green triangles) that have radial velocity measurements.
Four newly proposed members of 32 Ori are included (blue squares).}
\label{fig:uvw2}
\end{figure}

To compare BPMG to neighboring young associations (10--20 Myr), 
I have plotted $XYZ$ and $UVW$ for members of BPMG, TW Hya
\citep[TWA,][]{luh23twa}, 32 Ori \citep{luh22o}, and Sco Body 
(Section~\ref{sec:othergroups}) in Figure~\ref{fig:uvw2}. These diagrams
only include stars that have radial velocity measurements.
In addition, members that have discrepant velocities because of 
binarity have been omitted.
BPMG and 32 Ori are spatially adjacent but are offset in $U$ and $V$.
BPMG and Sco Body are adjacent in $X$ and $Y$, are offset in $Z$, and have 
overlapping $UVW$ velocities that are consistent with a single pattern of 
expansion, although their ages differ (Section~\ref{sec:ages}).
Previous studies have proposed that BPMG, TWA, and Sco Body are related to the
Sco-Cen OB association \citep{mam01,mam02,ort04,rat23}.

In Table~\ref{tab:mem}, I present a catalog of the 193 sources that I have 
adopted as members of BPMG, which contains source names from Gaia DR3 and 
previous studies; equatorial coordinates, proper motion, parallax, RUWE, and 
photometry from Gaia DR3; measurements of spectral types and the type adopted 
in this work; adopted measurement of the equivalent width of 
lithium; distance estimate based on the Gaia DR3 parallax \citep{bai21}; 
the adopted radial velocity measurement; the $UVW$ velocities calculated in
this section; the designations and angular separations of the closest
sources within $3\arcsec$ from 2MASS and WISE; flags indicating whether
the Gaia source is the closest match in DR3 for the 2MASS and WISE sources;
photometry from 2MASS and WISE (only for the Gaia source that is closest to 
the 2MASS/WISE source); and flags indicating whether excesses are detected 
in three WISE bands and a disk classification (Section~\ref{sec:disks}).

Among the 193 adopted members of BPMG, nine lack spectral classifications
and 58 lack accurate radial velocity measurements.
The former include seven close companions ($\lesssim2\arcsec$) to
classified stars and the latter include the six stars mentioned
earlier that have large spreads in their available velocities and seven
companions to stars that do have radial velocity data.
Both spectral classifications and radial velocities are unavailable for
six sources (five close companions). Measurements of spectroscopic age 
diagnostics and radial velocities for these sources would help to better 
constrain their membership. 

\subsection{Comparison to Previous Studies}
\label{sec:prev}

\begin{figure}
\epsscale{1.2}
\plotone{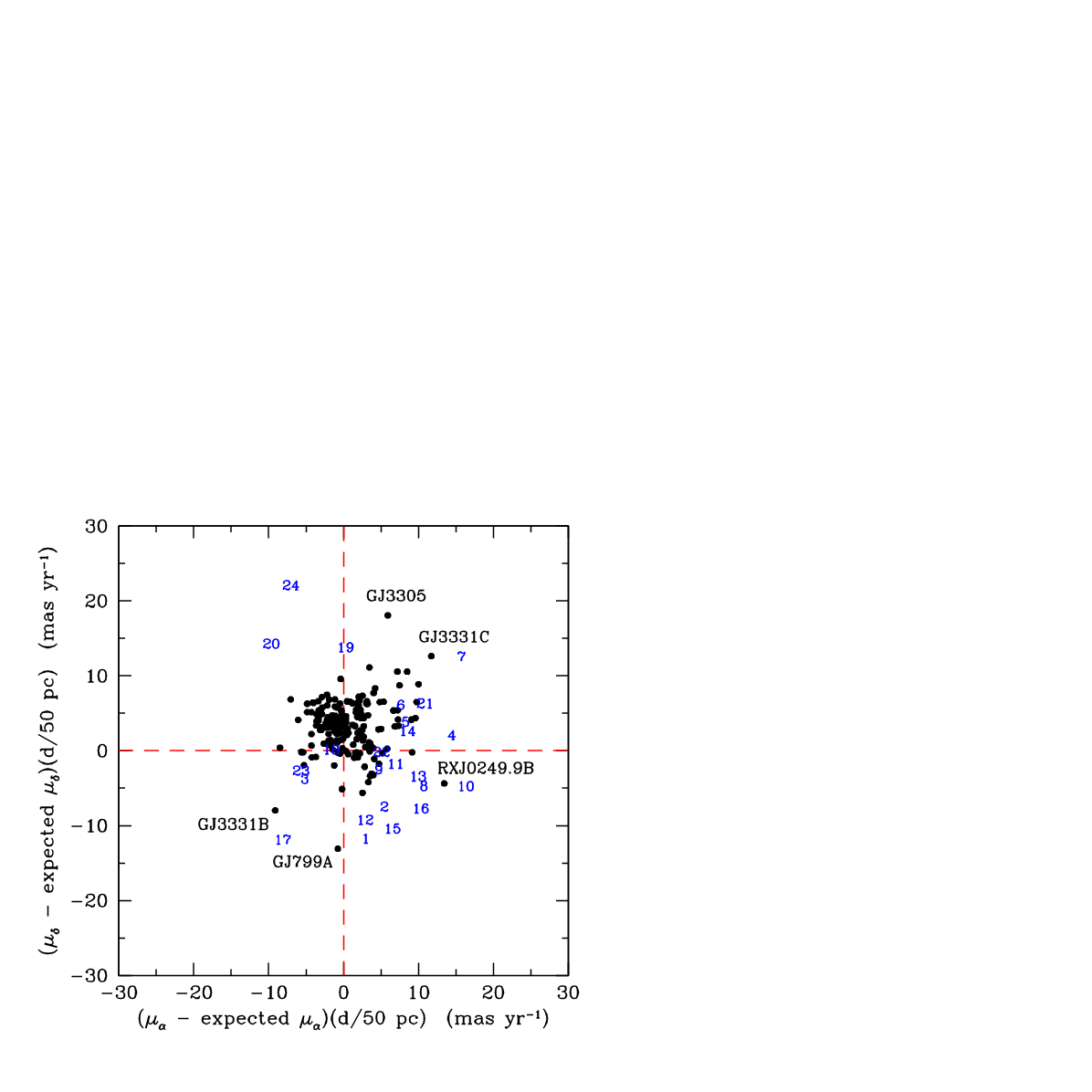}
\caption{
Proper motion offsets based on Gaia DR3 for adopted members of BPMG
and a sample of previously identified candidates that are excluded from
my catalog (blue numerals).  Discrepant measurements 
for adopted members are labeled with the source names.}
\label{fig:pp}
\end{figure}

\begin{figure}
\epsscale{1.2}
\plotone{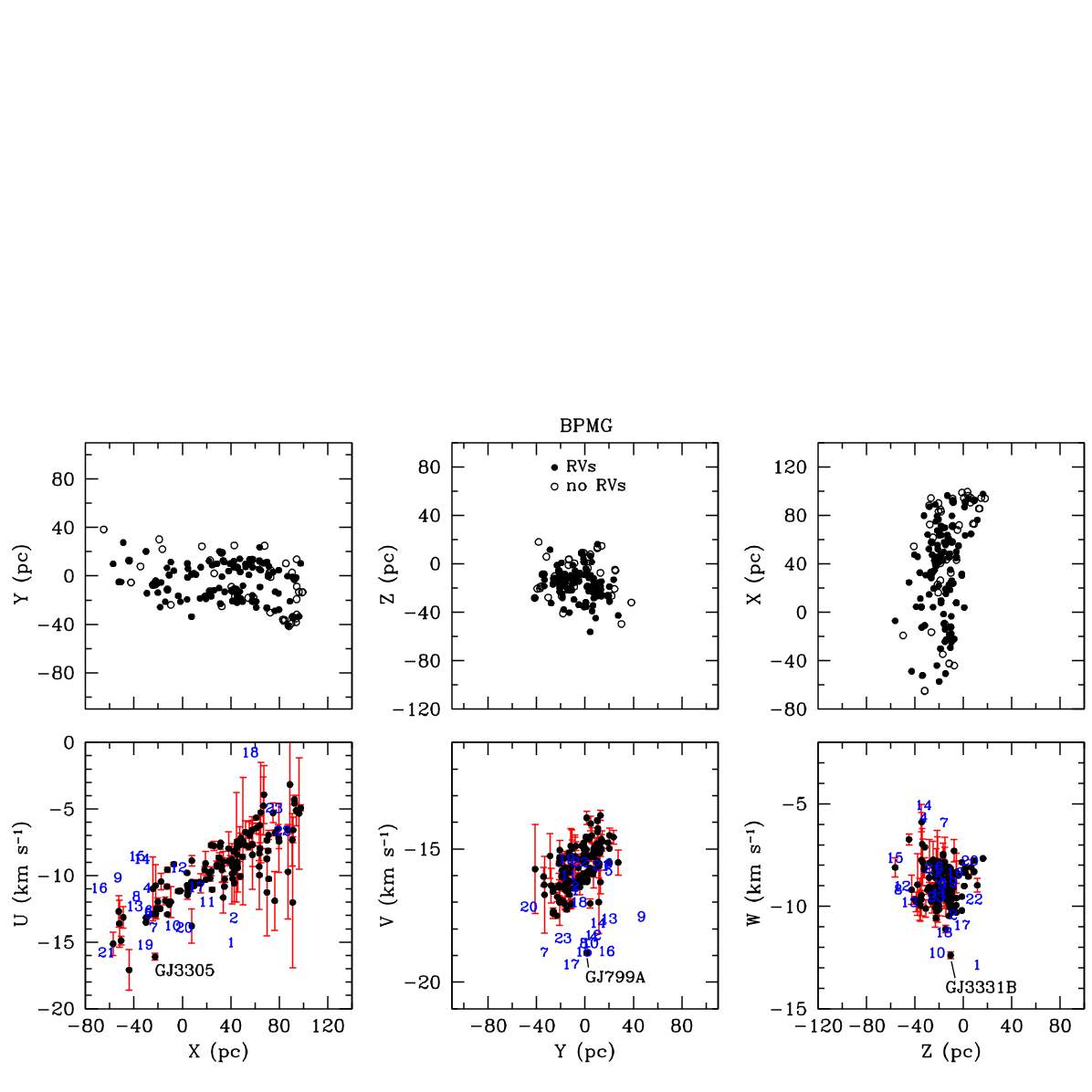}
\caption{
Galactic Cartesian coordinates and $UVW$ velocities for the adopted members 
of BPMG and a sample of
previously identified candidates that are excluded from my catalog 
(blue numerals; Table~\ref{tab:rej}). Discrepant measurements for 
BPMG members are labeled with the source names.}
\label{fig:uvw}
\end{figure}

Many of the previously proposed members of BPMG are included in my catalog,
which contains 22/24 from \citet{zuc01c},
30/34 from \citet{zuc04}, 38/41 from \citet{tor06}, 24/30 from \citet{mam14},
70/89 from \citet{bel15}, 48/56 from \citet{gag18b}, 93/113 from \citet{lee19},
24/25 (core) and 39/51 (extended) from \citet{cou23}, and 69/86 from
\citet{lee24}. These numbers omit companions that are not detected by Gaia.
Among the previous candidates that have been excluded
from my catalog, I discuss those that (1) have been classified as
members by several studies (HR 6070, HR 6749, HR 6750, HIP 12545, HD 14082A/B), 
(2) are notable for having a late spectral type
\citep[2MASS J06085283-2753583,][]{kir08,ric10,fah12}, a substellar companion 
\citep[2MASS 02495639$-$0557352,][]{dup18,chi21}, or
an edge-on disk \citep[BD+45 598,][]{hin21}, or
(3) have been classified as members by two of the more recent membership 
studies of BPMG \citep{gag18b,lee24}, which corresponds to a total of 24
sources. I have compiled their Gaia data, radial velocities, and $UVW$ 
velocities in Table~\ref{tab:rej} and I have plotted their proper motion 
offsets in Figure~\ref{fig:pp} and $UVW$ velocities in Figure~\ref{fig:uvw} for 
comparison to the adopted members.
To enable the identification of individual sources in those diagrams,
they are plotted with numerals from 1 through 24, which are listed
in Table~\ref{tab:rej}. For most velocities, the errors are smaller than 
the sizes of the numerals in Figure~\ref{fig:uvw}.

Relative to the adopted members of BPMG, the A star HR 6070 (1) is an outlier 
in proper motion offsets (Figure~\ref{fig:pp}) and in $U$ and $W$
(Figure~\ref{fig:uvw}).
The A stars HR 6749 and HR 6750 (2 and 3) form a $1\farcs8$ binary system.
Both components are modest outliers in proper motion offsets and the former 
is a modest outlier in $U$ (the latter lacks a radial velocity measurement).
The two stars are overluminous relative to the single-star sequence
for BPMG in Gaia CMDs \citep{cou23}, so membership in BPMG would require 
that each is an unresolved binary if the Gaia photometry is reliable.  
The K star HIP 12545 (4) is an outlier in proper motion offsets, $V$, and $W$.
HD 14082A and B (5 and 6) are F/G stars in a $14\arcsec$ binary system.
Their $UVW$ velocities are consistent with membership in BPMG and the primary
is near the single-star sequence of BPMG in Gaia CMDs (which is close to the 
zero age main sequence at that spectral type), but the secondary appears 
below BPMG's sequence and coincides with the sequence of Tuc-Hor
($\sim$40~Myr, Section~\ref{sec:othergroups}). The Li data for these stars are 
consistent with BPMG as well as older ages \citep{men08,jef23}.
The Gaia photometry for HD 14082B agrees well between DR2 and DR3, so there
is no indication that its photometry is erroneous. It also appears 
underluminous relative to BPMG in other CMDs like $M_{K_s}$ versus 
$G_{\rm RP}-K_s$. The late-type object 2MASS J06085283-2753583 (7) 
is a modest outlier in $V$ and $W$.  2MASS J02495436$-$0558015 (8) is an 
outlier in proper motion offsets and $V$. BD+45 598 (9) is near the edge 
of the distribution of proper motion offsets of adopted members and is an 
outlier in $U$ and $V$. All of these stars have low values of RUWE that 
suggest good astrometric fits.

Beyond the aforementioned stars, two proposed members from
\citet{gag18b} are omitted from my catalog for BPMG.
Gaia DR3 2478001486169801216 (10) is an outlier in proper motion
offsets, $V$, and $W$ and is located below the single-star sequence for BPMG.
The $UVW$ velocity of Gaia DR3 6400160947954197888 (11) is consistent
with membership using DR3 data, but it would be rejected with DR2.
For both data releases, the star has a large value of RUWE that indicates
a poor astrometric fit and it is slightly below the sequence for BPMG.

The 13 remaining stars in Table~\ref{tab:rej} were assigned membership
by \citet{lee24} but are absent from my catalog. Stars 12 through 18 are 
outliers in their space velocities, as shown in Figure~\ref{fig:uvw}.
V1311 Ori Aa+Ab (19) is a close binary \citep{jan12} that is unresolved by 
Gaia. \citet{tok22} has proposed that it is a member of a multiple system 
with four other stars, two of which form a second pair that is unresolved by 
Gaia (Ba+Bb).  The two close pairs have poor fits from Gaia based on their 
RUWEs, leading to significant differences in proper motions among DR2, DR3, 
and previous astrometric surveys. Given the uncertainties in those
motions and the available parallax and radial velocity measurements 
\citep{tok22}, the six stars could be comoving to the degree expected
for a multiple system. In addition, their space velocities are
consistent with membership in BPMG. However, a nondetection of lithium
in an unresolved spectrum of the B and C components, which are
separated by $5\farcs3$, and strong lithium in the D component
are not consistent with the same age \citep{bel17,jef23}.
In addition, the D component appears slightly below the sequence for BPMG. 
Given the uncertainties in the astrometry for A and B and these 
inconsistencies, I have omitted all of the components from my catalog.

Gaia DR3 5412403269717562240 (20) is a modest outlier in $U$.
It is spatially and kinematically closer to TWA than BPMG
and was adopted as a member of TWA in \citet{luh23twa}. 
The $UVW$ velocity of Gaia DR3 66245408072670336 (21)
supports membership but it is located slightly below the BPMG sequence
and the nondetection of lithium in data from the ninth
data release of the Large Sky Area Multi-Object Fiber Spectroscopic Telescope 
survey \citep[$\lesssim$0.03 \AA,][]{cui12,zha12} seems inconsistent with
adopted members of BPMG (Section~\ref{sec:ages}).
CD$-$27 11535 (22) is a close binary \citep{ell15} that is unresolved by 
Gaia. It has a parallax measurement from DR2 but not DR3.
The available radial velocities span more than 10 km s$^{-1}$.
If a value near the middle of that range is adopted, the space velocity 
is consistent with membership. The binary appears 1.2 mag above the
single-star sequence for BPMG, which would require more than two unresolved
components for membership. HD 169405 (23) is a K giant \citep{hou78}, which 
is incompatible with membership in a young association. 
Gaia DR3 244734765608363136 (24) is an outlier in proper motion offsets 
and appears slightly below the sequence for BPMG. Since it has a large 
range of measured radial velocities and no measurement of the system velocity,
I have not calculated its $UVW$ velocity.

One of the stars excluded from the catalog for BPMG, 
Gaia DR3 175329120598595200 (16), was previously classified as a member of 
32 Ori \citep{luh22o}.  Four additional stars also have kinematics and CMD 
positions that indicate membership in 32 Ori, consisting of 
2MASS J02495639$-$0557352 (8), BD+45 598 (9), Gaia DR3 73034991155555456 
(13), and Gaia DR3 5177677603263978880 (15).
These stars are located outside of the survey volume for 32 Ori in 
\citet{luh22o}. They are plotted as blue squares in Figure~\ref{fig:uvw2}, 
where they represent an extension of the previously identified members of 
32 Ori to higher values of $X$, overlapping spatially with members of BPMG.
I adopt these four stars as members of 32 Ori.

In addition to proposed members of BPMG, \citet{lee24} presented a sample
of rejected candidates, which include five sources that are adopted 
as members in this work: SCR J0017$-$6645, GJ3331B, GJ2006A, Gaia DR3 
6760846563417053056, and Gaia DR3 6652273015676968832\footnote{\citet{lee24}
did assign membership to companions of two of these stars, GJ3331A and 
GJ2006B.}. The kinematics of GJ3331B are discrepant (Figures~\ref{fig:pp} 
and \ref{fig:uvw}), but that is attributable to the presence of a close 
companion (GJ3331C). The $UVW$ velocities for SCR J0017$-$6645, GJ2006A, and 
Gaia DR3 6760846563417053056 support membership. 
The remaining source, Gaia DR3 6652273015676968832, lacks 
a radial velocity measurement. \citet{gag18d} identified 
it as a candidate L-type member of BPMG using data from Gaia DR2.
I find that it continues to be a promising candidate with Gaia DR3 
and I have spectroscopically confirmed that it is a young L dwarf
(Section~\ref{sec:spec}).
It is below the sequence of BPMG members in $M_{G_{\rm RP}}$ versus
$G-G_{\rm RP}$ (below the bottom of the CMD in Figure~\ref{fig:cmd1}),
but its positions in other CMDs that use IR photometry from 2MASS and WISE 
are consistent with membership (see the CMDs in Section~\ref{sec:othergroups}).
This L dwarf is the second coolest object in my catalog for BPMG.

\section{New Catalogs for Other Associations}
\label{sec:othergroups}

To facilitate the interpretation of the disk population in BPMG
(Section~\ref{sec:disks}), I have used Gaia data to
produce new catalogs for associations and clusters that bracket the age of
BPMG (20--50 Myr) and that have the best available constraints
on disk fractions for low-mass stars at those ages given their proximity,
numbers of members, and mid-IR imaging, which consist of Sco Body, Carina, 
Columba, $\chi^1$ For (also known as Alessi 13), Tuc-Hor, IC 2602, IC 2391, and
NGC 2547.  The adopted members of NGC 2547 are compiled in Table~\ref{tab:ngc}.
My catalogs for the other associations are included with BPMG in 
Table~\ref{tab:mem}. In the disk analysis, I also will make use of 
Gaia-derived membership catalogs for additional associations near that age 
range \citep{luh22sc,luh22o,luh23tau}.

\subsection{Sco Body}
\label{sec:sco}

I have not attempted to identify new members of Sco Body and have
only revised the catalog from \citet{rat23} to reduce contamination by 
nonmembers. That catalog contained 373 sources from Gaia DR3.
Five of those stars are among my adopted members of BPMG.  Among the remaining 
368 stars, I have excluded those that have proper motion offsets or $UVW$ 
velocities that are inconsistent with membership in Sco Body, CMD positions 
that are below the sequence for the association (unless mid-IR excess emission 
is present), or $\sigma_{\pi}/\pi>0.1$.
Since multiple bands of Gaia photometry are needed for an age assessment
in CMDs, I have discarded stars from \citet{rat23} that have only a 
single band of Gaia photometry unless they are probable companions to viable
candidates. Through these steps, I have retained 299 of the 373 candidate
members of Sco Body. 
In Figure~\ref{fig:cmd1}, I have included Gaia CMDs for the adopted members of
Sco Body. Based on these CMDs, Sco Body is slightly younger than BPMG 
(Section~\ref{sec:ages}). Gaia DR3 6027365242759490688 is underluminous in 
the $G_{\rm BP}-G_{\rm RP}$ CMD for Sco Body. It exhibits mid-IR excess
emission that indicates the presence of a disk \citep{luh22disks}, so its
anomalous photometry may be caused by accretion-related emission in 
$G_{\rm BP}$ or an edge-on disk.

\subsection{Carina, Columba, $\chi^1$ For, Tuc-Hor, IC 2602, IC 2391, and 
NGC 2547}

I have performed searches for members of several additional associations
with methods similar to those that I have 
applied to BPMG.  As with the latter, I have characterized the clustering of 
members in proper motion offsets, spatial positions, and CMDs
using previously proposed members of 
Carina, Columba \citep{pla98,tor08,zuc11,bel15,gag18b,woo23},
$\chi^1$ For \citep{dia02,yen18,can18,zuc19,gal21},
Tuc-Hor \citep{tor00,zuc00,zuc01b,kra14,bel15,gag18b}, 
IC 2602, IC 2391 \citep{ran01,bar04,dob10,jac20,nis22}, 
and NGC 2547 \citep{nay02,jef04,jac20}. Large-scale surveys for 
young populations in the Milky Way also have identified portions of
these associations \citep{can18,kou19}.
I then searched for additional candidate members with Gaia DR3 and 
further refined their membership using space velocities (for stars that have 
measured radial velocities) and the available spectroscopy. 
I analyzed new and archival spectra of 49 candidate members of
Tuc-Hor, Carina, IC 2602, and IC 2391.
The archival data were taken with GMOS at Gemini South through programs 
GS-2012B-Q-92, GS-2013B-Q-83 (D. Rodriguez), GS-2012B-Q-70, GS-2013A-Q-66, and 
GS-2014A-Q-55 (J. Gagn{\'e}), with the RC spectrograph at the CTIO 1.5~m 
telescope through program STSI 08b-03, and with the Goodman spectrograph at 
SOAR. The spectroscopic observations are included in 
Tables~\ref{tab:log}--\ref{tab:mem}.
\citet{pal20} obtained optical spectra for one of the members of $\chi^1$ For, 
Gaia DR3 5093945085525624448. I have measured a spectral type of M0: from 
those data.

In Carina, Columba, $\chi^1$ For, Tuc-Hor, and IC 2602, I expanded the survey
volumes until few additional candidates were found, as done with BPMG.
In my analysis of IC 2391, candidate members are present out to large 
distances from the cluster in all directions in the $XY$ plane, 
forming an extended population that has been referred to as a ``corona"
\citep{kou19,gag21,mei21}. I have truncated my catalog at a radius of 30 pc,
which includes part of the extended population.
For NGC 2547, I searched for members that are within a radius of
$0\fdg7$ from the cluster center and that fall within
24~\micron\ images from the Spitzer Space Telescope \citep{wer04}, which
are used for the disk census in that cluster (Section~\ref{sec:disks}).

Candidate members of Tuc-Hor from previous studies span distances of 
30--80~pc while the IC 2602 cluster has a distance near 150 pc, so the 
populations have appeared to be well separated.  However, they share similar 
space velocities and ages, leading \citet{gag21} to propose that they 
are physically connected. Indeed, I find that Tuc-Hor and IC 2602 do form a 
single population that is spatially and kinematically continuous. 
\citet{mei21} identified an extended population of stars associated
with IC 2602, many of which are recovered in my census.
For the purposes of the catalogs in Table~\ref{tab:mem}, members of 
Tuc-Hor/IC 2602 that are within 16 pc from the center of IC 2602 are assigned 
to that cluster while the remaining stars are assigned to Tuc-Hor. 
That radius approximates the boundary of IC 2602's density enhancement
relative to the entire population.

\citet{gag21} noted that Columba, Carina, and other more distant populations 
\citep{pla98,kou19} are spatially adjacent and appear to have similar ages and 
kinematics, indicating that they may be physically connected.
In my analysis, I do find that those groups form a single large 
population that is spatially and kinematically continuous. 
There is enough of a spatial gap between Columba and the other
stars that I continue to list Columba as a separate group. The remaining stars 
are assigned to a second association, which I have named 
Carina-Extended (Car-Ext) since its closest fringe corresponds to Carina.
The more distant cluster that forms the densest concentration in Car-Ext
is known as ``a Carinae" \citep{pla98} and Platais 8. 
The star AB Pic is located at the interface of 
Columba and Car-Ext (which have similar velocities near that position), so it 
could be assigned to either one. I have placed it in the catalog for Columba.

Outside of the $\chi^1$ For cluster as defined in previous studies
\citep{zuc19,gal21}, I have identified a sparse, extended distribution of
members. Those outer members include the stars i Eri and HD 37852, which have
earlier spectral types (B6 and B8) than the star $\chi^1$ For (A1).
Nevertheless, I refer to this entire population as the $\chi^1$ For association 
for continuity with previous work.

The Tuc-Hor/IC 2602 and Car-Ext/Columba complexes overlap spatially
but have sufficiently distinct kinematics that they are readily
separated with Gaia astrometry. They also have different ages, as discussed
in Section~\ref{sec:ages}.

\begin{figure}
\epsscale{1.2}
\plotone{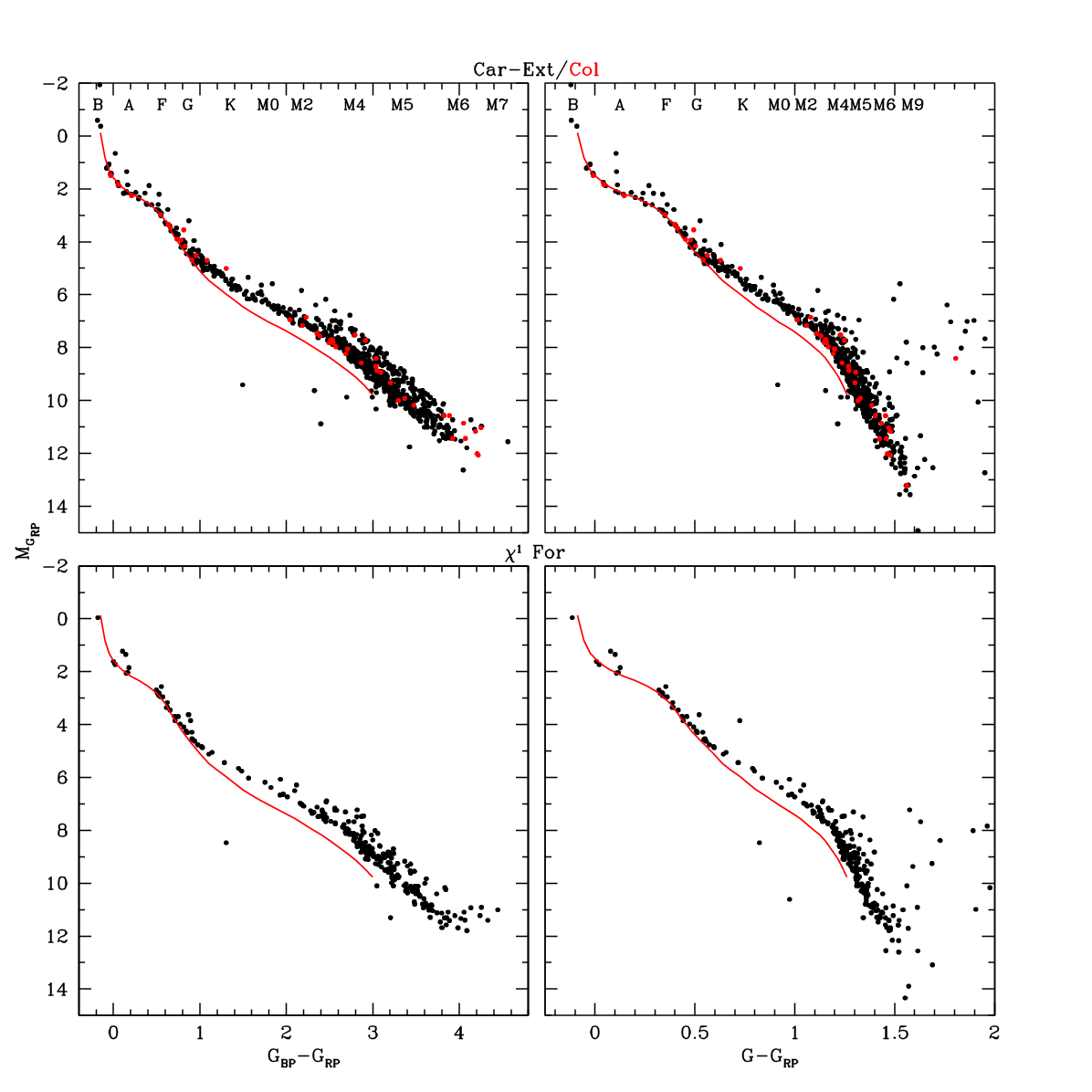}
\caption{
CMDs for adopted members of Car-Ext, Columba, and $\chi^1$ For.
}
\label{fig:cmd2}
\end{figure}

\begin{figure}
\epsscale{1.2}
\plotone{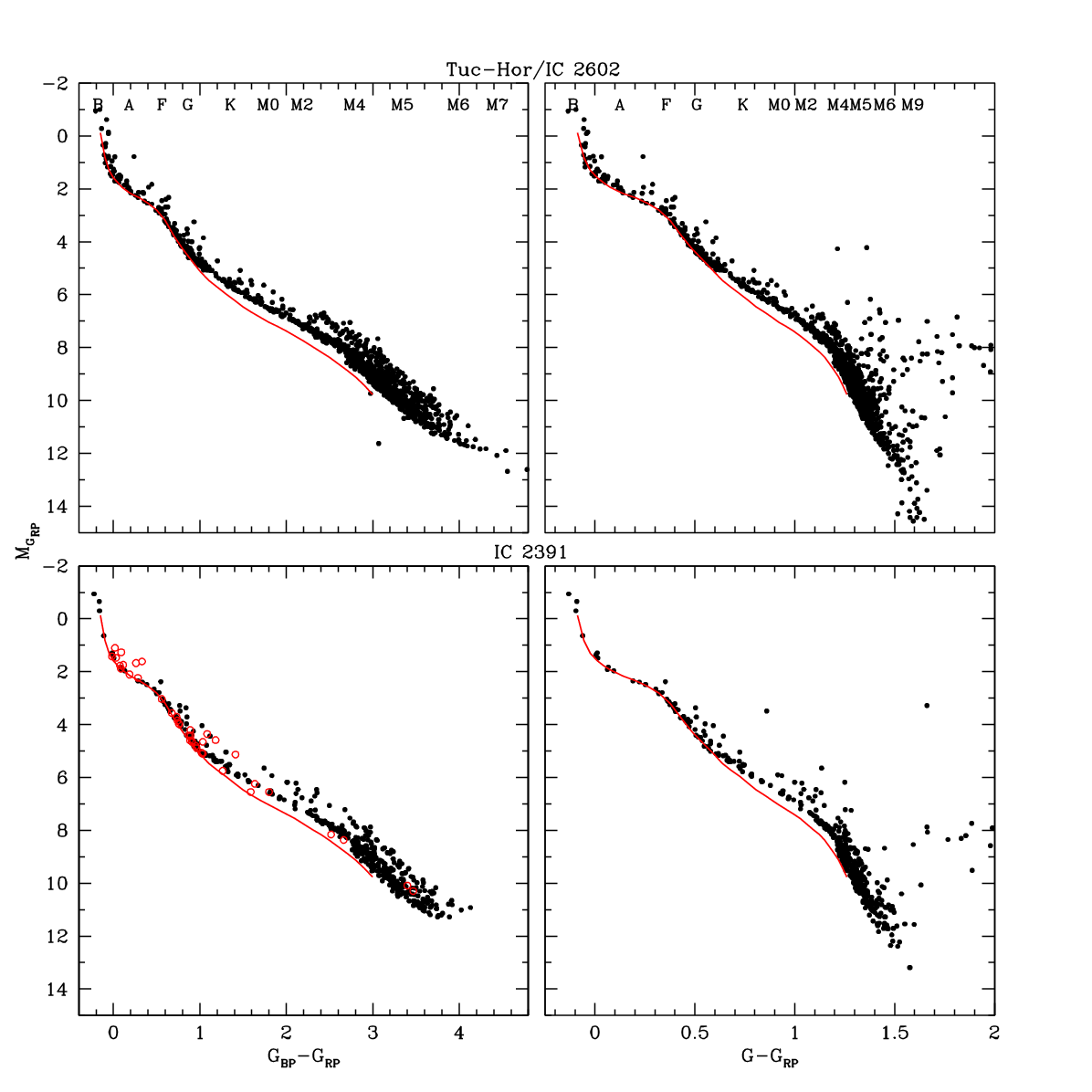}
\caption{
CMDs for adopted members of Tuc-Hor, IC 2602, and IC 2391.
The left CMD for IC 2391 includes proposed members of the Argus 
association from \citet{zuc19c} 
(red circles).}
\label{fig:cmd3}
\end{figure}

\begin{figure}
\epsscale{1.2}
\plotone{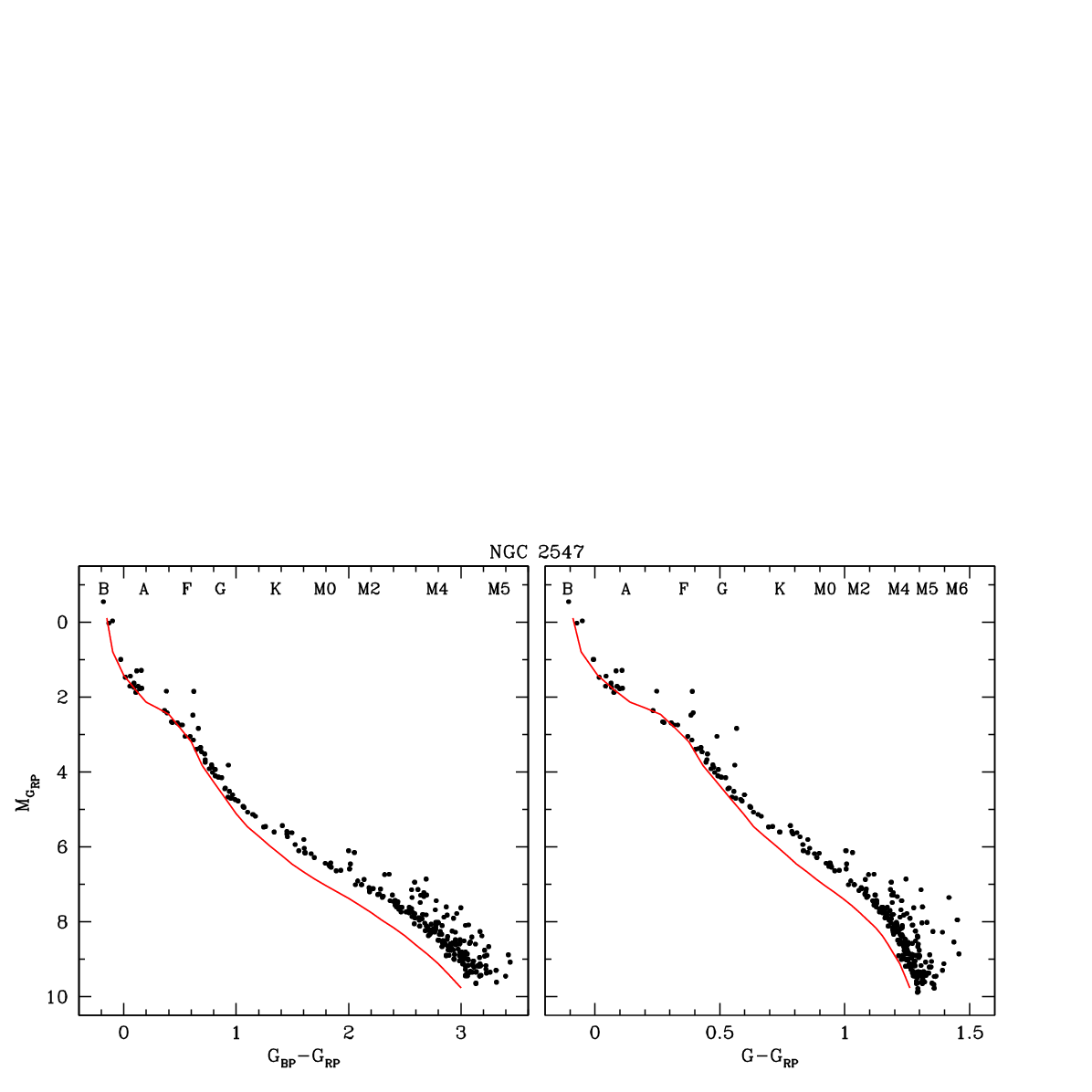}
\caption{
CMDs for adopted members of NGC 2547.}
\label{fig:cmd4}
\end{figure}

For the associations discussed in this section, the 
numbers of adopted members are as follows:
845 in Car-Ext, 51 in Columba, 287 in $\chi^1$ For,
815 in Tuc-Hor, 516 in IC 2602, 486 in IC 2391,
and 300 in NGC 2547. Gaia CMDs for these samples 
are presented in Figures~\ref{fig:cmd2}--\ref{fig:cmd4}.
Several stars in the CMDs are underluminous, 
appearing below the sequences for their associations.
Most of them have IR excesses (which is why they were not rejected),
so their anomalous photometry may be related to the presence of disks,
as mentioned for a similar star in Sco Body (Section~\ref{sec:sco}).
One underlumimous star, Gaia DR3 5176144780975151104 
(2MASS J02155892-0929121 C), lacks an IR excess. 
It is a $3\farcs4$ companion with a spectral type of M7 \citep{bow15}. 
It is much bluer in $G_{\rm BP}-G_{\rm RP}$ than expected for its spectral 
type, indicating that the $G_{\rm BP}$ photometry is probably erroneous.

\begin{figure}
\epsscale{1.2}
\plotone{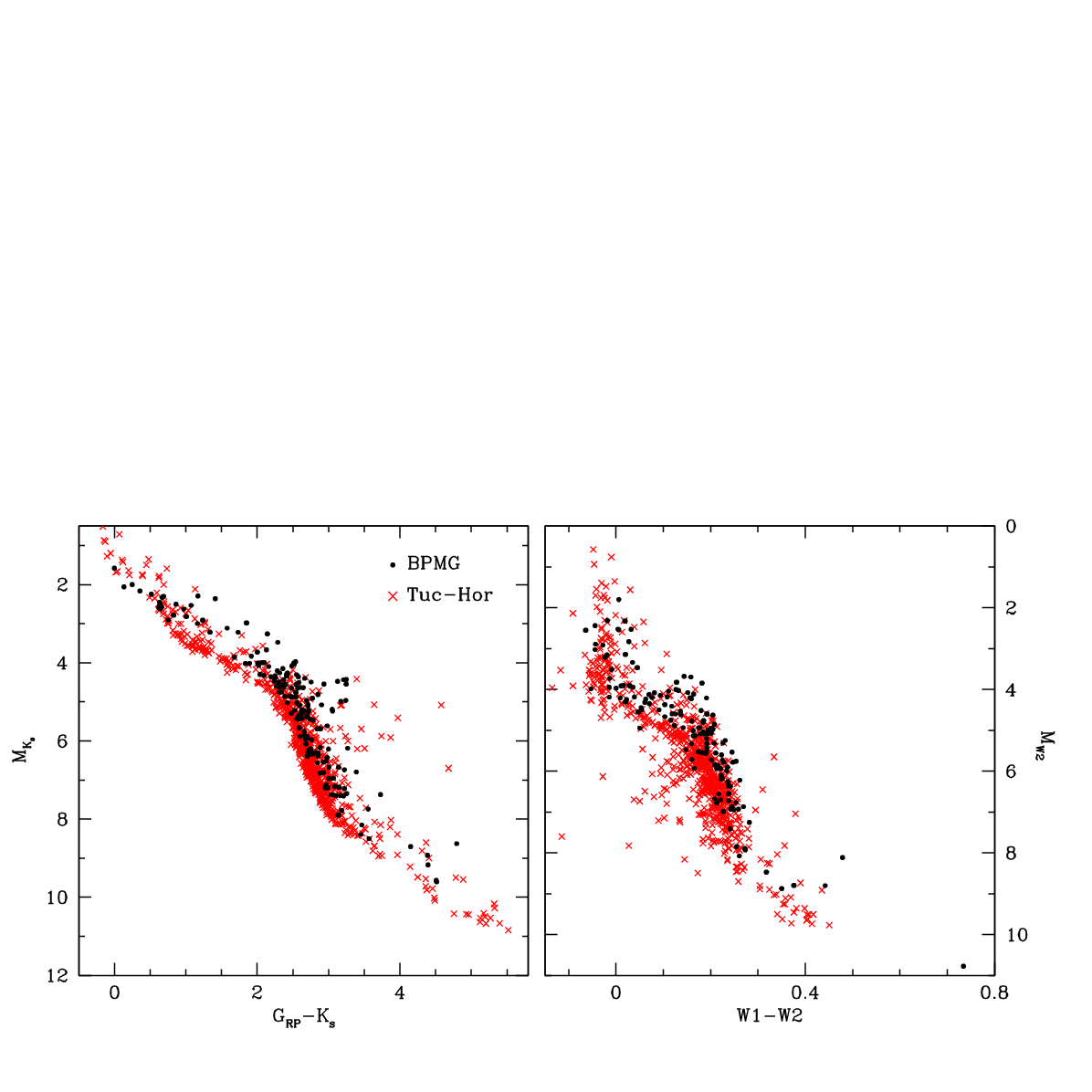}
\caption{
$M_{K_s}$ versus $G_{\rm RP}-K_s$ and $M_{W2}$ versus W1$-$W2 for
adopted members of BPMG and Tuc-Hor. The faintest known member of BPMG in
$M_{W2}$, PSO J318.5338$-$22.8603 \citep{liu13}, is absent from the left 
diagram since it lacks a detection with Gaia.}
\label{fig:cmdir}
\end{figure}

BPMG and Tuc-Hor extend to the closest distances among the populations
surveyed in this work, so their samples reach the lowest masses.
For those associations, I have plotted two additional CMDs in 
Figure~\ref{fig:cmdir} that use IR photometry from 2MASS and WISE, 
consisting of $M_{K_s}$ versus $G_{\rm RP}-K_s$ and $M_{W2}$ versus W1$-$W2.
The sequences have well-defined lower envelopes with the exception
of the WISE CMD for Tuc-Hor, where some of the faintest and most distant
members are subject to larger photometric errors and blending with other stars.

\begin{figure}
\epsscale{1.2}
\plotone{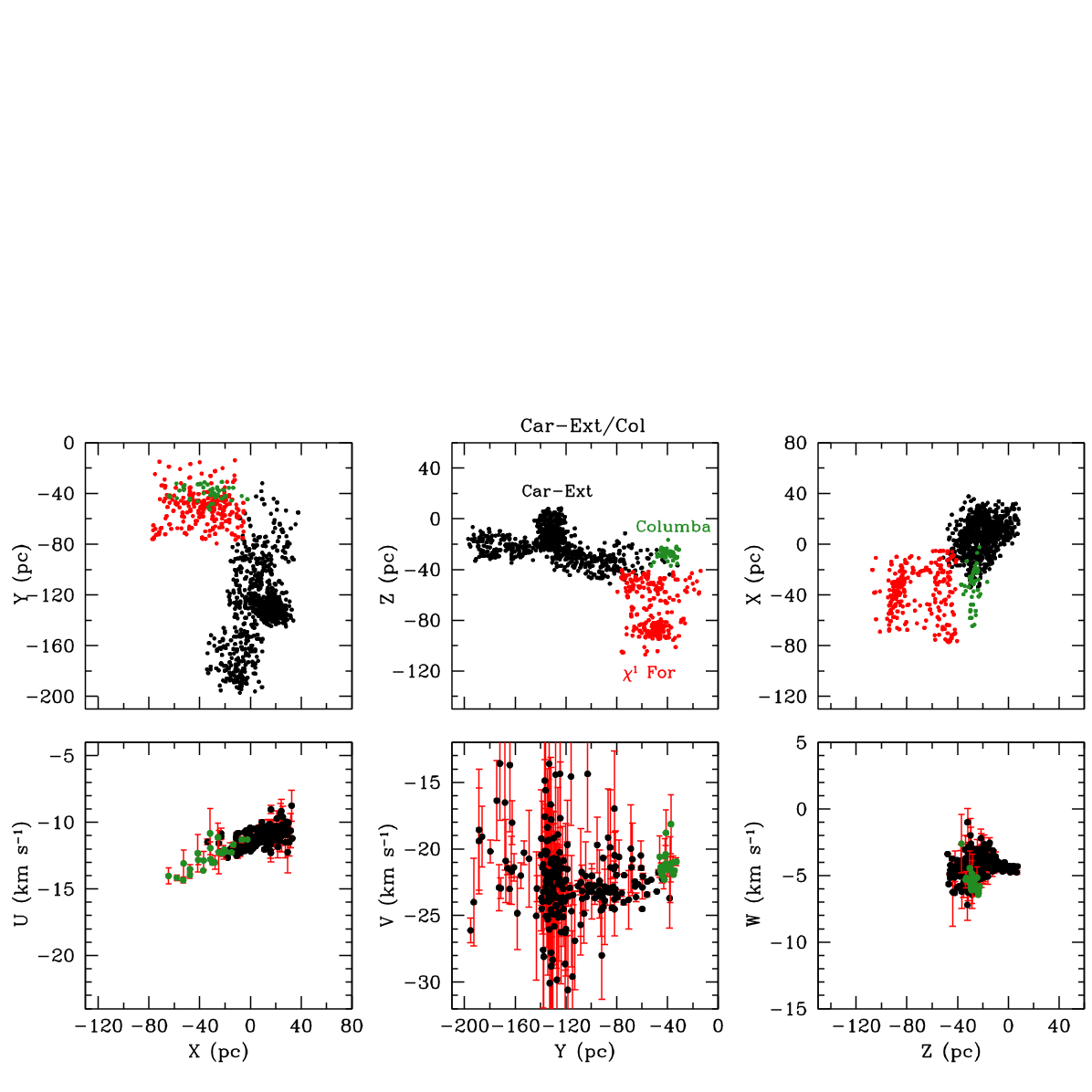}
\caption{
Galactic Cartesian coordinates and $UVW$ velocities for the adopted members 
of Car-Ext and Columba. The coordinates for members of $\chi^1$ For are
included in the top diagrams.}
\label{fig:uvwcar}
\end{figure}

\begin{figure}
\epsscale{1.2}
\plotone{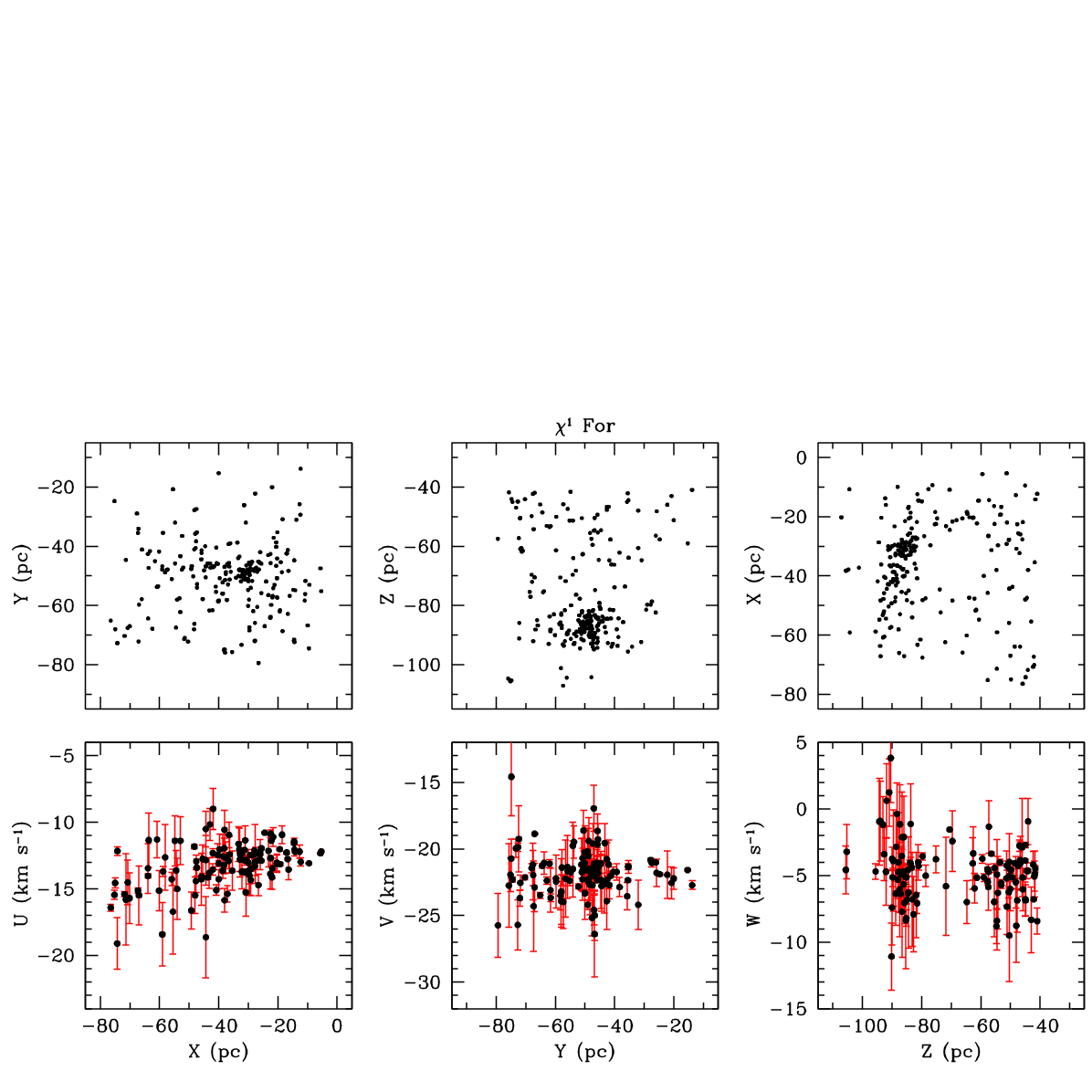}
\caption{
Galactic Cartesian coordinates and $UVW$ velocities for the adopted members 
of $\chi^1$ For.}
\label{fig:uvwfor}
\end{figure}

\begin{figure}
\epsscale{1.2}
\plotone{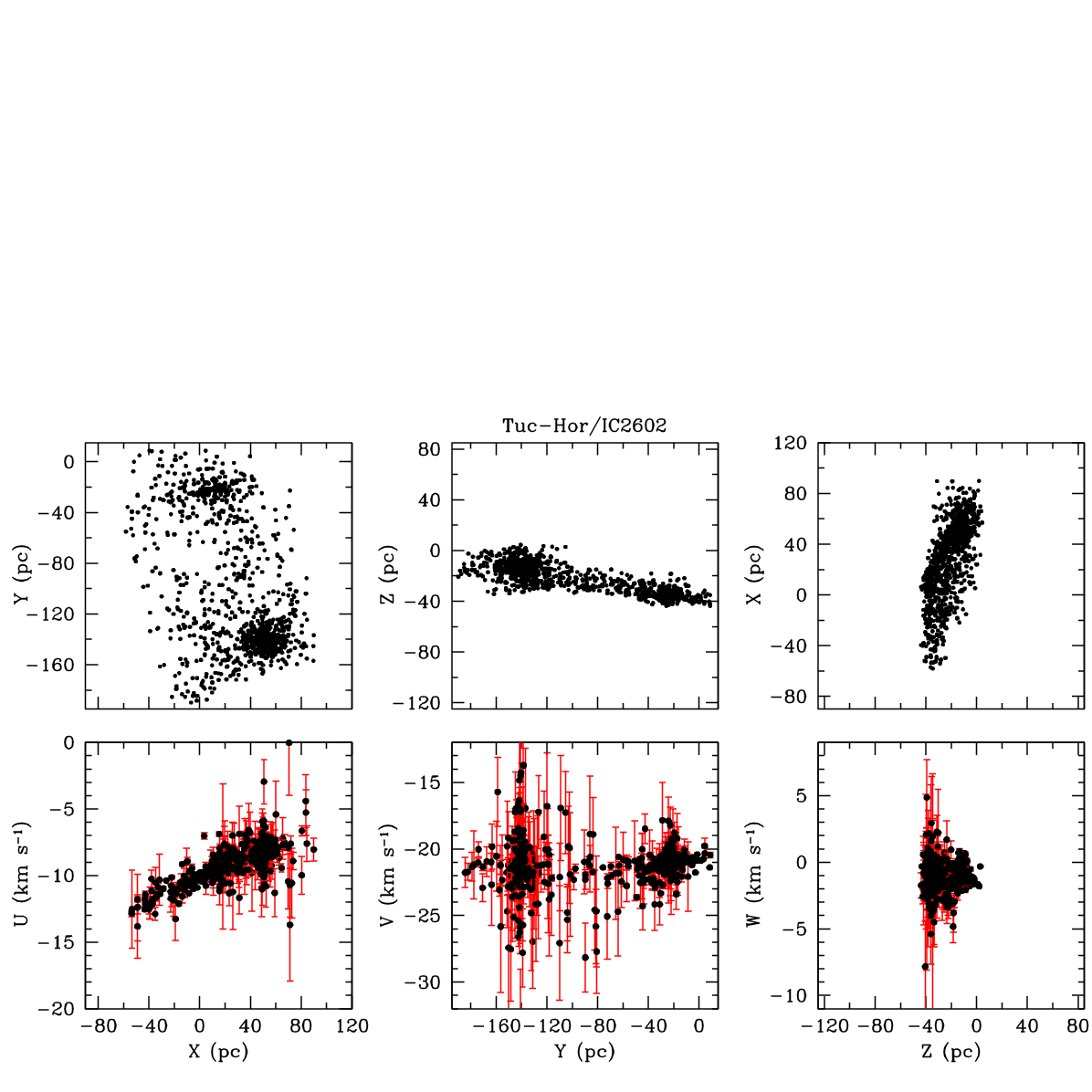}
\caption{
Galactic Cartesian coordinates and $UVW$ velocities for the adopted members 
of Tuc-Hor and IC 2602.}
\label{fig:uvwtuc}
\end{figure}

\begin{figure}
\epsscale{1.2}
\plotone{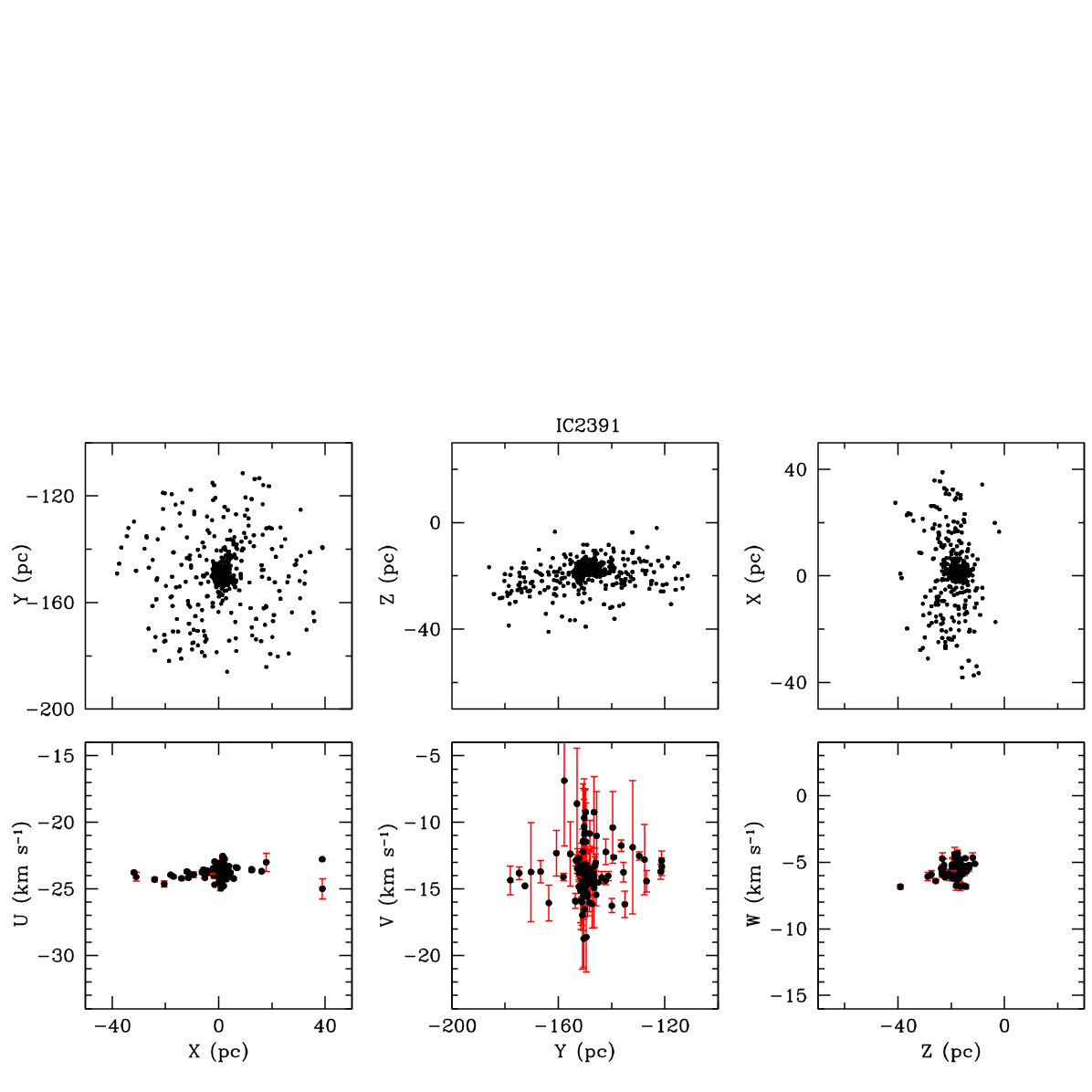}
\caption{
Galactic Cartesian coordinates and $UVW$ velocities for the adopted members 
of IC 2391.}
\label{fig:uvw2391}
\end{figure}

In Figures~\ref{fig:uvwcar}--\ref{fig:uvw2391}, I have plotted $XYZ$ and
$UVW$ for the adopted members of Car-Ext, Columba, $\chi^1$ For, Tuc-Hor,
IC 2602, and IC 2391. The first three associations are spatially adjacent 
and follow the same correlations of $UVW$ as a function $XYZ$. 
The CMDs and Li data for their low-mass stars are consistent with
the same age (Section~\ref{sec:ages}). Similarly, the spatial and kinematic
data for Tuc-Hor and IC 2602 and their CMD sequences (Section~\ref{sec:ages})
demonstrate that they represent a single population. As mentioned earlier,
most of the previously proposed members of Tuc-Hor had distances within
80 pc, or $Y>-60$ pc in Figure \ref{fig:uvwtuc}. The available measurements
of spectral types and radial velocities are concentrated among those
closer members of Tuc-Hor and within IC 2602.

\subsection{Argus}

An age of 40--50 Myr has been reported for the Argus association 
\citep{tor08,zuc19c}, so I have considered it for inclusion in this work. 
\citet{bel15} found that available membership samples for
Argus had significant contamination from nonmembers, leading them to
question the existence of the association. \citet{zuc19c} used data 
from Gaia DR2 to refine the membership for Argus and concluded that the
association does exist. 

\begin{figure}
\epsscale{1.2}
\plotone{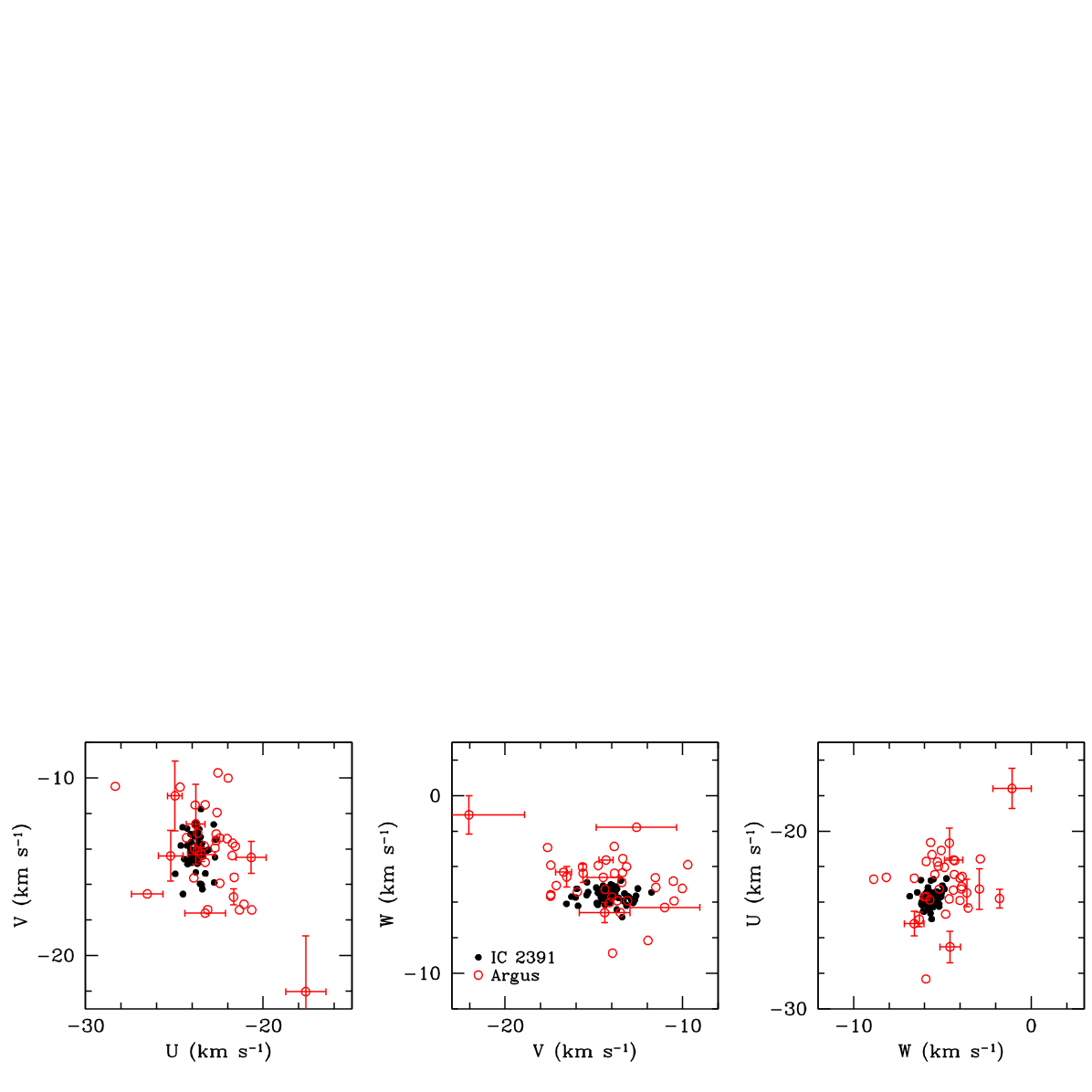}
\caption{
$UVW$ velocities for members of IC 2391 with errors less than 0.8 
km~s$^{-1}$ and proposed members of the Argus association from \citet{zuc19c}.}
\label{fig:uvwzuc}
\end{figure}

The catalog from \citet{zuc19c} contains 40 proposed members, 
39 of which have parallax data from Gaia DR3.
Radial velocity measurements are available for 37 of those 39 stars.
I have used the most accurate radial velocities for those 37 stars
\citep[Gaia DR3,][]{wil53,pou04,rie11,mal14a,sou18,bud21,zun21a}
and the astrometry from Gaia DR3 to calculate $UVW$ velocities, which are 
plotted in Figure~\ref{fig:uvwzuc}. The errors are omitted when they are 
smaller than the symbols. Since previous studies of Argus have noted a
possible relationship with IC 2391, I have included in Figure~\ref{fig:uvwzuc}
the adopted members of IC 2391 that have small velocity errors 
($<0.8$ km~s$^{-1}$). The Argus members do not show the tight clustering
in velocities that is found in IC 2391 and the other associations surveyed
in this work. To examine the ages of the Argus stars,
I have plotted them in one of the CMDs for IC 2391 in Figure~\ref{fig:cmd3}.
Most of the Argus stars are earlier than K and appear near the main
sequence, so the CMD provides poor constraints on their coevality.
Ten of the stars are redder than the main sequence turn-on for IC 2391
($G_{\rm BP}-G_{\rm RP}>1$), and they do not exhibit a well-defined sequence.
Approximately eight of the Argus stars have kinematics and CMD positions
that are consistent with IC 2391, all of which are located beyond
the 30 pc truncation radius for my IC 2391 catalog. I conclude that those
stars may be members of the extended population associated with IC 2391 
while the remaining Argus candidates from \citet{zuc19c} are field stars that 
are unrelated to IC 2391 or each other.

\subsection{Old M Stars with IR Excesses}

\begin{figure}
\epsscale{1.2}
\plotone{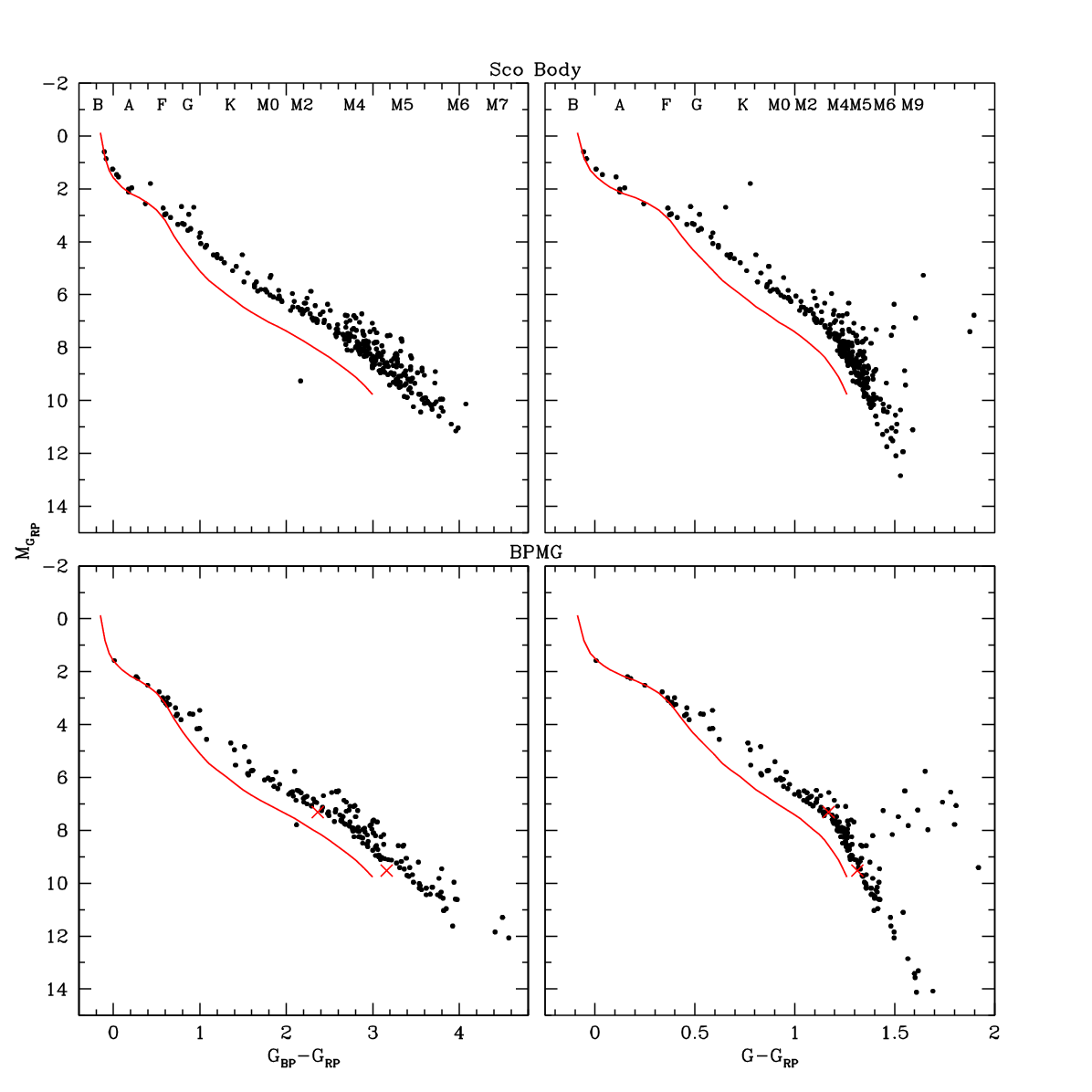}
\caption{
$M_{G_{\rm RP}}$ versus $G_{\rm BP}-G_{\rm RP}$ and $G-G_{\rm RP}$ for
adopted members of Sco Body and BPMG. In addition, 
the disk-bearing stars StH$\alpha$34 
\citep{whi05,har05} and 2MASS 15460752$-$6258042 
\citep{lee20} are plotted in the diagrams for BPMG (red crosses).
Each CMD includes a fit to the single-star sequence of the 
Pleiades. The spectral types that correspond
to the colors of young stars are indicated \citep{luh22sc}.}
\label{fig:cmd1}
\end{figure}

A growing number of low-mass stars with ages of $>$20 Myr have been found
to have mid-IR excesses that indicate the presence of disks
\citep{whi05,har05,gor07,bal09,bou16,sil16,sil20,mur18,zuc19,lee20,gai22,liu22,luh23tau}.
Some of those stars have been previously assigned membership to nearby 
associations, one of which is 2MASS J15460752$-$6258042. \citet{lee20}
classified it as a member of the Argus association based on its space velocity.
\citet{lee20} quoted an age of 55 Myr for Argus while \citet{zuc19c} estimated
an age of 40--50 Myr. As noted by \citet{lee20}, the star's Li absorption
is too strong for those ages given its spectral type (M5). Instead, its
Li measurement falls near the upper envelope of values for BPMG 
(Section~\ref{sec:ages}).
The star is near the single-star sequence for BPMG in the $G-G_{\rm RP}$
CMD and appears slightly below BPMG in $G_{\rm BP}-G_{\rm RP}$ 
(Figure~\ref{fig:cmd1}).  Given the presence of strong hydrogen emission 
lines \citep{lee20}, the older implied age in the latter CMD may be 
caused by accretion-related emission in $G_{\rm BP}$. In that case, both the 
Li data and CMDs would indicate that the star is roughly coeval with BPMG,
and therefore is not as old as reported in \citet{lee20}. The kinematics
do not match those of any of the associations considered in this work.

StH$\alpha$34 is a spectroscopic binary in which the components have similar
spectral types (M3) and are surrounded by a circumbinary disk
\citep{whi05,har05}. The system was treated as a member of the Taurus
star-forming region in some early studies, but the absence of Li absorption 
indicated an age older than expected for Taurus members.
Gaia data have confirmed that its space velocity is inconsistent with 
membership in Taurus \citep{luh18,luh23tau}. As with 2MASS J15460752$-$6258042, 
I have included StH$\alpha$34 in the CMDs for BPMG in Figure~\ref{fig:cmd1} 
and in a diagram of Li data for BPMG (Section~\ref{sec:ages}).
In the CMDs, the unresolved photometry for the system has been reduced by a 
factor of two. Based on those data, the age of StH$\alpha$34 is similar to 
that of BPMG.

The disk-bearing M stars WISEA J080822.18$-$644357.3 
and 2MASS J06320799$-$6810419 have been previously classified as members
of Carina \citep{sil16,gai22}, and they appear in my census of Car-Ext as well.
\citet{sil20} assigned two additional systems of this kind, 
WISEA 044634.16$-$262756.1 and WISEA 094900.65$-$713803.1, to Columba and
Carina, respectively. In my analysis, the former is identified as a
member of $\chi^1$ For, which is adjacent to Columba.
WISEA 094900.65$-$713803.1 is a $1\farcs5$ binary system in which only one
of the components has a proper motion offset that is consistent with Car-Ext
membership. The component that is inconsistent with membership 
appears to have a better astrometric fit based on its lower value of RUWE 
(1.0 versus 2.6), so the system is not included in my catalog for Car-Ext.

\section{Properties of Stellar Populations}

\subsection{Initial Mass Functions}
\label{sec:imf}

\begin{figure}
\epsscale{1.1} 
\plotone{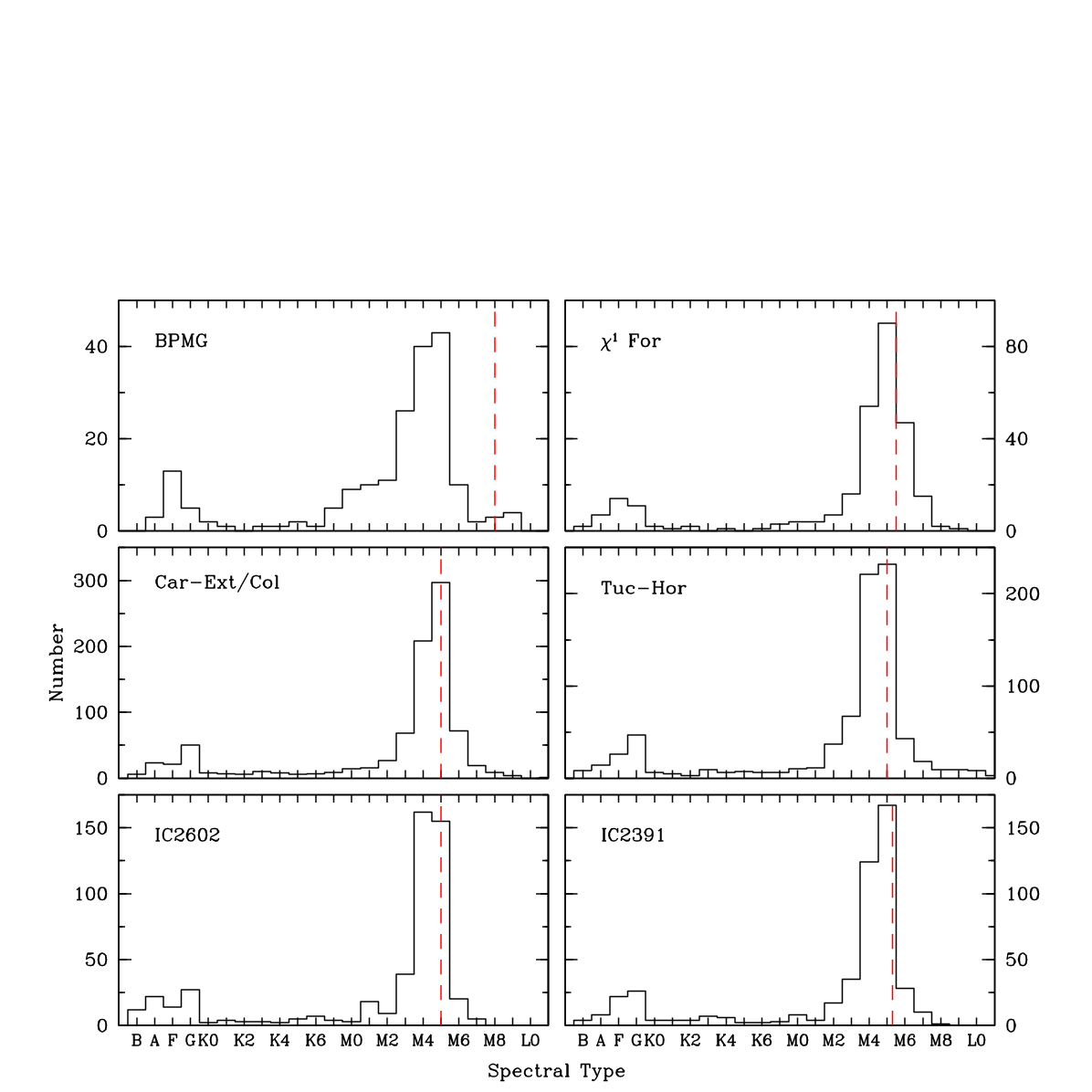}
\caption{
Histograms of spectral types for adopted members of BPMG and other young
associations. Completeness limits are indicated (dashed lines).}
\label{fig:histo}
\end{figure}

As done in my previous work, I can characterize the IMFs in the associations
that I have surveyed in terms of their histograms of spectral types.
In Figure~\ref{fig:histo}, I have plotted such histograms for the adopted
members of BPMG, Car-Ext/Columba, $\chi^1$ For, Tuc-Hor, IC 2602, and IC 2391.
For stars that lack spectroscopy, I have estimated spectral types from
Gaia and 2MASS photometry in conjunction with the typical intrinsic colors
of young stars \citep{luh22sc}. 
Five members of BPMG and Tuc-Hor are too cool ($>$L1)
to appear within the range of types shown in Figure~\ref{fig:histo}.
Gaia DR3 has a high level of completeness at
$G\lesssim19$--20 for most of the sky \citep{bou20,fab21}. 
In each histogram, I have marked the spectral type that corresponds to $G=19$
for the most distant members of the association. 
All of the associations have similar distributions of spectral types,
each reaching a maximum near M5 ($\sim0.15$~$M_\odot$), which resembles
other nearby associations and star-forming regions \citep[e.g.,][]{luh22sc}.

\subsection{X-ray Emission}

The ratio of X-ray to bolometric luminosity is high for newborn stars
and decays on a timescale that varies with stellar mass 
\citep{fei99,gud04,joh21,get22}. As a result, X-ray emission can serve
as a signature of youth in surveys for members of young populations
\citep{pre01,pre03,get02,get17,fei04,ste04,gud07,fei13}.
By imaging the entire sky with good sensitivity and angular resolution, 
the extended Roentgen Survey with an Imaging Telescope Array (eROSITA)
on board the Spectrum-Roentgen-Gamma mission \citep{pre21} is potentially
valuable for assessing the youth of candidate members of nearby associations.
eROSITA's first all-sky survey (eRASS1) has provided data
for the half of the sky between Galactic longitudes of 
$179\fdg9442$ and $359\fdg9442$ \citep{mer24}.
Within eRASS1, the single-band catalog (0.2--2.3 keV) offers the best
sensitivity, so I have identified the single-band sources that are
the closest matches to the adopted members of BPMG selected from
Gaia in Section~\ref{sec:identbp}. The Gaia positions were calculated
for an epoch of 2020 and the matching threshold was roughly twice
the positional error from eRASS1 ({\tt POS\_ERR}). For close pairs of Gaia
sources, only the component closest to the eRASS1 source was matched.
The eRASS1 coverage encompasses 110 of the 193 adopted members of BPMG. 
Among those 110 members, 66 are matched to sources from eRASS1, 13 are 
companions to matched members, and 31 lack counterparts in eRASS1.
For the latter members, I have retrieved the 3~$\sigma$ upper limits
for count rates that have been estimated at their celestial coordinates 
\citep{tub24}. I have performed the same matching process to eRASS1
for proposed members of other young associations discussed in this study,
consisting of TWA \citep{luh23twa}, 32 Ori \citep{luh22o}, 93 Tau 
\citep{luh23tau}, Sco Body, Car-Ext, Columba, $\chi^1$ For, Tuc-Hor, IC 2602,
and IC 2391 (Table~\ref{tab:mem}).

\begin{figure}
\epsscale{1.2}
\plotone{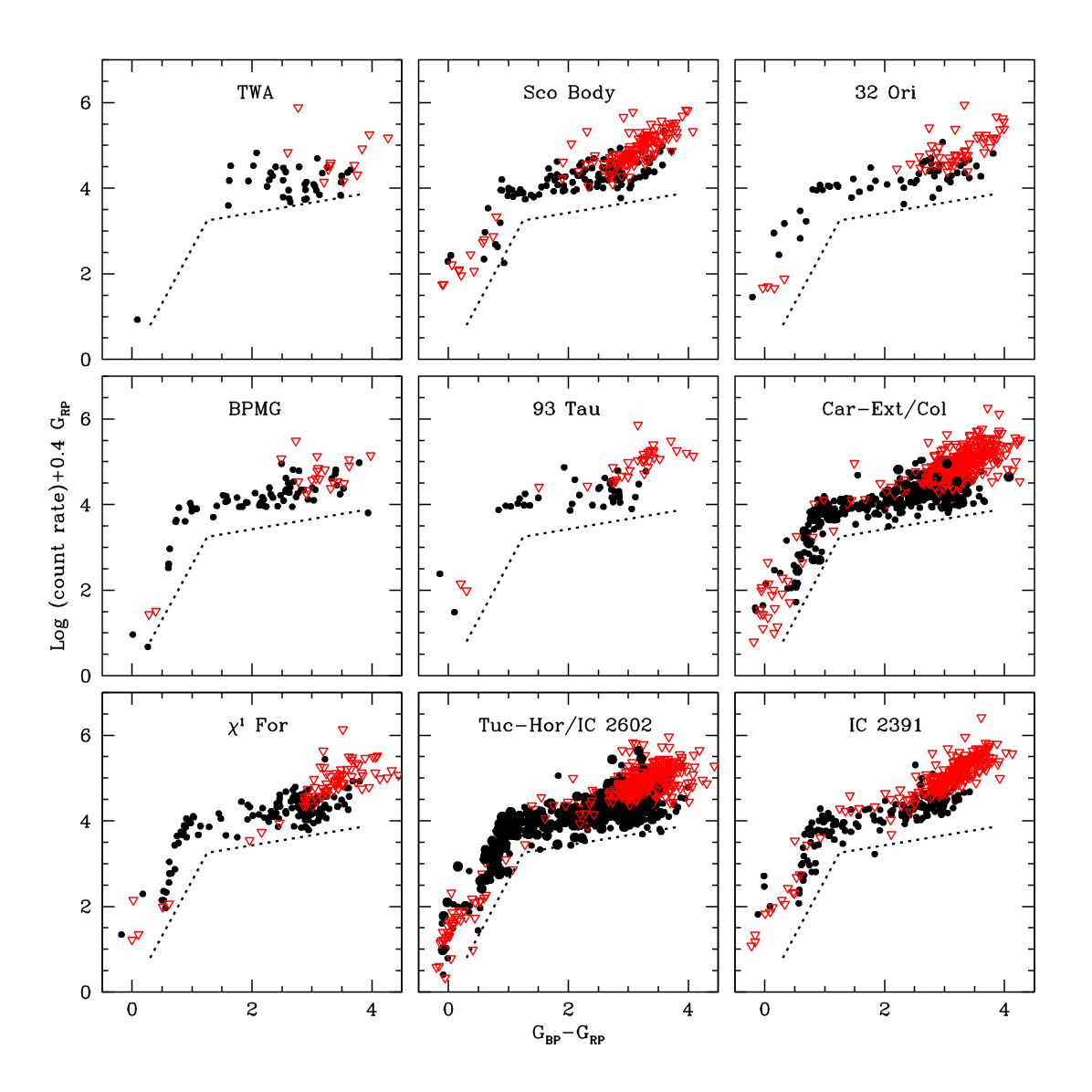}
\caption{Ratio of X-ray and $G_{\rm RP}$ fluxes versus $G_{\rm BP}-G_{\rm RP}$
for adopted members of BPMG and other young associations based on
data from eRASS1 and Gaia DR3. Upper limits for stars that are not
detected in eRASS1 are indicated (3 $\sigma$, red triangles).
A boundary that follows the lower envelope of the data in Car-Ext/Columba
is included in each diagram to facilitate comparison of the associations
(dotted line).}
\label{fig:xray}
\end{figure}

X-ray data for populations of young stars are often analyzed in terms of 
$L_{\rm X}/L_{\rm bol}$ as a function of optical color or spectral type.
As done in \citet{luh22sc} for eRASS1 data in Sco-Cen, I have calculated an 
observational parameter for the BPMG members that is analogous to 
$L_{\rm X}/L_{\rm bol}$ and that describes the ratio of X-ray and 
$G_{\rm RP}$-band fluxes, consisting of [log(count rate) + 0.4 $G_{\rm RP}$].
In Figure~\ref{fig:xray}, I have plotted that parameter versus
$G_{\rm BP}-G_{\rm RP}$ for members of BPMG and the other young associations
for which I have compiled eRASS1 data.
In each association, the eRASS1 detections are well clustered and contain
no faint outliers that would represent old nonmembers. None of the upper
limits on count rates are significantly below the loci of detections, 
indicating that the nondetections are consistent with the young ages
of the associations. There is little variation among the sequences in 
Figure~\ref{fig:xray}, which is a reflection of the fact that
$L_{\rm X}/L_{\rm bol}$ for low-mass stars ($\lesssim1$~$M_\odot$,
$G_{\rm BP}-G_{\rm RP}\gtrsim1$) remains near the saturation level across the 
age range of these associations \citep[e.g.,][]{gud04,get22}.

In the catalogs in Table~\ref{tab:mem}, I have included designations, position 
uncertainties, angular separations from Gaia counterparts, and count rates 
of the matching sources from eRASS1. The 3 $\sigma$ upper limits are provided 
for nondetections in eRASS1. A given eRASS1 source is listed only for the 
closest matching Gaia source. eRASS1 count rates and upper limits are absent 
in Table~\ref{tab:mem} for Gaia-resolved companions that were not the closest 
match to the X-ray source and for members that are located in the half of 
the sky that was not covered by eRASS1.

\subsection{Ages}
\label{sec:ages}

Ages of young associations can be estimated from their 
internal kinematics, Li depletion boundaries (LDBs), and positions in CMDs 
\citep{sod14}. Several studies have derived kinematic ages for BPMG 
\citep{ort02,ort04,mam14,cru19,mir20,cou23}, and such analyses should benefit
from the larger size and improved reliability of my catalog. 
However, I focus on the use of Li and CMDs since they
are more readily applied to the associations in this study.
For associations that have sufficient Li data, I adopt the ages produced 
by the empirical model for Li depletion from \citet{jef23},
which is called Empirical AGes from Lithium
Equivalent widthS (EAGLES)\footnote{\url{https://github.com/robdjeff/eagles}}.
For the remaining associations, I derive ages from their sequences of low-mass
stars in Gaia CMDs in a way that is calibrated to the LDB ages.

The input for EAGLES consists of effective
temperatures and Li equivalent widths for a sample of stars with
temperatures between 3000--6500 K. To estimate temperatures for stars
in my catalogs in a manner that is consistent with EAGLES,
I have derived a relation between $G_{\rm BP}-G_{\rm RP}$ and temperature 
from the young stars in \citet{jef23} that were used to develop EAGLES.
The Gaia colors were corrected for reddening using the
extinctions from \citet{jac22} and the appropriate reddening 
coefficients\footnote{\url{https://www.cosmos.esa.int/web/gaia/edr3-extinction-law}}.
My adopted relation is defined by lines connecting the following points: 
(log $T$, $G_{\rm BP}-G_{\rm RP}$)=(3.81, 0.61), (3.71, 1.03),
(3.67, 1.25), (3.59,1.91), (3.55, 2.41), (3.50, 2.92), (3.477, 3.3).

\begin{figure}
\epsscale{1.1}
\plotone{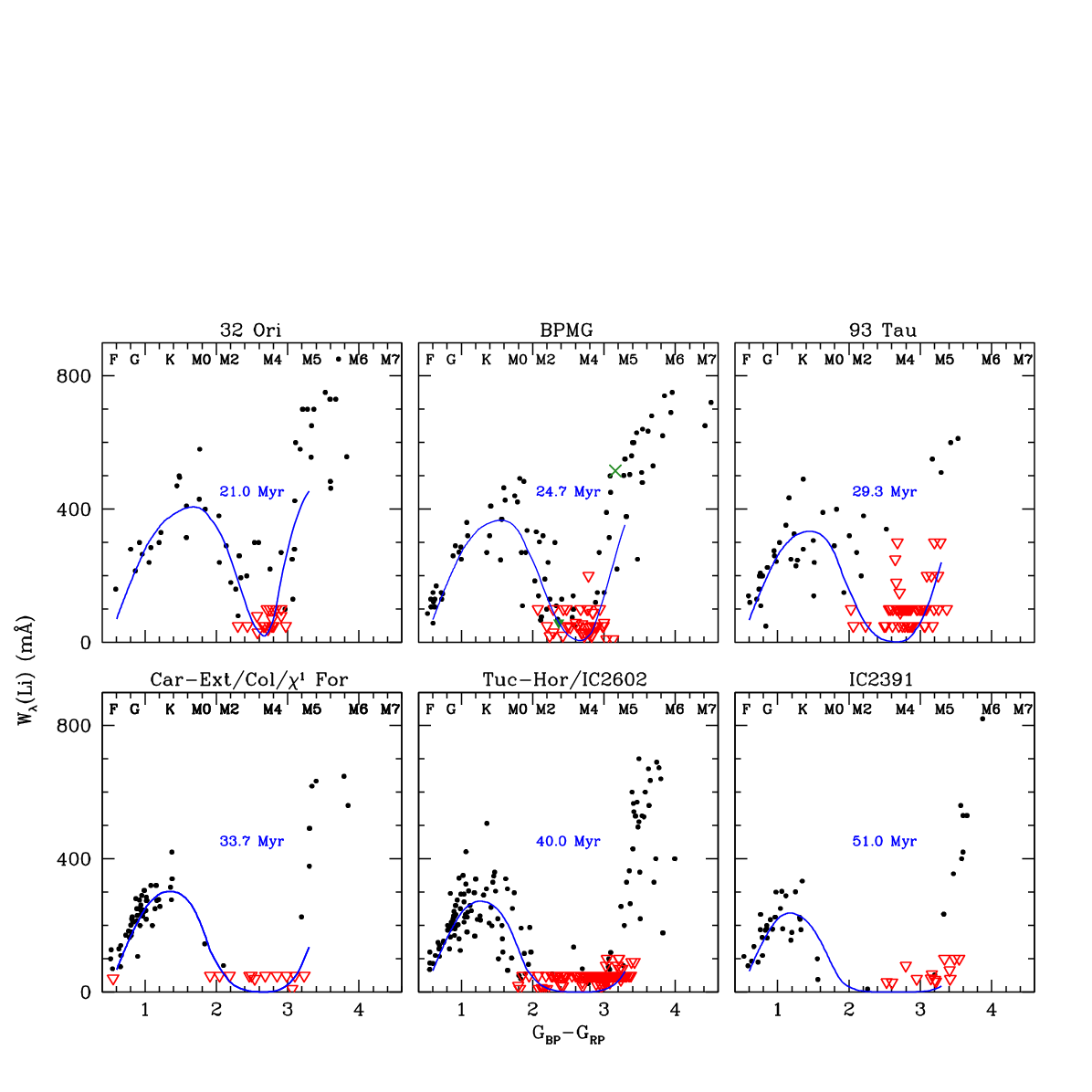}
\caption{Equivalent widths of Li versus $G_{\rm BP}-G_{\rm RP}$ for
adopted members of BPMG and other young associations
(black points and red triangles) and the values produced 
by the EAGLES model for Li depletion \citep{jef23} for the best-fitting ages
(blue lines). The disk-bearing stars StH$\alpha$34 \citep{whi05,har05} and 
2MASS 15460752$-$6258042 \citep{lee20} are plotted in the diagram for BPMG 
(green triangle and green cross).  The spectral types that correspond to the 
colors of young stars are indicated \citep{luh22sc}.}
\label{fig:li}
\end{figure}

I have applied EAGLES to the associations in this study
that have sufficient Li measurements.
The kinematics of Car-Ext, Columba, and $\chi^1$ For indicate that they
have a physical relationship, and their sequences of low-mass stars in
the Gaia CMDs are consistent with coevality. The same is true for
Tuc-Hor and IC 2602. Therefore, I have derived a LDB age from
the combined Li data for Car-Ext, Columba, and $\chi^1$ For, and I have done
the same for Tuc-Hor and IC 2602. I have excluded Li data for stars
that have IR excesses from disks (Section~\ref{sec:disks})
since $G_{\rm BP}-G_{\rm RP}$ can be affected by accretion-related emission. 
In Figure~\ref{fig:li}, I have plotted Li equivalent widths versus
$G_{\rm BP}-G_{\rm RP}$ for members of BPMG and other associations.
The membership samples for 32 Ori and 93 Tau are from \citet{luh22o} and
\citet{luh23tau} and their Li data are from previous studies and unpublished
spectra (K. Luhman, in preparation). The best-fitting ages from EAGLES are 
21.0$^{+1.0}_{-0.7}$ (32 Ori), 24.7$^{+0.9}_{-0.6}$ (BPMG),
29.3$\pm$1 (93 Tau), 33.7$^{+2.0}_{-1.9}$ (Car-Ext/Columba/$\chi^1$ For),
40.0$^{+1.9}_{-1.4}$ (Tuc-Hor/IC 2602), and 51.0$^{+5.6}_{-3.9}$ Myr (IC 2391).
The Li equivalent widths predicted by EAGLES for those ages are indicated by
the curves in Figure~\ref{fig:li}. In addition, I have applied EAGLES to 
my adopted members of NGC 2547 using Li data from \citet{jef23}, 
arriving at an age of 34.9$\pm$1.2 Myr.
The LDB age for IC 2391 has a large uncertainty because of the small
number of Li measurements at $G_{\rm BP}-G_{\rm RP}=1.4$--2 (K4--M1).
In addition, the two stars with the lowest Li at $G_{\rm BP}-G_{\rm RP}=1$--2
(one of which is a low outlier for its color) have a large effect on the
age. If they are excluded, the LDB age is 48.7 Myr. Li measurements for 
additional K and early M stars are needed for a reliable LDB age in IC 2391.

Previous studies have estimated LDB ages for BPMG and some of the other 
associations in my analysis 
\citep[e.g.,][]{men08,yee10,bin14,mam14,shk17,gal22}. 
Comparison of my LDB ages to those estimates is complicated by the fact the
LDB age for a 
given association depends on the sample of adopted members, the choice of 
evolutionary models, and the details of how the models and data are compared.
Nevertheless, my LDB ages tend to be similar to earlier estimates. I note
that the samples of members can differ substantially among studies.
For instance, 60 of the stars adopted as BPMG members by \citet{gal22}
are absent from my catalog.

\begin{figure}
\epsscale{1.1}
\plotone{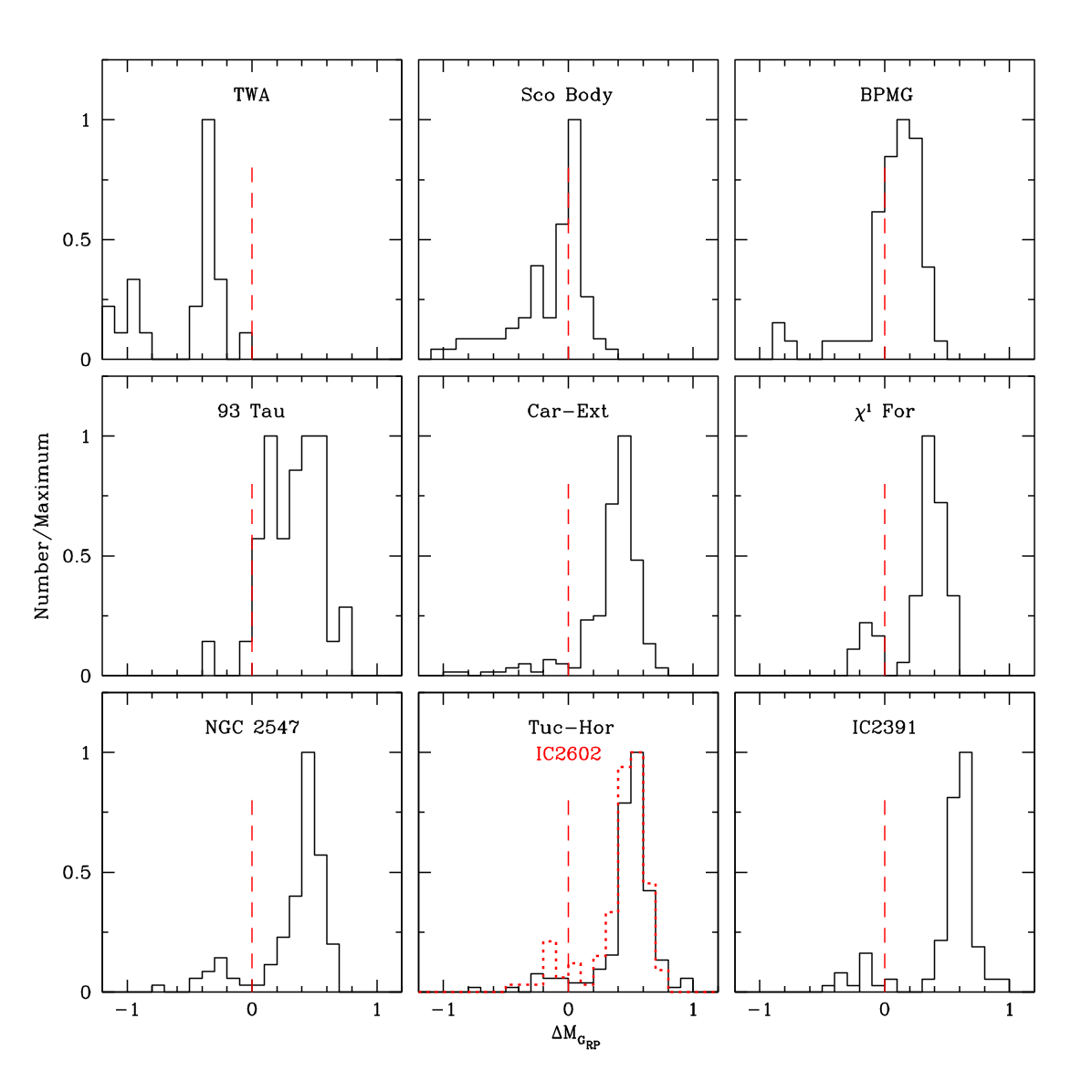}
\caption{
Histograms of offsets in $M_{G_{\rm RP}}$ from the median CMD sequence for
UCL/LCC for low-mass stars in BPMG and other young associations
\citep[Figures~\ref{fig:cmd1} and 
\ref{fig:cmd2}--\ref{fig:cmd4},][]{luh23tau,luh23twa}.
Negative values correspond to brighter magnitudes and younger ages.}
\label{fig:ages}
\end{figure}

The high-precision photometry and astrometry from Gaia have reduced
contamination from nonmembers in catalogs of young associations and have
provided accurate measurements of absolute magnitudes for members, both
of which have resulted in tight sequences of low-mass stars
in CMDs that can be used to estimate relative ages among associations.
As done in \citet{luh23tau}, I plot in Figure~\ref{fig:ages}
histograms of offsets in $M_{G_{\rm RP}}$ ($\Delta M_{G_{\rm RP}}$)
from the median CMD sequence for 
UCL/LCC for low-mass stars ($G_{\rm BP}-G_{\rm RP}=1.4$--2.8)
in TWA, Sco Body, BPMG, 93 Tau, Car-Ext, $\chi^1$ For, NGC 2547, Tuc-Hor,
IC 2602, and IC 2391. Some of the results from these data are as follows:
(1) Sco Body is coeval with UCL/LCC and
is younger than BPMG; (2) Car-Ext and $\chi^1$ For are coeval; 
(3) Tuc-Hor and IC 2602 are coeval; and (4) Car-Ext/$\chi^1$ For are 
slightly younger than Tuc-Hor/IC 2602. 

\begin{figure}
\epsscale{1.1}
\plotone{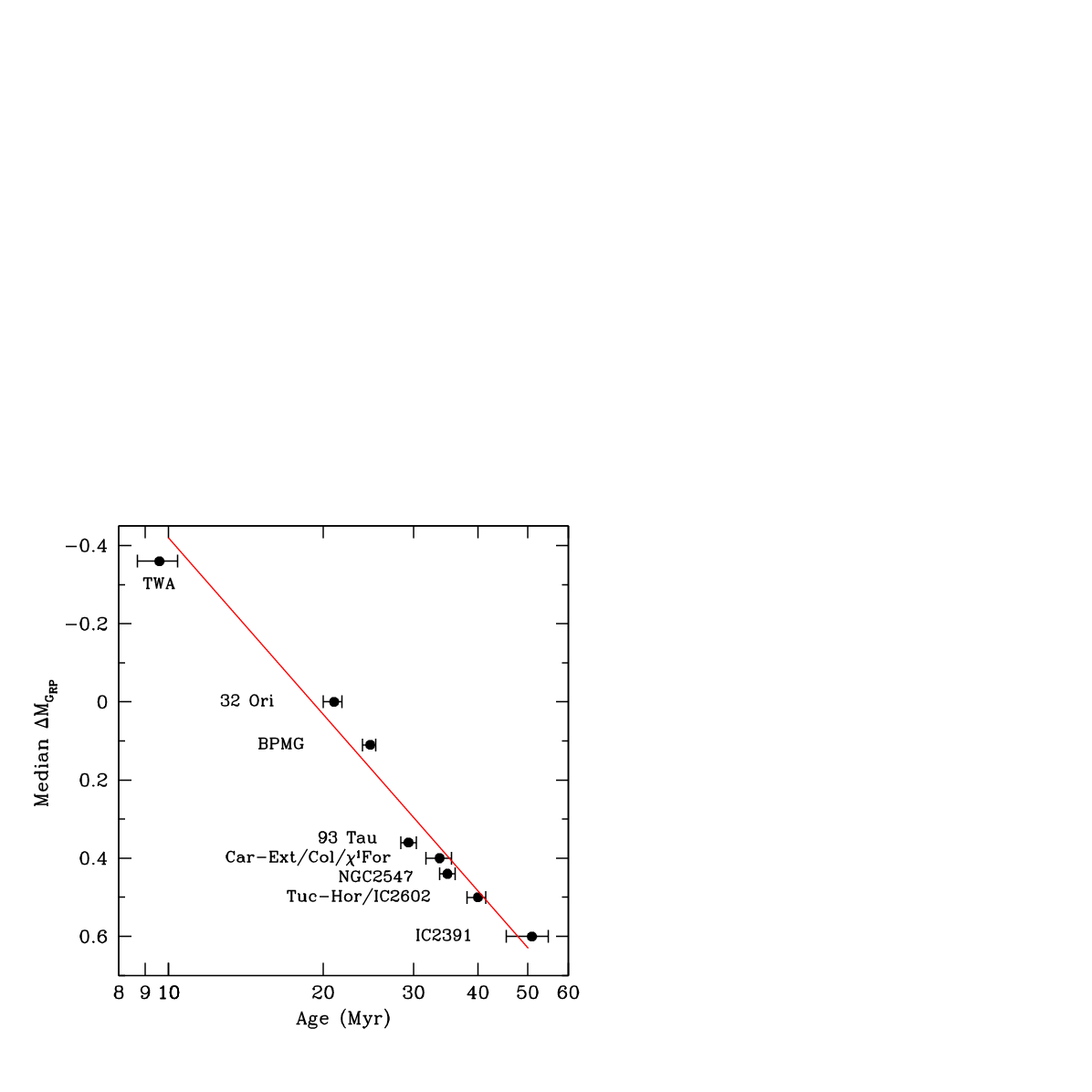}
\caption{
Median offset in $M_{G_{\rm RP}}$ from the median CMD sequence for
UCL/LCC (Figure~\ref{fig:ages}) versus age for low-mass stars in BPMG
and other young associations. The age for TWA is based on its expansion 
\citep{luh23twa} while the other ages are from the LDB (Figure~\ref{fig:li}).
A linear relation with a slope of 1.5 and an arbitrary intercept is
indicated (red line).}
\label{fig:ages2}
\end{figure}

In Figure~\ref{fig:ages2}, I have plotted the median values of
$\Delta M_{G_{\rm RP}}$ versus the LDB ages for 32 Ori, BPMG, 
93 Tau, Car-Ext/Columba/$\chi^1$ For, NGC 2547, Tuc-Hor/IC 2602, and IC 2391. 
In addition, I have included TWA using its expansion age \citep{luh23twa}.
For the temperatures and ages in question, theoretical evolutionary 
models predict that isochrones should fade at a rate approximated
by $\Delta$M$_{\rm bol}$/$\Delta$log~age$=1.5$ \citep{bar15,cho16,dot16,fei16}, 
so I have marked a line in Figure~\ref{fig:ages2} with that slope and an 
arbitrary intercept. The LDB ages exhibit a steeper slope of $\sim$2. 
Meanwhile, TWA and the four populations at 30--40 Myr
are consistent with the slope predicted by evolutionary models.
It is unclear whether the departures from linearity between the LDB ages
and the median $\Delta M_{G_{\rm RP}}$ accurately reflect the luminosity 
evolution or whether they originate from an error in the LDB model.
Combining the LDB age for Tuc-Hor/IC 2602, the 0.1 mag offset in 
$M_{G_{\rm RP}}$ between Tuc-Hor/IC 2602 and IC 2391, and a slope of 1.5,
the resulting age for IC 2391 is 47 Myr, which is adopted for the
disk analysis (Section~\ref{sec:disks}).
Based on the LDB age of 32 Ori and the value of $\Delta M_{G_{\rm RP}}$
of Sco Body relative to 32 Ori, I adopt an age of 20 Myr for Sco Body.

\section{Properties of Disk Populations}
\label{sec:disks}

\subsection{Measurement of Infrared Excesses}

I have used mid-IR data from WISE to search for excess emission
from disks among the adopted members of BPMG and the other
populations from Section~\ref{sec:othergroups}.
If two Gaia sources from my catalogs have the same WISE source as their 
closest match, the WISE designation is assigned to both of them in 
Table~\ref{tab:mem}, but WISE data and disk measurements are listed only 
for the member that is closest to the WISE source. 
The same approach is taken for matching to
sources from 2MASS. I have visually inspected the AllWISE Atlas images 
and unWISE images \citep{lan14,mei19} to check for detections that are 
false or unreliable, which are flagged in Table~\ref{tab:mem}.
For a few stars, the detections in W3 or W4 are shifted noticeably
relative to the positions in W1 and W2, and thus are likely arising from
a disk-bearing companion or a red background source. If a companion is not
detected by Gaia, the latter scenario is more likely, and the shifted
long-wavelength detection is flagged as false.

NGC 2547 is more distant that the other populations considered in my disk
analysis (383 pc), but most of it was observed with Spitzer
\citep{you04,gor07,for08}, which provided better angular resolution and 
sensitivity than WISE. The Spitzer images were obtained at 3.6, 4.5, 5.8, 
and 8.0~\micron\ ([3.6], [4.5], [5.8], [8.0]) with the Infrared Array Camera
\citep[IRAC;][]{faz04} and at 24~\micron\ ([24]) with the Multiband Imaging
Photometer for Spitzer \citep[MIPS;][]{rie04}.
I have utilized the catalog of IRAC sources from \citet{gor07}
and a new catalog of sources that I have derived from the MIPS images.
The IRAC and MIPS data for the members of NGC 2547 are included in 
Table~\ref{tab:ngc}.

\begin{figure}
\epsscale{1.2}
\plotone{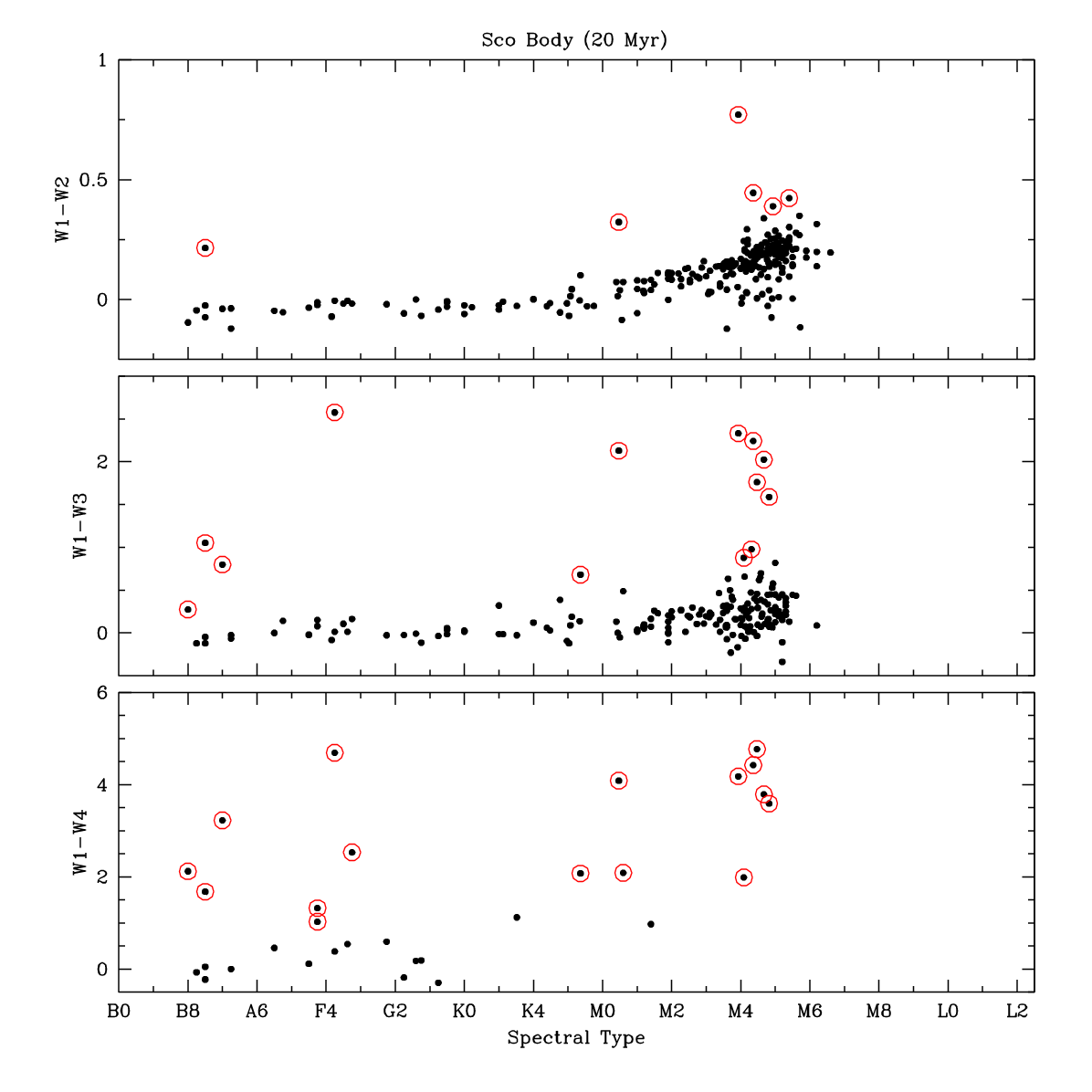}
\caption{ 
IR colors versus spectral type for adopted members of Sco Body. For stars that
lack spectroscopy, spectral types have been estimated from photometry.
In each diagram, the sequence of blue colors corresponds to stellar
photospheres. Color excesses from disks are indicated (red circles).
}
\label{fig:excsco}
\end{figure}

\begin{figure}
\epsscale{1.2}
\plotone{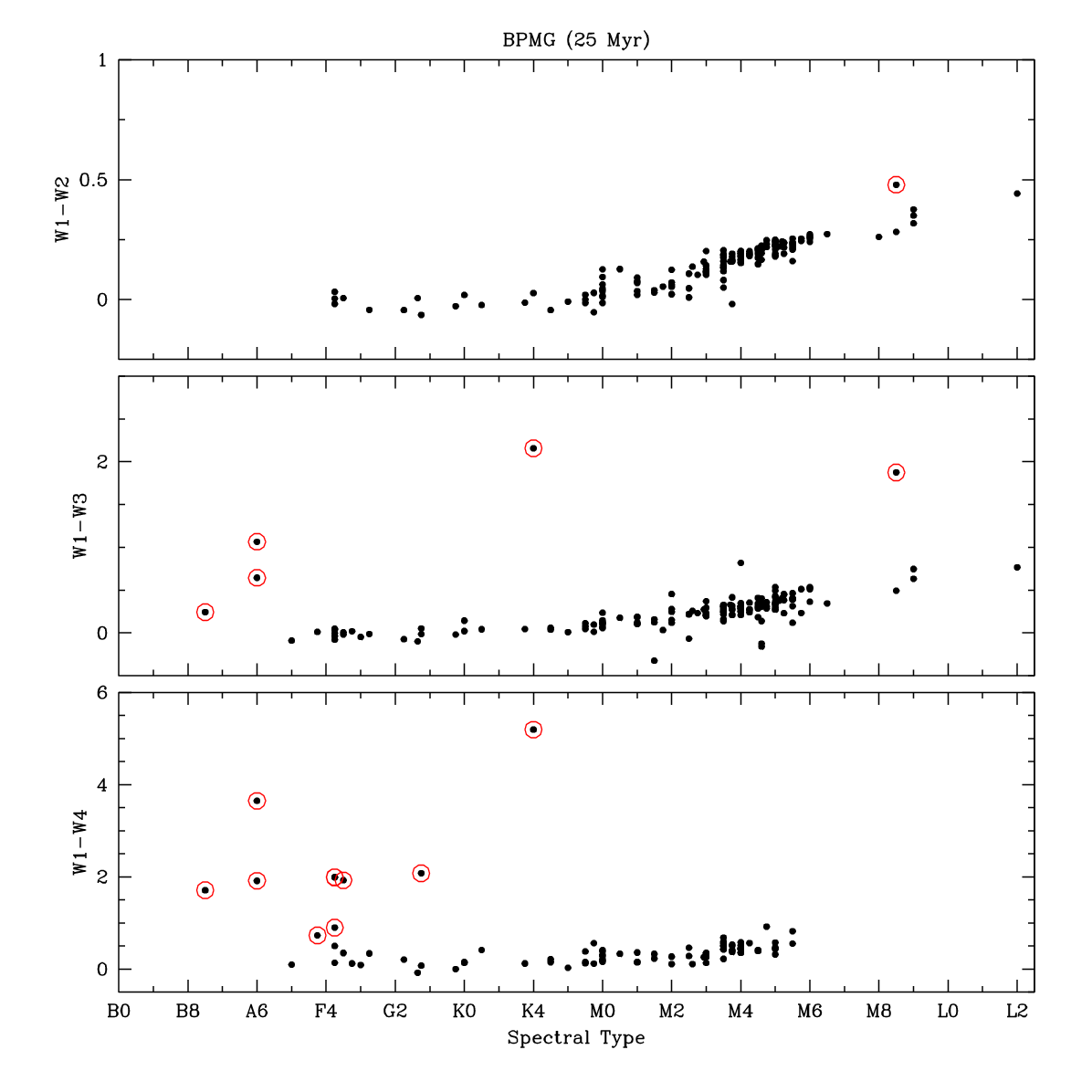}
\caption{
IR colors versus spectral type for adopted members of BPMG.}
\label{fig:excbp}
\end{figure}

\begin{figure}
\epsscale{1.2}
\plotone{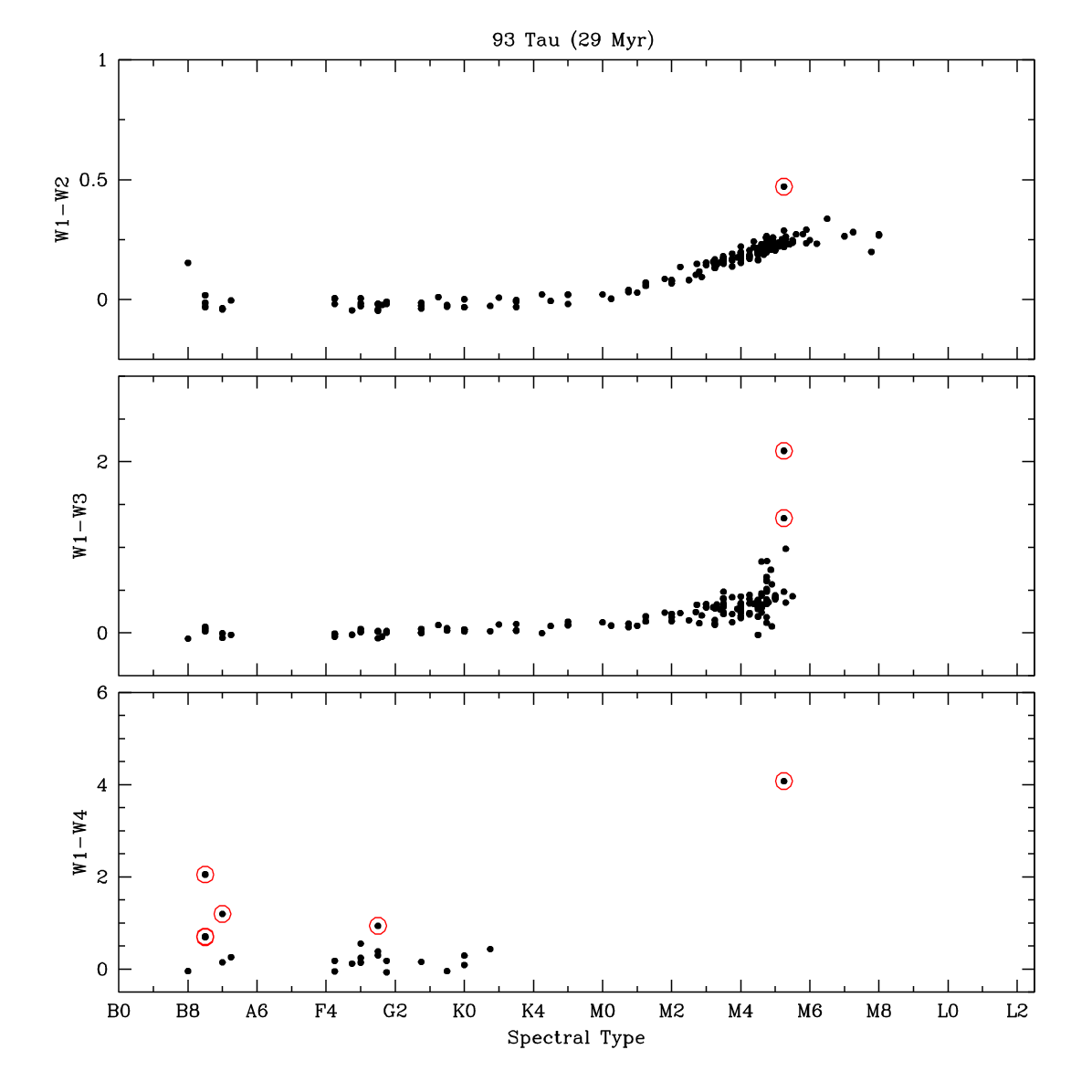}
\caption{
IR colors versus spectral type for adopted members of 93 Tau \citep{luh23tau}.}
\label{fig:exc93}
\end{figure}

\begin{figure}
\epsscale{1.2}
\plotone{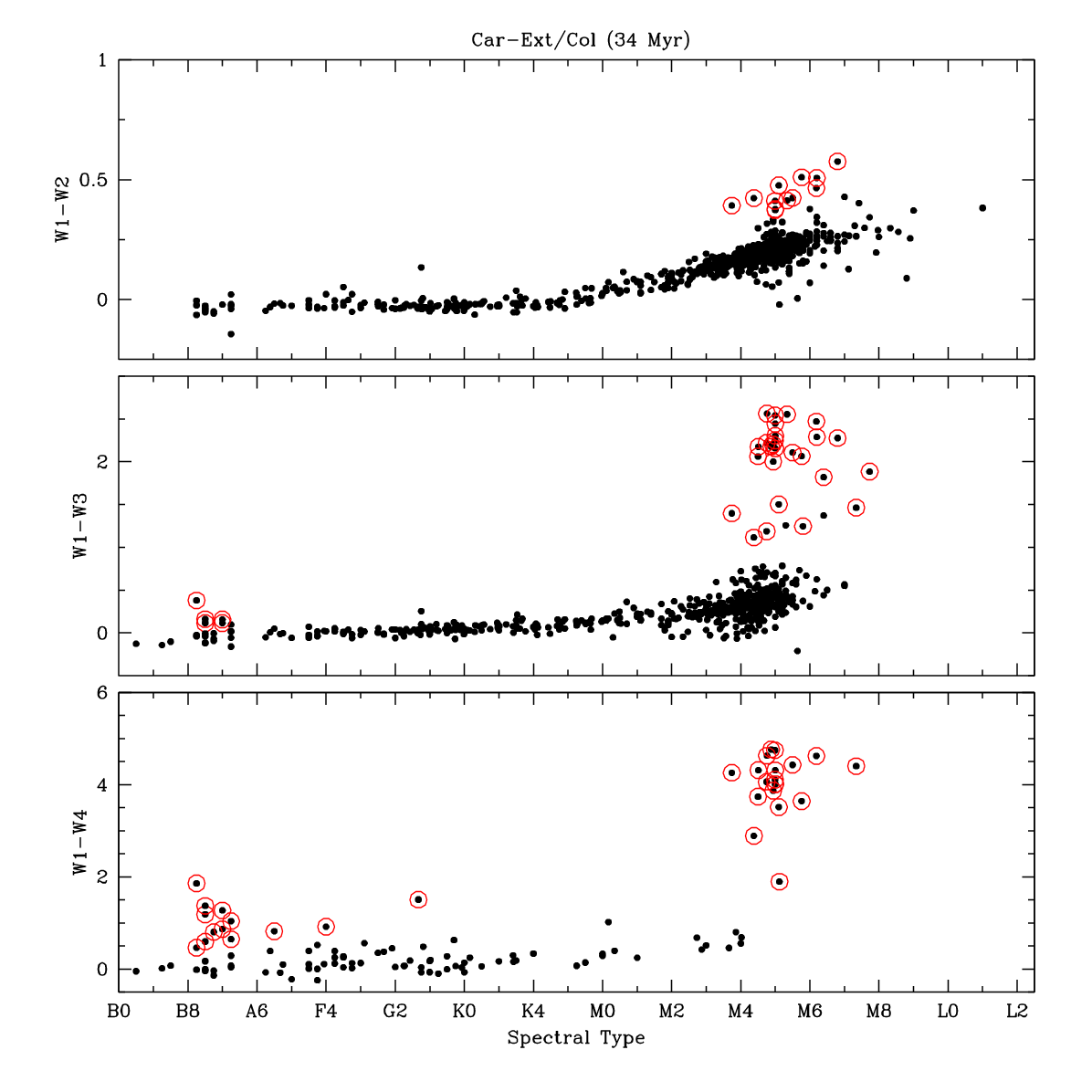}
\caption{
IR colors versus spectral type for adopted members of Car-Ext and Columba.}
\label{fig:exccar}
\end{figure}

\begin{figure}
\epsscale{1.2}
\plotone{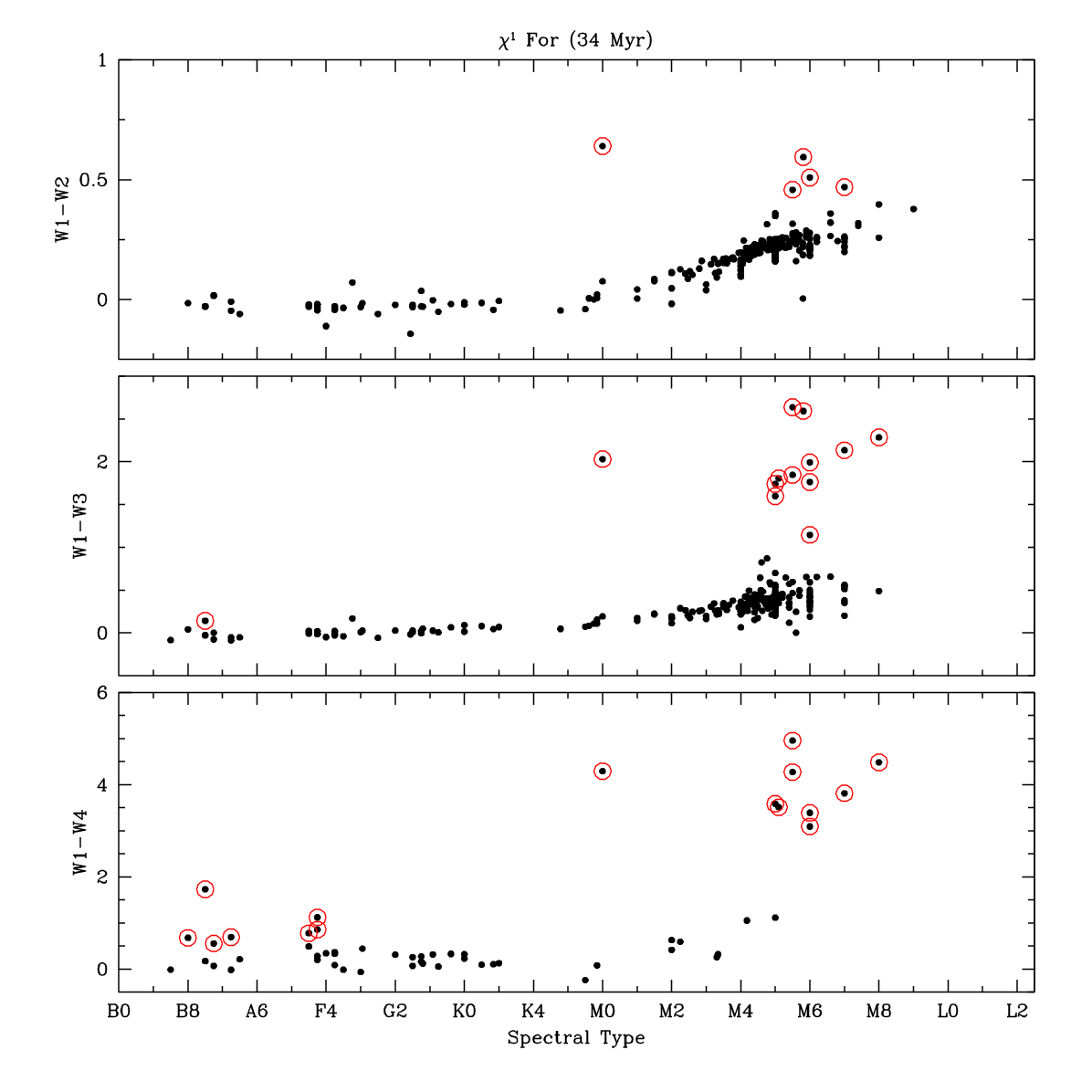}
\caption{
IR colors versus spectral type for adopted members of $\chi^1$ For.}
\label{fig:excfor}
\end{figure}

\begin{figure}
\epsscale{1.2}
\plotone{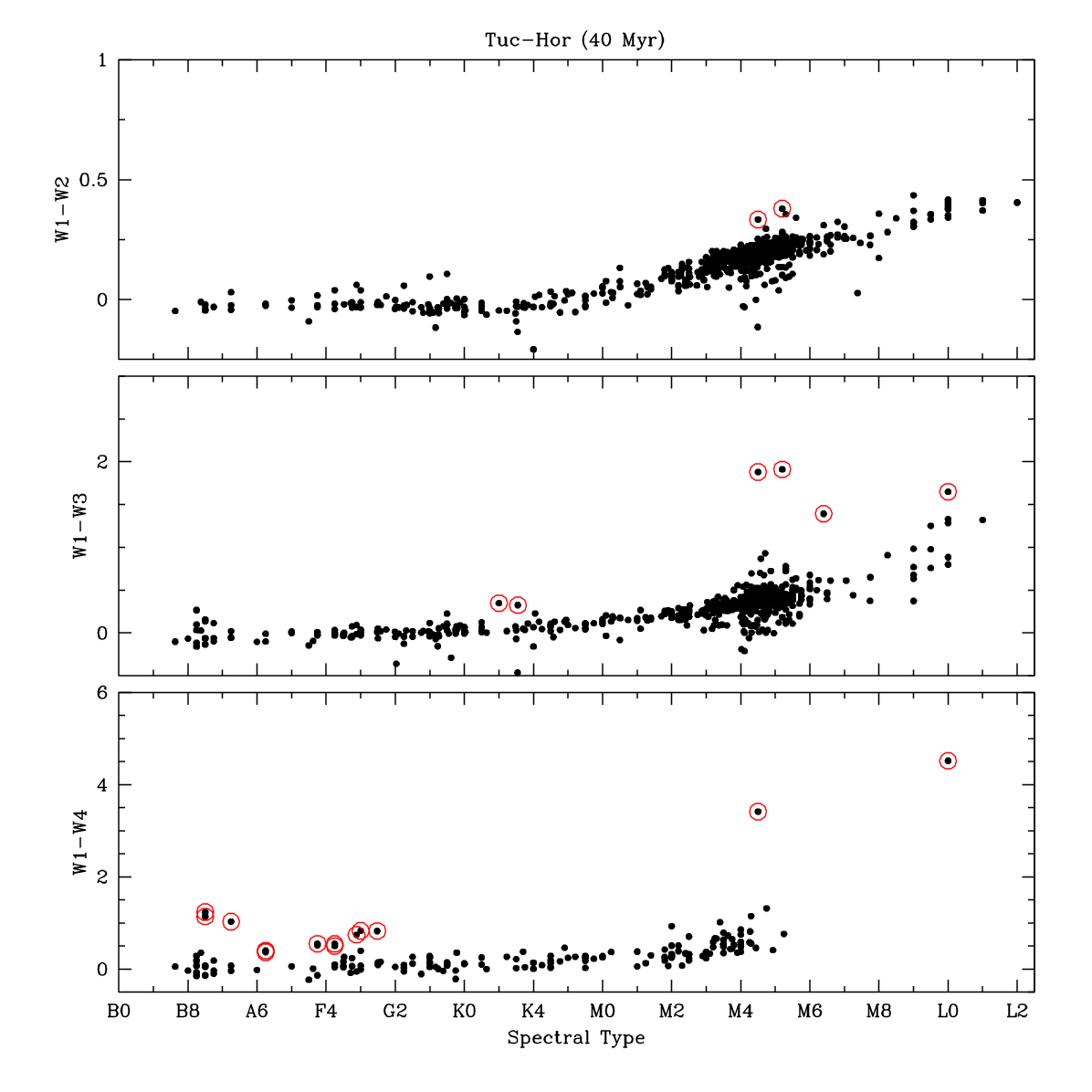}
\caption{
IR colors versus spectral type for adopted members of Tuc-Hor.}
\label{fig:exctuc}
\end{figure}

\begin{figure}
\epsscale{1.2}
\plotone{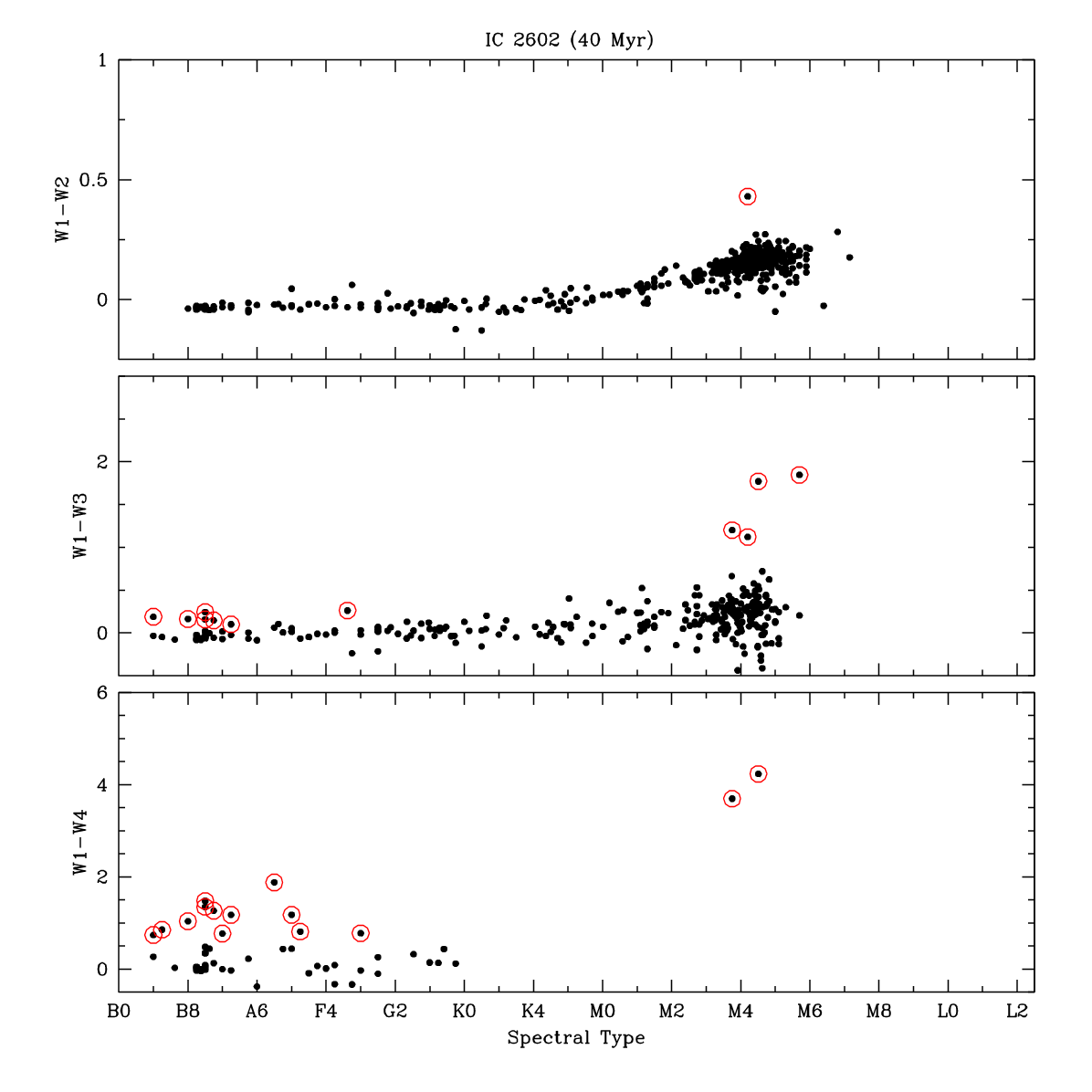}
\caption{
IR colors versus spectral type for adopted members of IC 2602.}
\label{fig:exc2602}
\end{figure}

\begin{figure}
\epsscale{1.2}
\plotone{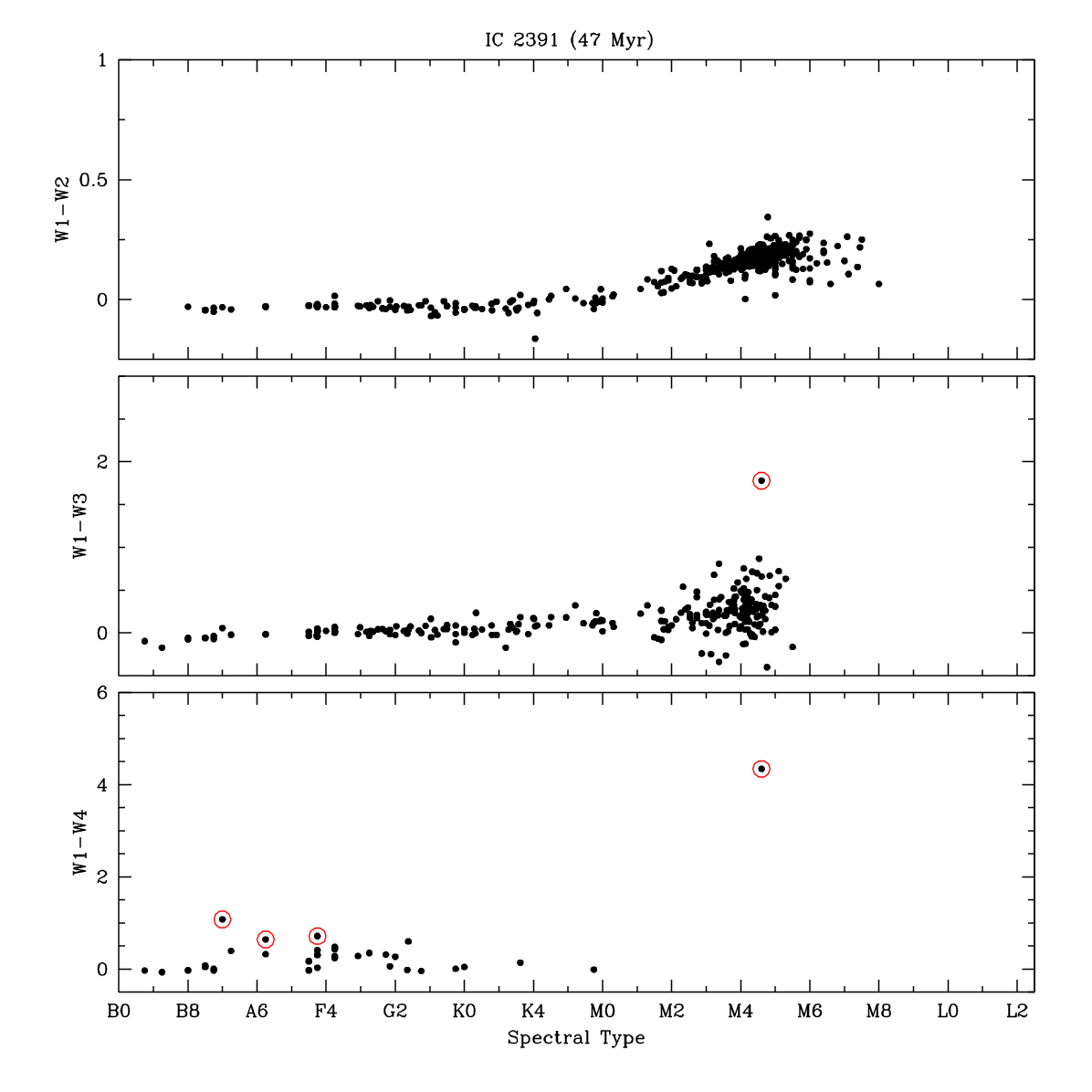}
\caption{
IR colors versus spectral type for adopted members of IC 2391.}
\label{fig:exc2391}
\end{figure}

\begin{figure}
\epsscale{1.2}
\plotone{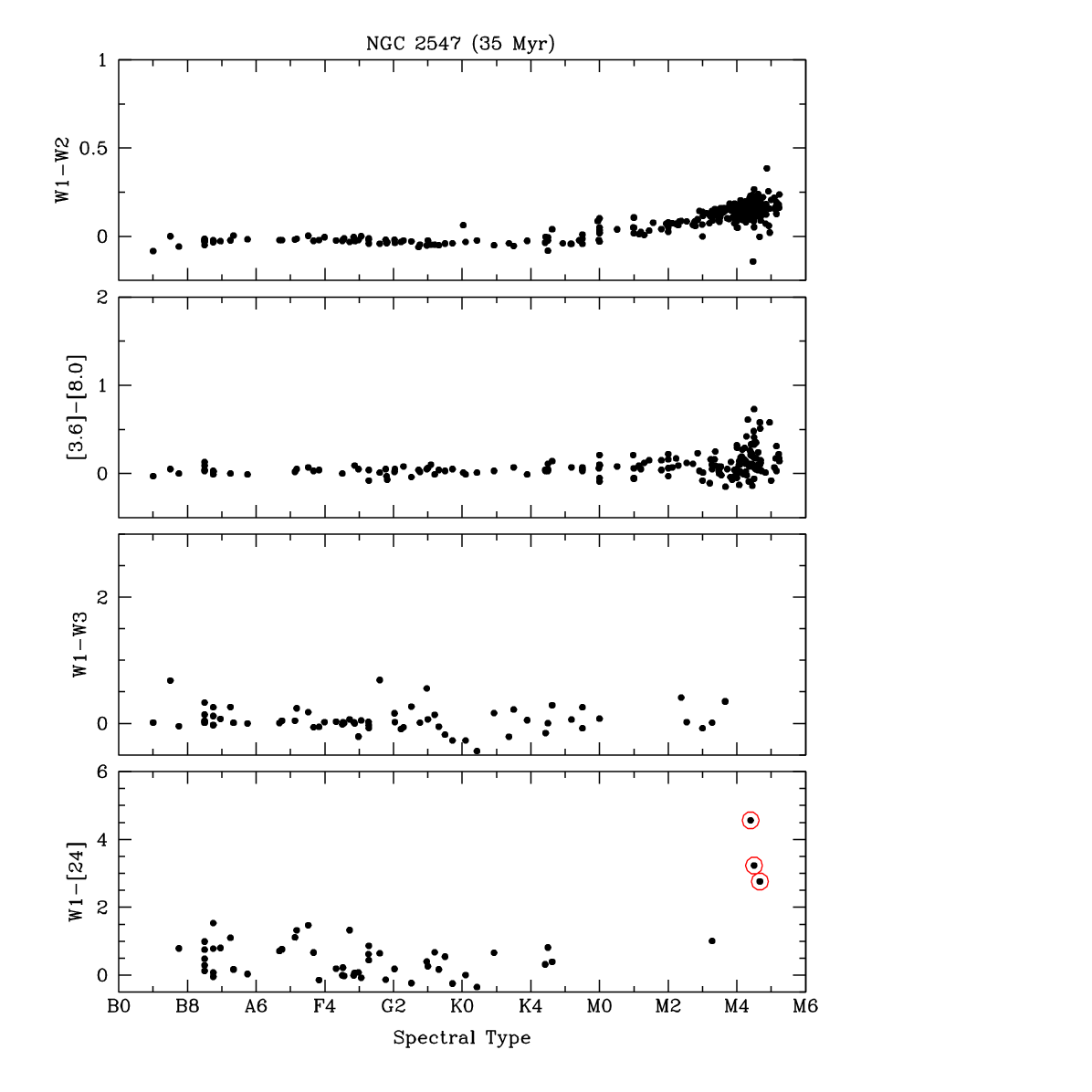}
\caption{
IR colors versus spectral type for adopted members of NGC 2547
\citep[][Table \ref{tab:ngc}]{gor07,for08}.}
\label{fig:excngc}
\end{figure}

As done in my previous surveys of young associations 
\citep[e.g.,][]{luh22disks}, I have used W1$-$W2, W1$-$W3, and W1$-$W4 to
detect excess emission from disks.  Those colors are plotted versus spectral 
type in Figures~\ref{fig:excsco}--\ref{fig:exc2391} for Sco Body, BPMG, 93 Tau, 
Car-Ext/Columba, $\chi^1$ For, Tuc-Hor, IC 2602, and IC 2391, which are in
order of age. For stars that lack spectroscopy, spectral types are estimated
from photometry. The excess measurements for 93 Tau were presented in 
\citet{luh23tau} and are included in this work for comparison.
Although Tuc-Hor and IC 2602 form a single population, their data are shown 
separately to illustrate how the limiting spectral type depends on distance.
Spitzer and WISE colors for members of NGC 2547 are shown in 
Figure~\ref{fig:excngc}.

In each mid-IR color in Figures~\ref{fig:excsco}--\ref{fig:excngc}, 
stellar photospheres appear as a blue sequence and stars with disk excesses 
have a wider range of redder colors. Significant color excesses are circled
in those diagrams. Flags indicating the presence or absence of excesses 
in W2, W3, and W4 are included in Table~\ref{tab:mem}.
A flag for [24] excesses in NGC 2547 is provided in Table~\ref{tab:ngc}.
An apparent color excess in a given band is flagged only if all detections
at longer wavelengths also show excesses. In other words, a few stars 
seem to have excesses at a shorter wavelength but lack excess emission
in detections at longer wavelengths, so those bands are not flagged for
excesses. A few stars in Figures~\ref{fig:excsco}--\ref{fig:excngc}
are significantly bluer or redder than the photospheric sequence and
are not circled as excesses (for the red colors). Most of those stars are
blended with other stars, so those deviations from photospheric
colors probably reflect erroneous photometry.

The sizes of mid-IR excesses can be used to estimate
the evolutionary stages of disks from among the following options:
full disk, transitional disk, evolved disk, evolved transitional disk, 
and debris disk \citep{ken05,rie05,her07,luh10,esp12}. The first four
classes apply to primordial disks while debris disks usually contain
little or no gas and consist primarily of second-generation dust that
is produced by collisions among planetesimals. Stars that lack any
detected excess emission are designated as class~III \citep{lw84,lad87}.
The mid-IR colors of primordial and debris disks do overlap, so they
do not provide definitive classifications.
The latter can be refined by measuring spectral energy distributions at
higher resolution and at longer wavelengths and by measuring tracers of gas.
For the stars that exhibit IR excesses, I initially classified their
disks based on the sizes of the excesses in $K_s-$W3 and $K_s-$W4
\citep{luh12u,esp14,esp18}. Those classifications were then revised
based on other available data. All B/A/F/G stars with excesses are
classified as debris systems unless they show evidence of accretion
from previous measurements \citep[AK Sco,][]{jen97}. 
For the M-type stars that have IR excesses in BPMG and the older populations, 
I list their disks as unclassified given the uncertainty in the nature of
disks around low-mass stars at those ages (Section~\ref{sec:fractions}).
Stars that lack mid-IR excesses but show disk emission at far-IR or
millimeter wavelengths in previous studies are labeled as debris disks.

\subsection{Sco Body}

Nineteen members of Sco Body exhibit excesses in the WISE data
(Figure~\ref{fig:excsco}). Evidence of disks has been previously 
reported for 15 of those stars \citep{cru14,liu14,cot16,van21,luh22disks},
which includes the well-studied binary system AK Sco \citep{jen97}.
Among the previously known disks, WISEAJ 164656.06$-$324254.3 is the 
counterpart for a $3\farcs9$ binary system in which the primary dominates 
in W1 and W2 and the disk-bearing secondary dominates in W4. 
One of the four stars with newly identified excesses, Gaia DR3 
5976613675832060288, has only a tentative detection of an excess in a single
band (W3). A second of those four stars, Gaia DR3 5962199593783949952,
has a tentative excess in W2, lacks detections in W3 and W4, and
has excesses at [4.5], [5.8], and [8.0] in archival IRAC data
\citep{ben03,chu09}. 

\subsection{BPMG}
\label{sec:bpexc}

The WISE data detect IR excesses for 11 members of BPMG 
(Figure~\ref{fig:excbp}). One of those systems is the K star V4046 Sgr, 
which has a well-known primordial disk \citep{del86,jen97,ros13}.
The late-M member Gaia 6752579812206787968 (WISEA J193555.98$-$284635.0) 
is known to have an excess in W2 \citep{liu16}, and I find
that it has an excess in W3 as well. With a distance of 57 pc, it
is one of the closest known brown dwarfs with a circumstellar disk.
The remaining nine stars with IR excesses have early spectral types (A/F/G)
and have been found to have debris disks in previous studies
\citep{smi84,aum85,cot87,pat91,man98,zuc04b,che08,reb08,mcd12,cot16}.

Ten members of BPMG lack mid-IR excess emission but have been classified
as debris systems in previous work based on observations at longer wavelengths, 
which consist of 51 Eri, GJ 182, AF Lep, AG Tri, AU Mic, AT Mic, GJ2006A, 
HD160305, CPD$-$72 2713, and GSC 07396-00759
\citep{spa01,kal04,liu04,zuc04b,che05,car08,reb08,pla09,riv14,moo16,sis18,moo20,tan20,ada21,cro22,cro23}.
For most of the stars, disk emission has been spatially resolved or detected
at high significance at multiple wavelengths. \citet{pla09} reported an 
excess in [24] MIPS images for AT Mic, but an excess is not present in the 
[24] photometry measured by \citet{che05} or in the W4 photometry 
(Figure~\ref{fig:excbp}). Excess emission has been detected from AT Mic 
in a single millimeter band with a significance of 8~$\sigma$ \citep{cro23}. 
\citet{pat14} identified an excess in W4 for 51 Eri while none 
was found in [24] MIPS images by \citet{reb08} nor in my analysis of the W4 
data. Photometry of 51 Eri at 70 and 100~\micron\ from the Herschel Space
Observatory \citep{pil10} has detected small excesses with a significance of
$\sim4$~$\sigma$ \citep{riv14}. GJ 182 has single 4~$\sigma$ detection
of excess emission at 850~\micron\ \citep{liu04}. Given the lower significance
of their excess measurements, 51 Eri and GJ 182 are assigned disk classes
of ``debris?" in Table~\ref{tab:mem}.

The M star Gaia DR3 4067828843907821824 appears to have excess emission
in W4, but [24] images from MIPS reveal the presence of extended emission,
which likely contaminates the W4 measurement \citep{gut15}.
A color excess is not present in W1$-$[24].

\subsection{93 Tau}

As found in \citet{luh23tau}, seven members of 93 Tau have excesses
in WISE bands (Figure~\ref{fig:exc93}).
Five B/A/G stars and one M star have W4 excesses. The M star also has
excess emission in W2 and W3. An additional M star exhibits a small
excess in W3 and a nondetection in W4.

\subsection{Carina-Extended and Columba}

Forty members of Car-Ext and Columba have excess emission in the WISE data
(Figure~\ref{fig:exccar}), 12 of which have had excesses reported in previous 
work 
\citep{man98,spa01,rie05,rhe07,zuc11,mcd12,mcd17,zuc12,cot16,sil16,sgr21,gai22}.
Excesses are present in W4 for 13 B/A/F/G stars, five of which also have
small W3 excesses. The remaining 27 stars with excesses have M types based
on spectroscopy or photometry, which include three stars that were
known to have excess emission:
WISEA J080822.18$-$644357.3 \citep{sil16,mur18},
WISEA J065648.22$-$582511.4 \citep{sgr21}, and
WISEA J063207.99$-$681041.6 \citep{gai22}.
Ten of the M stars with excesses are underluminous in the Gaia CMDs
(Figure~\ref{fig:cmd2}).
Two M stars appear to have W3 excesses in Figure~\ref{fig:exccar}
but are not flagged for excesses because their W3 detections have very low 
signal-to-noise ratios (S/Ns). Two of the new M-type WISE sources with excesses,
WISEA J095041.19$-$714659.9 and WISEA J084828.32$-$594114.1,
are resolved as $2\arcsec$ binaries by Gaia. In each system, the excess
emission appears to originate from the secondary based on inspection
of the WISE images. Since the primaries dominate the unresolved W1 photometry
for each system, the excesses implied by W1$-$W3 and W1$-$W4 for those 
secondaries are underestimated.

\subsection{$\chi^1$ For}
\label{sec:chi}

Nineteen members of $\chi^1$ For have IR excess emission in the WISE data
(Figure~\ref{fig:excfor}), 15 of which have been previously identified
as showing evidence of disks 
\citep{zuc04b,moo06,rhe07,mcd12,rod13,wu13,cot16,zuc19,sil20,gal21}.
Seven B/A/F stars have excesses in W4, one of which also has a tentative
excess in W3.
Twelve M stars exhibit excess emission in W3, some of which also have
excesses in W2 or W4. The four new disks have M types based on spectroscopy 
or photometry, two of which are underluminous in the Gaia CMDs 
(Figure~\ref{fig:cmd2}).
The latter two stars have UV detections from the Galaxy Evolution 
Explorer \citep{bia17} and one of them, Gaia DR3 5093945085525624448,
has intense emission lines from hydrogen and other species in an
optical spectrum from \citet{pal20} and in archival optical and IR
data from the X-shooter spectrograph \citep{ver11} at the Very Large Telescope,
which were obtained through program 109.234F (I. Pelisoli).
\citet{zuc19} reported IR excesses in WISE bands for seven additional
members, but they lack excesses in my analysis.

The F star HD 23380 has photospheric emission in the WISE bands but has
an excess in 70~\micron\ data from Herschel \citep{zuc19}.

\subsection{Tuc-Hor}

Seventeen members of Tuc-Hor have excess emission in WISE photometry
(Figure~\ref{fig:exctuc}).
Evidence of disks has been previously reported for seven of those stars
\citep{oud92,smi06,zuc11,don12,pat14,bou16,cot16,sil16,hig22}.  
Eleven A/F/G stars have small excesses in W4 and two K stars have small 
excesses in W3 and nondetections in W4. 
Three M stars have excesses, which are new detections of disks.
One of the M stars, Gaia DR3 5291374733442587520, is underluminous in the Gaia 
CMDs.
The L dwarf 2MASS J02265658$-$5327032 has a small and tentative excess
in W3 and a marginal detection in W4 (S/N$\sim$3), but a previous detection
of Pa$\beta$ emission supports the presence of an accretion disk \citep{bou16}. 

Seven members of Tuc-Hor lack excess emission in WISE bands but exhibit it in 
70~\micron\ data from Spitzer and Herschel \citep{zuc11,don12}.

\subsection{IC 2602}

Excesses are detected by WISE for 17 members of IC 2602 
(Figure~\ref{fig:exc2602}), 10 of which have had evidence of disks in 
previous studies \citep{rie05,su06,mcd12,riz12}.
Twelve B/A/F stars have small excesses in W4, six of which also have small 
W3 excesses. The F star HD 91042 has a small excess in W3 and an unreliable
detection in W4 due to extended emission.  
Four M stars have excesses in W3, some of which also have excesses in W2 or W4.
All of these M stars are among the new detections of disks.
One of the M stars, Gaia DR3 5237543979881486720, appears below the
sequence for IC 2602 in the Gaia CMDs (Figure~\ref{fig:cmd3}).

\subsection{IC 2391}

Two members of IC 2391 have excess emission in the WISE bands 
(Figure~\ref{fig:exc2391}). The A star HD 72323 has a small excess in W4, 
which has been reported in previous work \citep{cot16}. 
The M star Gaia DR3 5318492744638910848 has large excesses in both W3 and W4. 
In addition, it appears within the IRAC and MIPS images from \citet{sie07}. 
I find that excesses are present in the [8.0] and [24] bands from those
observations. The excess emission for this star is newly identified in 
this work.

The Spitzer images from \citet{sie07} encompass a small fraction of
the members of IC 2391 near the cluster center.  That study reported [24]
excesses for seven of the stars in my catalog. 
In my analysis of the W1$-$[24] colors using [24] measurements
from the Spitzer Enhanced Imaging Products Source List, I find excesses 
for two of those stars, which have spectral types of A7 and F3. 
Both lack significant excesses in W4, which has lower S/N than [24]. 
For the seven stars, I have assigned the W4 excess flags in 
Table~\ref{tab:mem} based on my results from [24] and I have used [24] in
place of W4 in Figure~\ref{fig:exc2391}.

\subsection{NGC 2547}
\label{sec:ngc}

Previous studies have used photometry from Spitzer to search for IR excesses
among members of NGC 2547 \citep{you04,gor07,for08}.
Those studies have identified several early-type members with excesses
in [24] data from MIPS. I have not measured [24] photometry for three
of those stars (HD 68451, HD 68496 HD 68396) because their emission
is extended \citep{for08}, so they are absent from the W1$-$[24] diagram 
in Figure~\ref{fig:excngc}. The remaining early-type members form
a broad locus in W1$-$[24] in which I cannot reliably distinguish between
stellar photospheres and stars with small excesses. Therefore, 
I do not report a measurement for the [24] excess fraction for those
spectral types in NGC 2547 (Section~\ref{sec:fractions}).
Large [24] excesses are present among three M stars, consisting of
2MASS J08093547$-$4913033 (ID7 in \citet{gor07}),
2MASS J08090344$-$4859210, and 2MASS J08085407$-$4921046.
The excess for the first star was identified in all three of the previous
Spitzer studies. The disk emission from that star was studied in more
detail through mid-IR spectroscopy with Spitzer \citep{tei09b}.
\citet{gor07} noted the [24] excess for 2MASS J08090344$-$4859210, who
discussed the uncertainty in its membership in NGC 2547. I find that
the astrometry from Gaia support membership. The third M star,
2MASS J08085407$-$4921046, was found to have a [24] excess in \citet{for08}.
The S/N for its [24] detection is low ($\sim$5).

The G star 2MASS J08090250$-$4858172 (ID8 from \citet{gor07}) 
has a large excess in [24] and excess emission at shorter wavelengths
\citep{gor07}. Photometric monitoring with Spitzer has revealed significant
variability in the emission from the disk \citep{men12,su19}.
Previous studies have treated the star as a member of NGC 2547, but
I find that the proper motion and parallax from Gaia DR3
and the radial velocity \citep{sac15} are inconsistent with membership.
The value of RUWE from Gaia DR3 is low, so there is no indication that
the discrepancy in the Gaia astrometry is caused by a poor astrometric fit.

\subsection{Excess Fractions}
\label{sec:fractions}

It has been difficult to measure accurate excess fractions for low-mass
stars (0.1--0.5~$M_\odot$) at ages of $>$20 Myr because the associations
that are close enough for their low-mass stars to be detected by WISE
have had only small samples of reliable members, as in the case of BPMG.
Data from Gaia have now made it possible to significantly
expand those membership catalogs and improve their reliability, enabling
better constraints on the excess fractions for low-mass stars at older ages.

Disk surveys in clusters and associations typically report excess fractions
for individual IR bands or disk fractions based on excesses across multiple
bands. In this work, I calculate excess fractions for W3 and W4.
[24] data from Spitzer are included with W4 when available (Upper Sco, BPMG).
Excess/disk fractions are often measured separately for primordial and
debris disks, particularly at early spectral types where the disks are more
likely to have adequate data (e.g., gas tracers) for discrimination
between the two disk types. I do not attempt such discrimination in my
excess fractions given the uncertainty in the nature of the M star disks
in this study. I calculate excess fractions for the following three
ranges of spectral types: B/A/F, G/K, and M0--M6.

With the exception of unresolved companions, all of the adopted members of
the associations in this study are detected in W1 and W2, but
the WISE data are increasingly incomplete at longer wavelengths,
later spectral types, smaller excesses, and larger distances.
To be meaningful, the excess fractions should be calculated in a way
that accounts for the incompleteness.
Ideally, one would consider only spectral type ranges in which nearly
all members of a given association are detected for the band in question,
but I have adopted more relaxed constraints to allow the calculation of
excess fractions for more bands and spectral types.
Namely, if a star is not detected in a given band and the typical magnitude
limit indicates that any excess is smaller than $\sim$1 mag, the star is
counted as not having an excess. I have adopted limits of W3=11.5 and W4=8.0,
which correspond to S/N=5 in the AllWISE Source Catalog \citep{cut13b}.
To calculate an excess fraction, I require that at least 75\% of the
members in a given spectral type range are detected in the WISE band
or have the aforementioned limits on excesses. That percentage was $>$95\% 
for most of the excess fractions. Members that are not detected by
WISE and that lack useful limits are excluded from the excess fractions.
The W3 and W4 excess fractions for BPMG and other associations at 10--50 Myr 
are presented in Table~\ref{tab:frac}. In addition to the associations surveyed
in this work, I have included TWA \citep{luh23twa}, 32 Ori \citep{luh22o},
Upper Sco, UCL/LCC \citep{luh22disks}, and 93 Tau \citep{luh23tau}. 
The statistical errors in the excess fractions are taken from \citet{geh86}. 

The ages in Table~\ref{tab:frac} consist of the LDB values from 
Section~\ref{sec:ages} when available. I have adopted 10, 11, and 20 Myr for 
TWA, Upper Sco and UCL/LCC, respectively \citep{luh20u,luh23twa}, and 
20 Myr for Sco Body (Section~\ref{sec:ages}).

I do not report excess fractions for NGC 2547 at earlier types for
reasons mentioned in Section~\ref{sec:ngc}. Too few M-type members
are detected in W3 and [24] for calculating excess fractions.
The detection of three M stars with [24] excesses when the sensitivity
to M stars is poor suggests that their excess fraction could be high enough
to be consistent with those in the closer associations at similar ages
like Car-Ext.

\begin{figure}
\epsscale{1.2}
\plotone{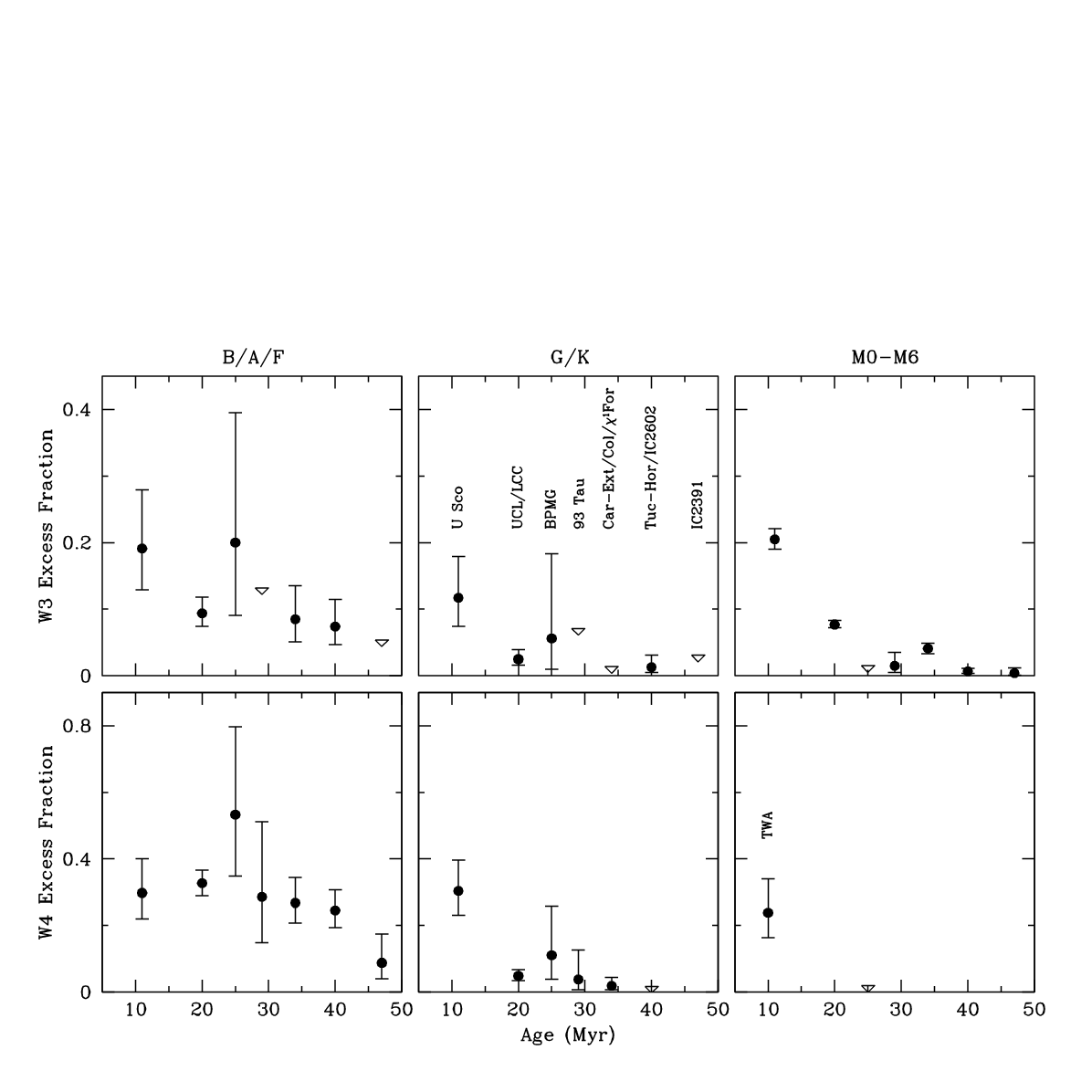}
\caption{Excess fractions in W3 and W4 versus age for BPMG and other young
associations (Table~\ref{tab:frac}).}
\label{fig:frac}
\end{figure}

Many studies have examined excess fractions as a function of age,
both for primordial disks and debris disks, in an attempt to study disk 
evolution, particularly at earlier spectral types 
\citep{hai01,rie05,her07,sie07,mey08,bal09,mam09,gas09,gas13,clo14,rib15,men17,ric18,che20}.
In Figure~\ref{fig:frac}, I have plotted excess fractions 
from Table~\ref{tab:frac} versus the ages of the associations.  For clarity, 
I have included only the fractions that have the smallest errors at a given 
age, which means omitting TWA, 32 Ori, and Sco Body in favor of Upper Sco and 
UCL/LCC in most of the diagrams in Figure~\ref{fig:frac}.  For W4 at M0--M6, 
only TWA and Tuc-Hor are close enough for a constraint on the excess fraction.
As a reminder, stars that lack excesses in the WISE bands can have excesses
at longer wavelengths, as mentioned for BPMG in Section~\ref{sec:bpexc}
(e.g., AU Mic).

Since some disks have excess emission in W4 but not W3, the former is
more sensitive to disks when it is available. 
Among B/A/F stars, the W4 excess fraction has a marginally significant
maximum in BPMG, is roughly constant for the other associations at 10--40 Myr, 
and is lower at the older age of IC 2391. Most of the disks for these 
early-type stars have been previously classified as debris disks.
At G/K types, a substantial number of the disks in Upper Sco
are likely to be primordial \citep{luh22disks} while debris disks should
dominate for the older associations. Among the G/K stars at $>$10 Myr,
a peak with marginal significance is once again found for BPMG.
For the M stars, the analysis of excess fractions versus age 
is restricted to W3. In that band, the excess fraction for M stars
exhibits a steady decline from Upper Sco to UCL/LCC to BPMG.
The latter does not have any M0--M6 members with W3 excesses, and
the resulting upper limit on their abundance is quite small.
However, the W3 fraction increases at older ages, reaching a peak in
Car-Ext/Columba/$\chi^1$ For, and drops to a lower level in the oldest two 
associations. This variation in an excess fraction for M stars at 25--50 Myr
has been observed for the first time in this study.
In their comparison of earlier membership samples for $\chi^1$ For
and Tuc-Hor (which they believed were coeval), \citet{zuc19} did note
that $\chi^1$ For appeared to have a significantly higher disk fraction,
providing a hint of the peak now detected in Car-Ext/Columba/$\chi^1$ For.
Meanwhile, BPMG is distinct from the other associations in that it
contains both the highest excess fraction for early-type stars and
the lowest fraction for M stars.

The W3 excess fractions for M stars are suggestive of two populations
of disks, decaying primordial disks at 10--20 Myr and a second class of disks
arising at older ages, possibly debris disks.  In this scenario,
the peak for M stars at 34 Myr could be a delayed version of the
peak among B/A/F stars at 25 Myr in BPMG.
That delay might be related to the longer lifetimes observed for primordial
disks of low-mass stars \citep{car06,luh22disks}.
However, several of the M stars with IR excesses are underluminous in CMDs,
indicating that they have blue excess emission from accretion or 
scattered light from edge-on disks, both of which would seem to 
require primordial disks rather than debris disks. In addition, some of these
stars have detections of UV emission and strong hydrogen emission lines
(Section~\ref{sec:chi}), which are not observed in debris disks 
\citep{wya08,mat14,hug18}.
WISEA J080822.18$-$644357 has one of the few disks among older M stars 
that has been studied in detail. Hydrogen emission lines
indicate the presence of accretion \citep{mur18} but millimeter observations 
of CO and dust emission are more suggestive of a debris disk \citep{fla19}.

Given the relatively large number of M star disks that are now available 
at ages of $>$20 Myr, it is possible to perform systematic studies of their
nature. However, most of these systems have few data available beyond 
Gaia and WISE. Studies of the disks would benefit from spectral
classifications of the stars, measurements of radial velocities for
membership confirmation, observations of gas and accretion tracers,
and millimeter continuum measurements for estimating dust
temperatures and masses. Observations at high angular resolution would also 
help to constrain the disk radii and better assess whether background objects 
are responsible for mid-IR excesses detected by WISE.

\section{Conclusions}

I have sought to improve the reliability and completeness of 
membership samples in BPMG and other nearby associations 
with ages of 20--50 Myr using photometry, astrometry, and radial velocities
from Gaia DR3 and data from other sources, including new
and archival spectra.
I have used the new catalogs to study these associations in terms
of their IMFs, X-ray emission, ages, and circumstellar disks.
The results are summarized as follows:

\begin{enumerate}

\item
The Carina association as defined in previous studies is spatially
and kinematically continuous with more distant populations, such
as the ``a Carinae" cluster (Platais 8). I have treated these stars as
a single association, which I have named Carina-Extended (Car-Ext). 
Columba, $\chi^1$ For, and Car-Ext are spatially adjacent and have similar
kinematics and ages, indicating that they are physically related.  
Similarly, I find that Tuc-Hor and IC 2602 form a coeval population that 
is spatially and kinematically continuous.  These results are consistent 
with hypotheses from \citet{gag21}.

\item
In the catalogs presented in this work, the numbers of adopted members are
193 in BPMG, 299 in Sco Body, 845 in Car-Ext, 51 in Columba, 
287 in $\chi^1$ For, 815 in Tuc-Hor, 516 in IC 2602, 486 in IC 2391,
and 300 in NGC 2547. Most of these catalogs are significantly larger
than previous samples of likely members (a factor of $\sim$2--4 for BPMG).

\item
The distances of the BPMG members range from 10 to 105 pc and have a median
value of 51 pc. The catalog includes a few objects that have non-Gaia 
parallaxes, such as the L dwarf PSO J318.5338$-$22.8603 \citep{liu13},
and several probable companions that lack parallax measurements.
One of the adopted members that has new spectroscopy
is Gaia DR3 6652273015676968832, which is classified as an early L dwarf,
making it the second coolest object in my catalog for BPMG.
$UVW$ velocities have been calculated for 134 adopted members, which
show evidence of expansion in $X$ and $Y$.
Among the 193 BPMG members, nine lack spectral classifications (seven are close
companions) and 58 lack accurate radial velocity measurements.
Their membership could be better constrained with measurements of
spectral signatures of youth and radial velocities.

\item
I classify four previous BPMG candidates as members of 32 Ori, which
include BD+45 598 and 2MASS J02495639$-$0557352. The former has an 
edge-on disk \citep{hin21} and the latter has a substellar companion
\citep{dup18} with strong H$\alpha$ emission \citep{chi21}.

\item
The proposed members of Argus from \citet{zuc19c} do not exhibit
the tight clustering of $UVW$ velocities expected for a young association. 
Also, the low-mass stars in that sample do not form a well-defined sequence
in Gaia CMDs. Roughly 20\% of those stars have kinematics and CMD positions
that are consistent with membership in the extended population associated with 
IC 2391 while the remaining candidates appear to be unrelated field stars.

\item
I have constructed histograms of spectral types based on spectroscopy
and photometry for BPMG and the other associations in this study.
For each association, the distribution reaches a maximum near M5
($\sim0.15$~$M_\odot$), which resembles data for other nearby associations
and star-forming regions.

\item
For the each association, I have compiled the available X-ray counterparts 
from eROSITA's eRASS1 catalog. As found in previous X-ray studies of
young stars, the ratio of X-ray and optical fluxes exhibits
a well-defined distribution as a function of optical color that is similar
among the associations. Within the half of the sky covered by eRASS1,
none of the adopted members of the associations have X-ray 
detections or limits that would indicate that they are too old to be members.

\item
I have used the EAGLES model for Li depletion \citep{jef23} to estimate
LDB ages for associations surveyed in this study and others that
are included in the disk analysis (32 Ori, 93 Tau). The resulting ages are 
21.0$^{+1.0}_{-0.7}$ (32 Ori), 24.7$^{+0.9}_{-0.6}$ (BPMG),
29.3$\pm$1 (93 Tau), 33.7$^{+2.0}_{-1.9}$ (Car-Ext/Columba/$\chi^1$ For),
34.9$\pm$1.2 (NGC 2547), 40.0$^{+1.9}_{-1.4}$ (Tuc-Hor/IC 2602), and 
51.0$^{+5.6}_{-3.9}$ Myr (IC 2391). The LDB age for IC 2391 is
poorly constrained, and its CMDs relative to those of Tuc-Hor/IC 2602 
suggests an age $\sim47$ Myr. A similar comparison to the CMDs of 32 Ori
indicates an age of $\sim20$ for Sco Body.

\item
I have used mid-IR photometry from WISE to check for excess emission
from circumstellar disks among the members of the associations.
Excesses are detected in the WISE photometry for 125 sources, 71 of
which have had evidence of disks in previous studies.
Most notably, this work has dramatically increased the number of known
M stars at ages of 30--50 Myr that exhibit IR excesses. For instance,
39 stars of this kind are present in Car-Ext/Columba/$\chi^1$ For.
In addition, I have performed similar analysis of Spitzer data in NGC 2547,
where I have recovered previously reported excess emission
for three M stars \citep{you04,gor07,for08}.

\item
Because of their sizes and reliability, the new membership samples
have enabled significantly improved measurements of IR excess fractions
for the associations in question, which is particularly important for
constraining the presence of disks among older M stars.
I have calculated excess fractions in W3 and W4 for each association
in three ranges of spectral types (B/A/F, G/K, M0--M6), as allowed
by the sensitivity of the WISE data.
I have included excess fractions measured from previous disk surveys
in other nearby associations with ages of 10--30 Myr, consisting of
TWA, 32 Ori, Upper Sco, UCL/LCC, and 93 Tau
\citep{luh22disks,luh22o,luh23tau,luh23twa}.
I have examined the excess fractions as a function of age.
Among B through K stars, the W4 excess fraction has a marginally significant
maximum in BPMG. Most of those disks have been previously classified as debris
disks. For the M stars, W4 excess fractions can be calculated in only
a few associations, but accurate fractions are available in W3.
The W3 excess fraction for M stars exhibits a steady decline from Upper Sco
to UCL/LCC to BPMG, where the latter has a stringent upper limit ($<$0.015).
The W3 fraction then increases at older ages, reaching a peak in
Car-Ext/Columba/$\chi^1$ For ($0.041^{+0.009}_{-0.007}$), and drops
to a lower level in the oldest two associations. This peak has not
been been previously detected. Its origin and the nature of the M star disks
at $>$20 Myr (primordial vs. debris) are unclear.

\end{enumerate}

\begin{acknowledgments}

I thank Robin Jeffries for advice regarding the use of his lithium
depletion model and I thank Konstantin Getman for helpful discussions. 
This work used data from the European Space Agency 
mission Gaia (\url{https://www.cosmos.esa.int/gaia}), processed by
the Gaia Data Processing and Analysis Consortium (DPAC,
\url{https://www.cosmos.esa.int/web/gaia/dpac/consortium}). Funding
for the DPAC has been provided by national institutions, in particular
the institutions participating in the Gaia Multilateral Agreement. 
The IRTF is operated by the University of Hawaii under contract 80HQTR19D0030
with NASA. The data at SOAR and CTIO were obtained through programs 
2020B-0049, 2022A-595242, and 2024A-657954 at NOIRLab. 
CTIO and NOIRLab are operated by the Association of Universities for 
Research in Astronomy under a cooperative agreement with the NSF. 
SOAR is a joint project of the Minist\'{e}rio da Ci\^{e}ncia, Tecnologia e 
Inova\c{c}\~{o}es do Brasil, the US National Science Foundation's 
NOIRLab, the University of North Carolina at Chapel Hill, and Michigan 
State University.
The Gemini data were obtained through programs GS-2012B-Q-70, GS-2012B-Q-92, 
GS-2013A-Q-66, GS-2013B-Q-83, GS-2014A-Q-55, and GS-2024A-FT-109.
Gemini Observatory is a program of NSF's NOIRLab, which is managed by the
Association of Universities for Research in Astronomy (AURA) under a
cooperative agreement with the National Science Foundation on behalf of the
Gemini Observatory partnership: the National Science Foundation (United States),
National Research Council (Canada), Agencia Nacional de Investigaci\'{o}n y
Desarrollo (Chile), Ministerio de Ciencia, Tecnolog\'{i}a e Innovaci\'{o}n
(Argentina), Minist\'{e}rio da Ci\^{e}ncia, Tecnologia, Inova\c{c}\~{o}es e
Comunica\c{c}\~{o}es (Brazil), and Korea Astronomy and Space Science Institute
(Republic of Korea). 2MASS is a joint project of the University of 
Massachusetts and IPAC at Caltech, funded by NASA and the NSF.
WISE is a joint project of the University of California, Los Angeles,
and the JPL/Caltech, funded by NASA.
The Spitzer Space Telescope was operated by JPL/Caltech under
contract with NASA. This work used data from the 
NASA/IPAC Infrared Science Archive, operated by Caltech under contract
with NASA, and the VizieR catalog access tool and the SIMBAD database, 
both operated at CDS, Strasbourg, France.
SRG is a joint Russian-German science mission supported by the Russian Space
Agency (Roskosmos), in the interests of the Russian Academy of Sciences
represented by its Space Research Institute (IKI), and the Deutsches Zentrum
f\"{u}r Luft- und Raumfahrt (DLR). The SRG spacecraft was built by Lavochkin
Association (NPOL) and its subcontractors, and is operated by NPOL with support
from the Max Planck Institute for Extraterrestrial Physics (MPE).
The development and construction of the eROSITA X-ray instrument was led by
MPE, with contributions from the Dr. Karl Remeis Observatory Bamberg \& ECAP
(FAU Erlangen-Nuernberg), the University of Hamburg Observatory, the
Leibniz Institute for Astrophysics Potsdam (AIP), and the Institute
for Astronomy and Astrophysics of the University of T\"{u}bingen, with
the support of DLR and the Max Planck Society. The Argelander Institute for
Astronomy of the University of Bonn and the Ludwig Maximilians Universität
Munich also participated in the science preparation for eROSITA.
The Center for Exoplanets and Habitable Worlds is supported by the
Pennsylvania State University, the Eberly College of Science, and the
Pennsylvania Space Grant Consortium.

\end{acknowledgments}

\clearpage

\begin{deluxetable}{llll}
\tabletypesize{\scriptsize}
\tablewidth{0pt}
\tablecaption{Summary of Spectroscopic Observations\label{tab:log}}
\tablehead{
\colhead{Telescope/Instrument} &
\colhead{Mode/Aperture} &
\colhead{Wavelengths/Resolution} &
\colhead{Targets}}
\startdata
IRTF/SpeX & prism/$0\farcs8$ slit & 0.8--2.5~\micron/R=150 & 56 \\
CTIO 4~m/COSMOS & red VPH/$1\farcs2$ slit & 0.55--0.95~\micron/4 \AA & 36 \\
CTIO 1.5~m/RC Spec & 47 Ib/$2\arcsec$ slit & 0.56--0.69~\micron/3 \AA & 2 \\
CTIO 1.5~m/RC Spec & 58 I/$2\arcsec$ slit & 0.62--0.88~\micron/6 \AA & 1 \\
Gemini South/GMOS & R400/$0\farcs5$ & 0.52--0.91 \micron/4 \AA & 4 \\
Gemini South/GMOS & R400/$0\farcs75$ & 0.6--1 \micron/5 \AA & 6 \\
Gemini South/GMOS & R400/$1\arcsec$ & 0.6--1 \micron/7 \AA & 2 \\
Gemini South/FLAMINGOS-2 & JH/HK/$0\farcs72$ & 0.9--2.5 \micron/R=700 & 1 \\
SOAR/Goodman & 400/$0\farcs45$ & 0.6--0.9 \micron/3 \AA & 4 \\
SOAR/Goodman & 600/$1\arcsec$ & 0.63--0.9 \micron/4 \AA & 40
\enddata
\end{deluxetable}

\begin{deluxetable}{ll}
\tabletypesize{\scriptsize}
\tablewidth{0pt}
\tablecaption{Spectroscopic Data for Adopted Members of BPMG and Other Associations\label{tab:spec}}
\tablehead{
\colhead{Column Label} &
\colhead{Description}}
\startdata
Gaia & Gaia DR3 source name \\
RAdeg & Gaia DR3 right ascension (ICRS at Epoch 2016.0)\\
DEdeg & Gaia DR3 declination (ICRS at Epoch 2016.0)\\
SpType & Spectral type\tablenotemark{a}\\
Instrument & Instrument for spectroscopy\\
Date & Date of spectroscopy
\enddata
\tablenotetext{a}{Uncertainties are 0.25 and 0.5~subclass for optical and
IR spectral types, respectively, unless indicated otherwise.}
\tablecomments{
The table is available in its entirety in machine-readable form.}
\end{deluxetable}

\clearpage

\startlongtable
\begin{deluxetable}{ll}
\tabletypesize{\scriptsize}
\tablewidth{0pt}
\tablecaption{Adopted Members of BPMG, Sco Body, Car-Ext, Columba, $\chi^1$ For,
Tuc-Hor, IC 2602, and IC 2391\label{tab:mem}}
\tablehead{
\colhead{Column Label} &
\colhead{Description}}
\startdata
GaiaDR3 & Gaia DR3 source name \\
Name & Other source name \\
RAdeg & Gaia DR3 right ascension (ICRS at Epoch 2016.0)\tablenotemark{a}\\
DEdeg & Gaia DR3 declination (ICRS at Epoch 2016.0)\tablenotemark{a}\\
SpType & Spectral type \\
r\_SpType & Spectral type reference\tablenotemark{b} \\
Adopt & Adopted spectral type \\
f\_EWLi & Flag for EWLi \\
EWLi & Equivalent width of Li \\
e\_EWLi & Error in EWLi\tablenotemark{c} \\
r\_EWLi & Li reference\tablenotemark{c} \\
pmRA & Gaia DR3 proper motion in right ascension\tablenotemark{a}\\
e\_pmRA & Error in pmRA \\
pmDec & Gaia DR3 proper motion in declination\tablenotemark{a}\\
e\_pmDec & Error in pmDec \\
plx & Gaia DR3 parallax\tablenotemark{a}\\
e\_plx & Error in plx \\
rmedgeo & Median of geometric distance posterior \citep{bai21}\tablenotemark{a}\\
rlogeo & 16th percentile of geometric distance posterior \citep{bai21}\tablenotemark{a}\\
rhigeo & 84th percentile of geometric distance posterior \citep{bai21}\tablenotemark{a}\\
RVel & Radial velocity \\
e\_RVel & Error in RVel \\
r\_RVel & Radial velocity reference\tablenotemark{d} \\
U & $U$ component of space velocity \\
e\_U & Error in U \\
V & $V$ component of space velocity \\
e\_V & Error in V \\
W & $W$ component of space velocity \\
e\_W & Error in W \\
Gmag & Gaia DR3 $G$ magnitude\\
e\_Gmag & Error in Gmag \\
GBPmag & Gaia DR3 $G_{\rm BP}$ magnitude\\
e\_GBPmag & Error in GBPmag \\
GRPmag & Gaia DR3 $G_{\rm RP}$ magnitude\\
e\_GRPmag & Error in GRPmag \\
RUWE & Gaia DR3 renormalized unit weight error\\
2m & Closest 2MASS source within $3\arcsec$ \\
2msep & Angular separation between Gaia DR3 (epoch 2000) and 2MASS \\
2mclosest & Is this Gaia source the closest match for the 2MASS source? \\
wise & Closest WISE source within $3\arcsec$\tablenotemark{e} \\
wisesep & Angular separation between Gaia DR3 (epoch 2010.5) and WISE \\
wiseclosest & Is this Gaia source the closest match for the WISE source?\\
Jmag & 2MASS $J$ magnitude \\
e\_Jmag & Error in Jmag \\
Hmag & 2MASS $H$ magnitude \\
e\_Hmag & Error in Hmag \\
Ksmag & 2MASS $K_s$ magnitude \\
e\_Ksmag & Error in Ksmag \\
W1mag & WISE W1 magnitude \\
e\_W1mag & Error in W1mag \\
W2mag & WISE W2 magnitude \\
e\_W2mag & Error in W2mag \\
f\_W2mag & Flag on W2mag\tablenotemark{f} \\
W3mag & WISE W3 magnitude \\
e\_W3mag & Error in W3mag \\
f\_W3mag & Flag on W3mag\tablenotemark{f} \\
W4mag & WISE W4 magnitude \\
e\_W4mag & Error in W4mag \\
f\_W4mag & Flag on W4mag\tablenotemark{f} \\
ExcW2 & Excess in W2? \\
ExcW3 & Excess in W3? \\
ExcW4 & Excess in W4? \\
DiskType & Disk type \\
erass1 & eRASS1 source name \\
poserr & Position error from eRASS1 1B Catalog \\
xsep & Angular separation between Gaia DR3 (epoch 2020) and eRASS1 \\
f\_rate & Flag for rate \\
rate & Count rate from eRASS1 1B Catalog \\
e\_rate & Error in rate \\
association & Association
\enddata
\tablenotetext{a}{The astrometry and distance for HD 161460 are from 
Gaia DR2 and \citet{bai18}. Those parameters are from \citet{liu16} for 
PSO J318.5338-22.8603.  The proper motion, parallax, and distance for 
GJ 3076 are from \citet{rie14}.}
\tablenotetext{b}{
(1) \citet{hou88};
(2) \citet{can93};
(3) \citet{tor06};
(4) \citet{pec13};
(5) \citet{ria06};
(6) \citet{rie17};
(7) this work;
(8) \citet{kra14};
(9) \citet{rie14};
(10) \citet{the74};
(11) \citet{shk17};
(12) \citet{bow19};
(13) \citet{ter15};
(14) \citet{gai14};
(15) \citet{rei95};
(16) \citet{new14};
(17) \citet{rei07};
(18) \citet{shk09};
(19) \citet{zuc04};
(20) \citet{vys56};
(21) \citet{schl12b};
(22) \citet{giz00};
(23) \citet{all13};
(24) \citet{zic98};
(25) \citet{cow69};
(26) \citet{gra89};
(27) \citet{hou99};
(28) \citet{ste86};
(29) \citet{yee10};
(30) \citet{cru03};
(31) \citet{kir08};
(32) \citet{des12};
(33) \citet{lep13};
(34) \citet{gra06};
(35) \citet{schl12c};
(36) \citet{rob84};
(37) \citet{hen94};
(38) \citet{alc96};
(39) \citet{gag15c};
(40) \citet{sch19};
(41) \citet{hou78};
(42) \citet{son03};
(43) \citet{hou75};
(44) \citet{kra97};
(45) \citet{pas89};
(46) \citet{mar95};
(47) \citet{hou82};
(48) \citet{sta22};
(49) \citet{pha17};
(50) \citet{nes95};
(51) \citet{gra87};
(52) \citet{low00};
(53) \citet{gue01};
(54) \citet{rei08};
(55) \citet{mar10};
(56) \citet{fah16};
(57) \citet{lep09};
(58) \citet{sch07};
(59) \citet{lee22};
(60) \citet{kee89};
(61) \citet{pri14};
(62) \citet{neu02};
(63) \citet{liu13};
(64) \citet{shk12};
(65) \citet{riz15};
(66) \citet{pec16};
(67) \citet{mam02};
(68) \citet{the94};
(69) \citet{pec12};
(70) \citet{gai22};
(71) \citet{sil16};
(72) \citet{mur18};
(73) \citet{hil69};
(74) \citet{cri05};
(75) \citet{cru07};
(76) \citet{cru18};
(77) \citet{abt95};
(78) \citet{sto91};
(79) \citet{gal21};
(80) \citet{rod13};
(81) \citet{jac55};
(82) \citet{cor84};
(83) \citet{sil20};
(84) \citet{koe17};
(85) \citet{bar17};
(86) \citet{kir11};
(87) \citet{gra17};
(88) \citet{cru09};
(89) \citet{pha06};
(90) \citet{kir06};
(91) \citet{bow15};
(92) \citet{gar94};
(93) \citet{art15};
(94) \citet{lod05};
(95) \citet{kir10};
(96) \citet{cow72};
(97) \citet{luh07};
(98) \citet{dob10};
(99) \citet{whi61};
(100) \citet{bou09};
(101) \citet{bar04};
(102) \citet{per69};
(103) \citet{pat96};
(104) \citet{bus65};
(105) \citet{gah83}.}
\tablenotetext{c}{
(1) \citet{tor06};
(2) this work;
(3) \citet{mal14b};
(4) \citet{shk17};
(5) \citet{kis11};
(6) \citet{mal13};
(7) \citet{bin16};
(8) \citet{men08};
(9) \citet{rei02};
(10) \citet{zic98};
(11) \citet{bow19};
(12) \citet{son03};
(13) \citet{das09};
(14) \citet{yee10};
(15) \citet{sta22};
(16) \citet{alc96};
(17) \citet{moo13};
(18) \citet{whi07};
(19) \citet{kra14};
(20) \citet{son02};
(21) \citet{gai22};
(22) \citet{woo23};
(23) \citet{mur18};
(24) \citet{rod13};
(25) \citet{pha17};
(26) \citet{hou23};
(27) \citet{pec16};
(28) \citet{nis22};
(29) \citet{jef23};
(30) \citet{gut20};
(31) \citet{dob10};
(32) \citet{bar04}.}
\tablenotetext{d}{
(1) Gaia DR3;
(2) \citet{mir20};
(3) \citet{mal14a};
(4) \citet{shk17};
(5) \citet{fou18};
(6) \citet{shk12};
(7) \citet{mac15};
(8) \citet{fah16};
(9) \citet{sou18};
(10) \citet{zun21a};
(11) \citet{laf20};
(12) \citet{sch19};
(13) \citet{gon06};
(14) \citet{str00};
(15) \citet{apo17};
(16) \citet{qua00};
(17) \citet{tor06};
(18) \citet{sta22};
(19) \citet{bud21};
(20) \citet{bur15};
(21) \citet{all16};
(22) \citet{kra14};
(23) Gaia DR2;
(24) \citet{bro22};
(25) \citet{gai22};
(26) \citet{mur18};
(27) \citet{moo13};
(28) \citet{igl18};
(29) \citet{whi07};
(30) \citet{rod13};
(31) \citet{gal21};
(32) \citet{gal23};
(33) \citet{bou16};
(34) \citet{hou23};
(35) \citet{nis22};
(36) \citet{mer09};
(37) \citet{ran22};
(38) \citet{ste20};
(39) \citet{ran18};
(40) \citet{pou04};
(41) \citet{jac20};
(42) \citet{bor23}.}
\tablenotetext{e}{
Source name from the AllWISE Source Catalog, the AllWISE Reject
Catalog, or the WISE All-Sky Source Catalog.}
\tablenotetext{f}{nodet = nondetection; false = detection from
WISE appears to be false or unreliable based on visual inspection.}
\tablecomments{
The table is available in its entirety in machine-readable form.}
\end{deluxetable}
\onecolumngrid

\clearpage

\begin{deluxetable}{ll}
\tabletypesize{\scriptsize}
\tablewidth{0pt}
\tablecaption{Previous Candidates Excluded from BPMG Catalog\label{tab:rej}}
\tablehead{
\colhead{Column Label} &
\colhead{Description}}
\startdata
ID & ID in Figures~\ref{fig:pp} and \ref{fig:uvw} \\
GaiaDR3 & Gaia DR3 source name \\
Name & Other source name \\
RAdeg & Gaia DR3 right ascension (ICRS at Epoch 2016.0)\tablenotemark{a}\\
DEdeg & Gaia DR3 declination (ICRS at Epoch 2016.0)\tablenotemark{a}\\
pmRA & Gaia DR3 proper motion in right ascension\tablenotemark{a}\\
e\_pmRA & Error in pmRA \\
pmDec & Gaia DR3 proper motion in declination\tablenotemark{a}\\
e\_pmDec & Error in pmDec \\
plx & Gaia DR3 parallax\tablenotemark{a} \\
e\_plx & Error in plx \\
rmedgeo & Median of geometric distance posterior \citep{bai21}\tablenotemark{a}\\
rlogeo & 16th percentile of geometric distance posterior \citep{bai21}\tablenotemark{a}\\
rhigeo & 84th percentile of geometric distance posterior \citep{bai21}\tablenotemark{a}\\
RVel & Radial velocity \\
e\_RVel & Error in RVel \\
r\_RVel & Radial velocity reference\tablenotemark{b} \\
U & $U$ component of space velocity \\
e\_U & Error in U \\
V & $V$ component of space velocity \\
e\_V & Error in V \\
W & $W$ component of space velocity \\
e\_W & Error in W \\
Gmag & Gaia DR3 $G$ magnitude\\
e\_Gmag & Error in Gmag \\
GBPmag & Gaia DR3 $G_{\rm BP}$ magnitude\\
e\_GBPmag & Error in GBPmag \\
GRPmag & Gaia DR3 $G_{\rm RP}$ magnitude\\
e\_GRPmag & Error in GRPmag \\
RUWE & Gaia DR3 renormalized unit weight error
\enddata
\tablenotetext{a}{The astrometry and distance for CD-27 11535 are from 
Gaia DR2 and \citet{bai18}.}
\tablenotetext{b}{
(1) Gaia DR3;
(2) \citet{gon06};
(3) \citet{bai12};
(4) \citet{sou18};
(5) \citet{ric10};
(6) \citet{shk17};
(7) \citet{fou18};
(8) \citet{sch19};
(9) \citet{mir20};
(10) \citet{tok22};
(11) \citet{jon20};
(12) \citet{ell14}.}
\tablecomments{
The table is available in its entirety in machine-readable form.}
\end{deluxetable}

\clearpage

\begin{deluxetable}{ll}
\tabletypesize{\scriptsize}
\tablewidth{0pt}
\tablecaption{Adopted Members of NGC 2547\label{tab:ngc}}
\tablehead{
\colhead{Column Label} &
\colhead{Description}}
\startdata
GaiaDR3 & Gaia DR3 source name \\
Name & Other source name \\
RAdeg & Gaia DR3 right ascension (ICRS at Epoch 2016.0)\\
DEdeg & Gaia DR3 declination (ICRS at Epoch 2016.0)\\
SpType & Spectral type \\
r\_SpType & Spectral type reference\tablenotemark{a} \\
Adopt & Adopted spectral type \\
pmRA & Gaia DR3 proper motion in right ascension\\
e\_pmRA & Error in pmRA \\
pmDec & Gaia DR3 proper motion in declination\\
e\_pmDec & Error in pmDec \\
plx & Gaia DR3 parallax \\
e\_plx & Error in plx \\
rmedgeo & Median of geometric distance posterior \citep{bai21}\\
rlogeo & 16th percentile of geometric distance posterior \citep{bai21}\\
rhigeo & 84th percentile of geometric distance posterior \citep{bai21}\\
RVel & Radial velocity \\
e\_RVel & Error in RVel \\
r\_RVel & Radial velocity reference\tablenotemark{b} \\
Gmag & Gaia DR3 $G$ magnitude\\
e\_Gmag & Error in Gmag \\
GBPmag & Gaia DR3 $G_{\rm BP}$ magnitude\\
e\_GBPmag & Error in GBPmag \\
GRPmag & Gaia DR3 $G_{\rm RP}$ magnitude\\
e\_GRPmag & Error in GRPmag \\
RUWE & Gaia DR3 renormalized unit weight error\\
wise & AllWISE source name \\
W1mag & WISE W1 magnitude \\
e\_W1mag & Error in W1mag \\
W2mag & WISE W2 magnitude \\
e\_W2mag & Error in W2mag \\
W3mag & WISE W3 magnitude \\
e\_W3mag & Error in W3mag \\
f\_W3mag & Flag on W3mag\tablenotemark{c} \\
W4mag & WISE W4 magnitude \\
e\_W4mag & Error in W4mag \\
f\_W4mag & Flag on W4mag\tablenotemark{c} \\
24mag & Spitzer [24] magnitude \\
e\_24mag & Error in 24mag \\
Exc24 & Excess in [24]?
\enddata
\tablenotetext{a}{
(1) \citet{jef03};
(2) \citet{can93};
(3) \citet{har76};
(4) \citet{hou78};
(5) \citet{jef05};
(6) \citet{tei09b}.}
\tablenotetext{b}{
(1) Gaia DR3;
(2) \citet{hou23};
(3) \citet{jac20};
(4) \citet{gon06}.}
\tablenotetext{c}{nodet = nondetection; false = detection from
AllWISE appears to be false or unreliable based on visual inspection.}
\tablecomments{
The table is available in its entirety in machine-readable form.}
\end{deluxetable}

\clearpage

\movetabledown=2in
\begin{rotatetable}
\begin{deluxetable}{lllllllll}
\tabletypesize{\scriptsize}
\tablewidth{0pt}
\tablecaption{Excess Fractions for Associations with Ages of 10--50 Myr\label{tab:frac}}
\tablehead{
\colhead{Association} &
\colhead{Age\tablenotemark{a}} &
\multicolumn{3}{c}{W3 Excess Fractions} & &
\multicolumn{3}{c}{W4 Excess Fractions}\\
\cline{3-5} \cline{7-9}
\colhead{} & \colhead{(Myr)} & \colhead{B/A/F} & \colhead{G/K} & \colhead{M0--M6} & & \colhead{B/A/F} & \colhead{G/K} & \colhead{M0--M6}}
\startdata
     TWA & 10 & 1/2 = $0.500^{+1.150}_{-0.413}$ & 0/1 & 10/50 = $0.200^{+0.085}_{-0.062}$ &  & 1/2 = $0.500^{+1.150}_{-0.413}$ & 0/1 & 10/42 = $0.238^{+0.102}_{-0.074}$ \\
   U Sco & 11 & 9/47 = $0.191^{+0.087}_{-0.063}$ & 7/60 = $0.117^{+0.063}_{-0.043}$ & 184/899 = $0.205^{+0.016}_{-0.015}$ &  & 14/47 = $0.298^{+0.103}_{-0.079}$ & 17/56 = $0.304^{+0.093}_{-0.073}$ & \nodata \\
 UCL/LCC & 20 & 23/245 = $0.094^{+0.024}_{-0.019}$ & 7/278 = $0.025^{+0.014}_{-0.009}$ & 234/3024 = $0.077\pm0.005$ &  & 80/245 = $0.327^{+0.041}_{-0.036}$ & 12/244 = $0.049^{+0.019}_{-0.014}$ & \nodata \\
Sco Body & 20 & 4/19 = $0.211^{+0.166}_{-0.101}$ & 1/27 = $0.037^{+0.085}_{-0.031}$ & 8/199 = $0.040^{+0.020}_{-0.014}$ &  & 7/19 = $0.368^{+0.198}_{-0.136}$ & 1/24 = $0.042^{+0.096}_{-0.034}$ & \nodata \\
  32 Ori & 21 & 1/15 = $0.067^{+0.153}_{-0.055}$ & 0/13 = $<$0.142 & 10/121 = $0.083^{+0.035}_{-0.026}$ &  & 4/16 = $0.250^{+0.198}_{-0.120}$ & 0/13 = $<$0.142 & \nodata \\
    BPMG & 25 & 3/15 = $0.200^{+0.195}_{-0.109}$ & 1/18 = $0.056^{+0.128}_{-0.046}$ & 0/126 = $<$0.015 &  & 8/15 = $0.533^{+0.263}_{-0.185}$ & 2/18 = $0.111^{+0.147}_{-0.072}$ & 0/98 = $<$0.019 \\
  93 Tau & 29 & 0/14 = $<$0.132 & 0/26 = $<$0.071 & 2/131 = $0.015^{+0.020}_{-0.010}$ &  & 4/14 = $0.286^{+0.226}_{-0.137}$ & 1/26 = $0.038^{+0.088}_{-0.032}$ & \nodata \\
Car-Ext/Col/$\chi^1$ For & 34 & 6/71 = $0.085^{+0.050}_{-0.034}$ & 0/132 = $<$0.014 & 31/764 = $0.041^{+0.009}_{-0.007}$ &  & 19/71 = $0.268^{+0.077}_{-0.061}$ & 2/106 = $0.019^{+0.025}_{-0.012}$ & \nodata \\
Tuc-Hor/IC 2602 & 40 & 7/94 = $0.074^{+0.040}_{-0.027}$ & 2/150 = $0.013^{+0.018}_{-0.009}$ & 6/835 = $0.007^{+0.004}_{-0.003}$ &  & 23/94 = $0.245^{+0.062}_{-0.051}$ & 0/111 = $<$0.017 & \nodata \\
 IC 2391 & 47 & 0/34 = $<$0.054 & 0/59 = $<$0.031 & 1/275 = $0.004^{+0.008}_{-0.003}$ &  & 3/34 = $0.088^{+0.086}_{-0.048}$ & \nodata & \nodata 
\enddata
\tablenotetext{a}{Sources of ages are described in Section~\ref{sec:fractions}.}
\end{deluxetable}
\end{rotatetable}

\clearpage

\clearpage

\end{document}